\documentclass[12pt]{article}
\usepackage{floatfig,epsfig,amsmath,amscd,amssymb,wrapfloat,supertabular}
\begin{document}
\pagestyle{empty}
\baselineskip=0.212in
\renewcommand{\theequation}{\arabic{section}.\arabic{equation}}
\renewcommand{\thefigure}{\arabic{section}.\arabic{figure}}
\renewcommand{\thetable}{\arabic{section}.\arabic{table}}

\initfloatingfigs

\begin{flushleft}
\normalsize 
{SAGA-HE-97-97, DOE/ER/40561-255-INT96-19-01 
\hfill Jan. 15, 1998}  \\
\end{flushleft}
 
\vspace{1.0cm}

\begin{center}

\Large{{\bf Flavor Asymmetry of}} \\

\vspace{0.3cm}

\Large{{\bf Antiquark Distributions in the Nucleon }} \\
 
\vspace{0.8cm}

\Large
{S. Kumano $^*$}         \\
 
\vspace{0.2cm}

{Department of Physics, Saga University, Saga 840, Japan} \\

\vspace{0.0cm}

{and}

\vspace{0.0cm}

{Institute for Nuclear Theory, University of Washington} \\

{Seattle, WA 98195, U.S.A.}

\vspace{1.2cm}

\Large{ABSTRACT}
  
\end{center}

Violation of the Gottfried sum rule was suggested by
the New Muon Collaboration in measuring
proton and deuteron $F_2$ structure functions. 
The finding triggered many theoretical studies
on physics mechanisms for explaining the antiquark
flavor asymmetry $\bar u-\bar d$ in the nucleon.
Various experimental results and proposed theoretical ideas
are summarized. Possibility of finding the flavor
asymmetry in Drell-Yan experiments is discussed together
with other processes, which are sensitive to the $\bar u/\bar d$ 
asymmetry. 

\vspace{0.5cm}

 
\noindent
{\rule{6.cm}{0.1mm}} \\
 
\vspace{-0.2cm}
\normalsize
\noindent
{* Email: kumanos@cc.saga-u.ac.jp. 
   Information on his research is available at}  \\

\vspace{-0.6cm}
\noindent
{http://www.cc.saga-u.ac.jp/saga-u/riko/physics/quantum1/structure.html.} \\

\vspace{0.1cm}

\vspace{-0.2cm}
\hfill
{submitted to Physics Reports}

\vspace{0.3cm}

\fbox{\parbox{14.0cm}{
\scriptsize
\noindent
{PREPARED FOR THE U.S. DEPARTMENT OF ENERGY 
    UNDER GRANT DE-FG06-90ER40561} \\

\scriptsize
This report was prepared as an account of work sponsored by the United
States Government. Neither the United States nor any agency thereof, nor
any of their employees, makes any warranty, express or implied, or assumes
any legal liability or responsibility for the accuracy, completeness, or
usefulness of any information, apparatus, product, or process disclosed, or
represents that its use would not infringe privately owned rights.
Reference herein to any specific commercial product, process, or service by
trade name, mark, manufacturer, or otherwise, does not necessarily
constitute or imply its endorsement, recommendation, or favoring by the
United States Government or any agency thereof. The views and opinions of
authors expressed herein do not necessarily state or reflect those of the
United States Government or any agency thereof.
}}

\vfill\eject
\normalsize
\pagestyle{empty}
\tableofcontents
\addtocontents{toc}{\protect\vspace{0.7cm}}

\vfill\eject
\pagestyle{plain}
\setcounter{page}{1}
\section{{\bf Introduction}}\label{INTRO}
\setcounter{equation}{0}
\setcounter{figure}{0}
\setcounter{table}{0}
\vspace{0.7cm}

{
\setcounter{enumi}{\value{figure}}
\addtocounter{enumi}{1}
\setcounter{figure}{0}
\renewcommand{\thefigure}{\arabic{section}.\theenumi}
\begin{floatingfigure}{7.0cm}
   \begin{center}
      \mbox{\epsfig{file=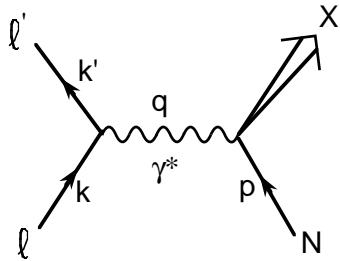,width=5.0cm}}
   \end{center}
 \vspace{-0.8cm}
 \caption{\footnotesize Lepton-nucleon scattering.}
 \label{fig:escat}
\end{floatingfigure}
\setcounter{figure}{\value{enumi}}
}
\quad
Nucleon substructure has been investigated through
various high-energy experiments. Electron or muon projectile
is ideal for probing minute internal structure of the nucleon.
The reaction is illustrated in Fig. \ref{fig:escat},
where the virtual photon from the lepton interacts
with the target nucleon.
Its cross section is related to two structure functions
$F_1$ and $F_2$ depending on transverse and longitudinal
reactions for the photon. 
They depend in general on two kinematical variables $Q^2=-q^2$ 
and $x=Q^2/2p\cdot q$ where $q$ is the virtual photon momentum 
and $p$ is the nucleon momentum. These structure functions provide 
important clues to internal structure of the nucleon 
\cite{FEC,MUTA,RGRBOOK}.
It is known that the structure functions are almost independent
of $Q^2$, which is referred to as Bjorken scaling. 
It indicates that the photon scatters on structureless objects,
which are called partons.
The partons are now identified with quarks and gluons.
The cross section is calculated by the lepton scattering on individual
quarks with incoherent impulse approximation, 
then the structure functions are described by quark
distributions in the nucleon: for example
$F_2(x,Q^2)=\sum_i e_i^2 x [q_i(x,Q^2)+\bar q_i(x,Q^2)]$.
Because the variable $x$ is the light-cone momentum fraction carried
by the struck quark, the structure function $F_2$ suggests
quark-momentum distributions in the nucleon.

Quark-antiquark pairs are created perturbatively 
according to Quantum Chromodynamics (QCD) so that
there could be infinite number of quarks and antiquarks
in the nucleon. A meaningful quantity is, for example, the difference
between quark and antiquark numbers. It is certainly restricted
by the baryon number and charge of the proton.
The valence-quark distribution $q_v$ is defined by
$q_v \equiv q - \bar q$, then the quark distribution is split 
into two parts: valence and sea distributions. 
With the definition of the valence quark,
the sea-quark distribution is given by $q_s=q-q_v=\bar q$. 
The valence quarks are the ``net'' quarks in
the nucleon. On the other hand, the sea quarks are thought to be 
produced mainly in the perturbative process of gluon
splitting into a $q\bar q$ pair. 
Because $u$, $d$, and $s$ quark masses are fairly small compared with
a typical energy scale in the deep inelastic scatting,
the splitting processes are expected to occur almost equally
for these quarks. 
Therefore, it was assumed until rather recently that the sea was 
flavor symmetric ($\bar u=\bar d=\bar s$) \cite{SU3}.

Both the valence and sea contribute to the electron or muon 
cross section, so that other processes have to be used
in addition for studying the details of the sea. 
Valence-quark distributions are obtained in neutrino interactions. 
Once the valence distributions are fixed,
the antiquark distributions in the nucleon are estimated 
from electron and muon scattering data or 
independently from Drell-Yan processes. 
The flavor-symmetric antiquark distributions
in $\bar u$, $\bar d$, and $\bar s$
had been used for a while; however, neutrino induced
dimuon events revealed that the strange sea is roughly half of
the $u$-quark or $d$-quark sea \cite{CDHS,CCFR}.
It had been, however, assumed that antiquark distributions
$\bar u$ and $\bar d$ are same. If they are different,
it should appear as a failure of the Gottfried sum rule \cite{GOTT}.
The sum rule was obtained by integrating the difference
between the proton and neutron $F_2$ structure functions over $x$,
$I_G\equiv \int dx (F_2^p-F_2^n)/x=1/3$.
There is an important assumption in this sum rule, and
it is the light antiquark flavor symmetry $\bar u=\bar d$.
If it is not satisfied, the Gottfried sum rule is violated.
However, it should be noted that the sum rule is not an ``exact" one,
which can be derived by using current algebra without a serious
assumption. 
Therefore, the fundamental theory of strong interaction, QCD, is not
in danger even if the sum-rule violation is confirmed.

There is an earlier indication of the sum-rule violation in the data at
the Stanford Linear Accelerator Center (SLAC)
in the 1970's [$I_G$=$0.200\pm 0.040$] \cite{SLAC75}.
The analysis in 1975 showed a significant deviation from
the Gottfried value 1/3.
However, no serious discussion could be made on the possible
violation because the smallest accessible $x$ point 
in the experiment was $x$=0.02 and there could be a significant
contribution to the sum from the smaller $x$ region.
Nevertheless, it is interesting to
conjecture a possible physics mechanism of the sum-rule violation.
Because the integral is given by 
$I_G=\int dx (u+\bar u-d-\bar d)=1/3+(2/3)\int dx(\bar u-\bar d)$,
the fact that the measured value $I_G$=0.200 is smaller than 1/3 
suggests a $\bar d$ excess over $\bar u$ in the nucleon.
Proposed ideas for creating the flavor asymmetry in the 1970's  
are, as far as the author is aware, 
a diquark model \cite{PAV76} and a Pauli blocking mechanism 
\cite{FF,OTHEREXC}.
The details of these models are discussed in sections 
\ref{PAULI} and \ref{DIQUARK}.
Other experimental information came from Drell-Yan processes.
Fermi National Accelerator Laboratory 
(Fermilab) E288 Drell-Yan data in 1981 \cite{E288}
suggested also a flavor asymmetric sea: $\bar u=\bar d(1-x)^{3.48}$.
Later, the sum rule was tested by the European Muon Collaboration (EMC)
in 1983 and 1987 \cite{EMC80S}.
The 1987 analysis indicates that 
the sum is $0.197 \pm 0.011 (stat.) \pm 0.083 (syst.)$
in the region 0.02$\le x \le$0.8, 
and the extrapolated value is 0.235$_{-0.099}^{+0.110}$.
Again, the data suggested a significant deficit in the sum rule.
It was, however, not strong enough to surprise our community
because the measured difference was still within the standard deviation.
The another muon group at the European Organization for Nuclear 
Research (CERN), Bologna-CERN-Dubna-Munich-Saclay
(BCDMS) collaboration, also investigated
the sum rule in muon scattering on the hydrogen and deuterium 
\cite{BCDMS}.
The BCDMS result in 1990 is $0.197\pm 0.006(stat.) \pm 0.036(syst.)$
in the region 0.06$\le x \le$0.8 at $Q^2$=20 GeV$^2$. Their estimate
of the small $x$ contribution is between 0.07 and 0.22,
so that the result could be consistent with 1/3. 

Although the sum rule was proposed in 1967,
there is little progress in 1970's and 1980's.
The crucial point was, as it is common in most sum rules,
the lack of small $x$ data with good accuracy.
The first clear indication of the sum-rule breaking was suggested
by the New Muon Collaboration (NMC) in 1991 \cite{NMC9194}.
They obtained data with $x$ as small as 0.004 by using a CERN muon beam.
They fitted $F_2^p(x)-F_2^n(x)$ data
by a smooth curve and extrapolated it into the unmeasured
small $x$ region. According to the NMC, the integral $I_G$ became
0.240$\pm$0.016, which is approximately 28\% smaller than
the Gottfried sum. Their reanalysis in 1994 indicates
a similar value 0.235$\pm$0.026. 
Considering the small errors, we conclude that the light antiquark
distributions are not flavor symmetric and we have a $\bar d$ excess
over $\bar u$ in the proton.

Recent measurements of $F_2^p/F_2^n$ by the Fermilab-E665 \cite{E665} and
the HERMES \cite{HERMES} collaborations agree with the NMC results.
Estimate of the Gottfried sum is not reported yet;
however, the agreement of $F_2^p/F_2^n$ suggests a violation of the sum.
Moreover, the charged-hadron-production data by the HERMES
support the NMC flavor asymmetry.

On the other hand, there are existing Drell-Yan data.
As it was mentioned, the Fermilab-E288 in 1981 suggested a $\bar d$ 
excess over $\bar u$ \cite{E288}.
However, later Fermilab-E772 collaboration data
showed no significant flavor asymmetry \cite{E772}
in 800 GeV proton-induced Drell-Yan measurements
for the deuteron, carbon, and tungsten.
Strictly speaking, these nuclear data cannot be compared with the NMC
results because nobody knows how large nuclear modification is.
A possible nuclear modification of $\bar u-\bar d$ is
discussed in section \ref{NUCLEI}.
There are data from p-p and p-d
Drell-Yan processes by the NA51 collaboration \cite{NA51} at CERN.
The data indicated large flavor asymmetry 
$\bar u/\bar d$=0.51$\pm$0.04$\pm$0.05 at $x$=0.18.
It is again a clear indication of the flavor asymmetry
in the light antiquark distributions.
In order to get more information for the asymmetry, 
the E866 experiment is in progress at Fermilab 
by measuring the Drell-Yan processes \cite{E866}.
Preliminary data also indicate $\bar u<\bar d$ which
could be consistent with the NMC.
The existing E288, E772, and NA51 Drell-Yan results
and the details of the Drell-Yan processes are explained
in section \ref{OTHEREXP} together with other processes, 
which are sensitive to the flavor asymmetry $\bar u-\bar d$.

Next, we discuss a brief outline of theoretical studies.
First, there is a conservative view that 
the Gottfried sum is satisfied without the $\bar u/\bar d$ asymmetry
by including a significant contribution from the small $x$ region
($x\le 0.004$) \cite{MRS90}. However, this idea is not consistent
with the NA51 data. 
We also note that perturbative corrections to $I_G$
are fairly small and it is of the order of
0.3\% at $Q^2$=4 GeV$^2$ \cite{RS,HK,KKPS,ST8461}.
The small correction is from the $\bar q\rightarrow q$ splitting
process which can occur in the next-to-leading-order (NLO) case.
If the sum-rule violation or the flavor
asymmetry is confirmed, it should be explained 
by a nonperturbative mechanism.

A reliable way of treating nonperturbative phenomena is
to use lattice QCD. Although real lattice calculation of
the Gottfried sum is not available at this stage, 
scalar matrix elements were evaluated in Ref. \cite{LD}.
The studies of the isoscalar-isovector ratio 
indicated significant flavor asymmetry when the quarks are light.
The difference comes from the process with quarks propagating
backward in time.
In order to understand the meaning of the sum-rule violation,
we should rely on quark-parton models.

Proposed theoretical ideas in the 1970's and 1980's
are the diquark model \cite{PAV76}
and the Pauli exclusion effect \cite{FF,OTHEREXC,ST}
which were originally intended to explain the old SLAC data.
In the diquark model, the violation is expected due to
the vector-diquark admixture.
Even though earlier results \cite{PAV76,AP}
seemed to be in agreement with the SLAC and NMC data, 
the deviation from the Gottfried sum
becomes very small if the virtual-photon
interaction with a quark inside the diquark is taken into account 
with a realistic mixing factor between vector and scalar
diquarks \cite{ABCP}. 
According to the Pauli blocking model, 
$u\bar u$ pair creations are more suppressed than $d\bar d$ creations
because of the valence $u$ quark excess over valence $d$ in the proton.
However, the effect would not be large enough to explain
the NMC result because a naive counting estimate is
$\bar u/\bar d=4/5$.
Furthermore, it was found recently \cite{ST} that antisymmetrization
between quarks could change the situation. If its effects are combined
with the Pauli-exclusion ones, the $\bar u$ distribution could be larger
than the $\bar u$.

On the other hand, mesonic models seem to be the most popular idea for
explaining the NMC result and the flavor asymmetry, at least by judging
from number of publications. Because of the difference between
$\bar u$ and $\bar d$ in virtual pion
clouds in the nucleon, we have the flavor asymmetry 
\cite{HM,INDIANA,ADELAIDE}.
For example, the proton decays into $\pi^+ n$ or $\pi^0 p$.
Because the $\pi^+$ has a valence $\bar d$ quark,
these processes produce an excess of $\bar d$ over $\bar u$ in the proton.
This model was further developed by including many virtual
states \cite{JULICH}. Combined mesonic and nuclear-shadowing effects
were studied in Ref. \cite{ZPION}.
In the early stage of these models, about a half of the NMC violation
was explained by the virtual states. In the Adelaide model \cite{ADELAIDE},
the NMC deficit was explained by adding the Pauli exclusion effect.
On the other hand, there is a possibility of explaining the
whole violation within the mesonic model by considering
different $\pi NN$ and $\pi N\Delta$ form factors \cite{KFS},
or by including many virtual states \cite{JULICH}.
Recently, off-shell pion effects were studied in Refs. \cite{SST,SS},
but they did not change the pionic contribution to $\bar u-\bar d$
significantly.
The mesonic mechanism can be described also in chiral models
\cite{SC,WAKA,EHQ,WH,LI,KRE,BPG,KOCH,SBF,CL}.
In the chiral field theory with quarks, gluons, and Goldstone
bosons \cite{EHQ,KRE,SBF,CL}, the flavor asymmetry comes from the virtual
photon interaction with the pions.
The obtained results also indicated
a significant deviation from the Gottfried sum.
In the chiral soliton models \cite{SC,WAKA,BPG}, a fraction
of the nucleon isospin is carried by the pions, and
the deviation from the sum is given by the ratio
of moments of inertia for the nucleon and pion.
A possible relation to the $\sigma$ term was also discussed
in the chiral models \cite{WH,LI,SF}.
The virtual mesons could modify not only the $x$ distribution 
$\bar u(x)-\bar d(x)$
but also $Q^2$ evolution of $I_G$ at relatively small $Q^2$ \cite{FBBG}.

Although it is usually thought to be very small, 
isospin-symmetry violation, e.g. $u_p\ne d_n$, was studied
in Ref. \cite{MSG,STNA51}.
In order to distinguish the isospin-symmetry violation
from the flavor asymmetry, we should investigate neutrino reactions,
the Drell-Yan p-n asymmetry, and charged-hadron production.
On the other hand, 
shadowing effects in the deuteron were investigated
\cite{BK,ZOLLER,KU,EFGS,BGNPZ,MST,PRW}
to find nuclear corrections in extracting the 
neutron $F_2$ from the deuteron data.
Although there are uncertain factors in nuclear potential,
the obtained correction to the sum is about $\delta I_G=-$0.02.
We should mention that it varies depending on the shadowing model.
However, the correction is a small negative number
(except for the pion excess model). 
If it is taken into account, the NMC deficit is magnified!
There are also papers on parton-transverse-motion corrections 
\cite{SV,MS96}.
It became possible to make flavor decomposition in
parametrization of antiquark distributions.
With the NMC and NA51 data, new parametrizations of parton distributions
were studied \cite{PRS,MRS,CTEQ,GRV}. 

The flavor asymmetry in the nucleon could be related to other
observables. Nuclear modification of the $\bar u-\bar d$ was investigated
in a parton-recombination model \cite{SK95}.
Because of the difference between
$u$ and $d$ quark numbers in neutron-excess nuclei,
$u\bar u$ and $d\bar d$ recombination rates are different.
This mechanism produces a finite $\bar u-\bar d$ distribution
in a nucleus even if it vanishes in the nucleon.
On the other hand, a relation to spin physics was studied
\cite{BS,EHQ,CL}. 
For example, if the Pauli blocking is the right mechanism for
producing the asymmetry, it also affects the spin content problem.
Because $u_v^\uparrow$ is larger than $u_v^\downarrow$
in the quark model, a $u_s^\downarrow$ excess over $u_s^\uparrow$
is expected. This could be one of the interpretations of
the proton spin problem.

We introduced various theoretical models. In order to distinguish
among these models, we need theoretical and experimental efforts,
in particular by studying consistency with other observables.

The NMC flavor asymmetry can be checked by other experimental
reactions. The best possibility is the aforementioned Drell-Yan process.
We have already explained the existing data.
Theoretical analyses of the Drell-Yan p-n asymmetry are discussed
in Refs. \cite{ES,KL,EHQDY,HGMP,MRSDY,JULDY,MSGDY}.
The asymmetry should become much clearer by the Fermilab-E866
experiment.
Charged-hadron-production data in muon scattering by the EMC 
\cite{EMCHPM,EMCPARA} were analyzed
for finding the $\bar u-\bar d$ \cite{LMS}.
At that time, experimental errors were not small enough to judge
whether or not the flavor distributions are symmetric.
However, the recent HERMES measurements show more clearly the NMC
type flavor asymmetry \cite{HERMES}.
On the other hand,
$W^\pm$ and $Z^0$ production can also be used \cite{MRS90,BS93,DHKS,PJ}.
Even though the W production is not very sensitive to the $\bar u/\bar d$
asymmetry in the $p+\bar p$ reaction, the $\bar u/\bar d$ ratio can be measured
in the $p+p$ \cite{PJ}.
Quarkonium production is usually dominated by the gluon-gluon fusion process;
however, the $\bar u/\bar d$ could be measured in the large $|x_F|$ region
\cite{PJC} if experimental data are accurate enough.
Neutrino scattering is another possibility. Combining neutral-current
and charged-current structure functions, or combining
different ones $F_1$, $F_2$, and $F_3$ for a practical purpose,
we could obtain the $\bar u-\bar d$ distribution
\cite{RS,INDIANA}.

In the following sections, we summarize theoretical and
experimental studies on the Gottfried sum rule and 
on the antiquark flavor asymmetry $\bar u-\bar d$ in the nucleon.
Future experimental possibilities are also discussed.

\vfill\eject
\section{{\bf Possible violation of the Gottfried sum rule}}\label{VIOLATION}
\setcounter{equation}{0}
\setcounter{figure}{0}
\setcounter{table}{0}

First, the Gottfried sum rule is derived in a naive parton model.
Earlier experimental results by the SLAC, EMC, and BCDMS are explained.
Then NMC experimental results are discussed.
We also comment on recent HERMES data.
As an independent experimental test of the NMC flavor asymmetry,
existing Drell-Yan data are shown.

\subsection{Gottfried sum rule}\label{GOTT}

The Gottfried sum rule is associated with the difference
between the proton and neutron $F_2$ structure functions
measured in unpolarized electron or muon scattering.  
Because there is no fixed neutron target, the deuteron
is usually used for obtaining the neutron $F_2$ by subtracting
out the proton part with nuclear corrections. 
 
The cross section of unpolarized electron or muon deep inelastic scattering
is calculated by assuming the one-photon exchange process in
Fig. \ref{fig:escat} \cite{FEC,MUTA,RGRBOOK}:
\begin{multline}
d\sigma = \frac{1}{4\sqrt{(k\cdot p)^2-m^2 M^2}} \ 
                 \overline{\sum_{pol}} \sum_X 
                  \, (2\pi)^4 \, \delta^4(k+p-k'-p_{_X})
 \\
        \times \    |{\mathcal M}(ep\rightarrow e'X)|^2    \,
               \frac{d^3 k'}{(2\pi)^3 2 E'} 
\ \ \ ,
\end{multline}
where the matrix element is
\begin{equation}
{\mathcal M}(ep\rightarrow e'X) \, = 
                \, \bar u(k',\lambda ')e\gamma_\mu u(k,\lambda )   \, 
                               \frac{g^{\mu\nu}}{(k-k')^2}  \,
                                <X \, | \, e J_\nu (0) \, | \, p,\sigma >
\ \ \ .
\end{equation}
$M$ and $m$ are the proton and lepton masses,
$k$ and $k'$ ($\lambda$ and $\lambda'$) 
are initial and final lepton momenta (helicities),
and $J_\nu$ is the electromagnetic current.
The proton momentum and spin are denoted by
$p$ and $\sigma$, and $p_{_X}$ is the momentum
of the hadron final state $X$.
The notation $\overline \sum_{pol}$ indicates that
spin average and summation are taken for the initial
and final states respectively.
From these equations, the cross section is expressed
by a leptonic current part $L^{\mu\nu}$ and a hadronic
one $W_{\mu\nu}$:
\begin{equation}
d\sigma = \frac{2M}{s-M^2} \, \frac{\alpha^2}{Q^4} 
               \, L^{\mu\nu} \ W_{\mu\nu} \ \frac{d^3 k'}{E'}   
\ \ \ ,
\label{eqn:SIGMA}
\end{equation}
where $\alpha$ is the fine structure 
constant, $s$ is given by $s=(p+k)^2$, 
$E'$ is the scattered lepton energy, 
and $Q^2$ is defined by $Q^2=-q^2$.
Throughout this paper, the convention $-g_{00}=g_{11}=g_{22}=g_{33}=+1$
is used so as to have $p^2=p_0^2-\vec p^{\, \, 2}=M^2$.
The lepton tensor can be calculated as
\begin{align}
L^{\mu\nu} &= \ \overline{\sum_{\lambda, \lambda '}}
                  \ [\bar u(k',\lambda') \gamma^\mu u(k,\lambda)]^*
                  \ [\bar u(k',\lambda') \gamma^\nu u(k,\lambda)]     
\nonumber \\
              &= 2\ (k^\mu {k'}^\nu + {k'}^\mu k^\nu 
                                     - k\cdot k' g^{\mu\nu})
\ \ \ ,
\label{eqn:LEPTON}
\end{align}
in the unpolarized case. 
The hadronic part is given by
\begin{align}
W_{\mu\nu} &= \frac{1}{4\pi M} \sum_X (2\pi)^4 \, \delta^4(p+q-p_{_X}) \,  
                \overline{\sum_\sigma} 
                   <p,\sigma |J_\mu (0)|X>
                                       <X|J_\nu (0)|p,\sigma >  
\nonumber \\
           &= \frac{1}{4\pi M}  \overline{\sum_\sigma} \int d^4 \xi
                               \, e^{iq\cdot \xi}
                 <p,\sigma |\, [J_\mu (\xi),J_\nu (0)]\, |p,\sigma>
\ \ \ .
\label{eqn:wmunu}
\end{align}
Using light-cone variables $q^\pm =(q^0\pm q^3)/\sqrt{2}$ with
$q=(\nu,0,0,-\sqrt{\nu^2+Q^2})$ and $\nu=E-E'$, 
we have $q^+=-Mx/\sqrt{2}=$finite and
$q^- = \sqrt{2} \nu\rightarrow \infty$ in the Bjorken scaling limit,
$Q^2\rightarrow \infty$ with finite $x$.
The exponential factor becomes 
$e^{iq\cdot \xi} = e^{iq^+ \xi^-} e^{iq^- \xi^+}$.
Because the $q^-$ part is a rapidly oscillating term,
the integral vanishes except for the singular region of the integrand
according to the Riemann-Lebesque theorem.
Therefore, the integral is dominated by the light-cone region 
$\xi^+ \approx 0$. In the deep inelastic lepton scattering,
we can probe light-cone momentum distributions of internal
charged constituents in the proton. 
The formal approach for analyzing the hadron tensor
is to use operator product expansion. It is discussed in section 
\ref{PQCD}
in explaining QCD corrections to the sum rule. Here, we do not step
into the details and simply discuss general properties.
Using parity conservation, time-reversal invariance,
symmetry under the exchange of the Lorentz indices $\mu$ and $\nu$,
and current conservation, we can express the hadron tensor
in term of two structure functions $W_1$ and $W_2$:
\begin{equation}
W_{\mu\nu} = - W_1 \left ( g_{\mu\nu} - \frac{q_\mu q_\nu}{q^2} \right )
             + W_2 \frac{1}{M^2} 
                \left ( p_\mu -\frac{p\cdot q}{q^2} q_\mu \right ) 
                \left ( p_\nu -\frac{p\cdot q}{q^2} q_\nu \right )
\ \ \ .
\label{eqn:HADRON}
\end{equation}
From Eqs. (\ref{eqn:SIGMA}), (\ref{eqn:LEPTON}), (\ref{eqn:HADRON}),
the cross section becomes
\begin{equation}
\frac{d\sigma}{d\Omega dE^{\prime}}
=\frac{\alpha^2}{4E^2\sin^4\frac{\theta}{2}}
\left[2W_1(\nu,Q^2)\sin^2\frac{\theta}{2}
+W_2(\nu,Q^2)\cos^2\frac{\theta}{2}\right] 
\ \ \ .
\end{equation}

Scaling structure functions $F_1$ and $F_2$ are defined
in terms of $W_1$ and $W_2$:
\begin{equation}
F_1 \equiv M \, W_1 \ \ \ , \ \ \ F_2 \equiv \nu \, W_2 \ \ \ . 
\end{equation}
The $F_1$ is associated with the transverse cross section, and
the $F_2$ is with the transverse and longitudinal ones.
In the Bjorken limit, two structure functions
are related by the Callan-Gross relation $2xF_1=F_2$.
In the parton picture, the deep inelastic process can be described
by virtual photon interactions with individual quarks with
incoherent impulse approximation. It is supposed to be valid 
at large $Q^2$ in the sense that virtual-photon-interaction time
with a quark is fairly small compared with the interaction time 
among quarks.
Then, the leading-order (LO) or DIS-scheme structure function
$F_2$ is given by quark-momentum distributions in the nucleon:
\begin{equation}
F_2(x,Q^2) = \sum_i e_i^2 \, x \, [ \, q_i(x,Q^2) \, 
                             + \, \bar q_i(x,Q^2) \, ]
\ \ \ ,
\end{equation}
where $i$ denotes the quark flavor.
In the next-to-leading order (NLO) except for the DIS scheme case, 
the gluon distribution also
contributes to $F_2$ through the splitting $g\rightarrow q\bar q$.
With the assumption of isospin symmetry in the nucleon,
parton distributions in the neutron could be related to 
those in the proton. The d-quark distribution
in the neutron is equal to the u-quark distribution
in the proton [$u_n(x,Q^2)=d_p(x,Q^2)$] and in the similar way 
for other partons [$d_n=u_p$, $\bar u_n=\bar d_p$, 
$\bar d_n=\bar u_p$, and etc.].
Hereafter, the parton distributions are assumed as
those in the proton except for section \ref{ISOSPIN} where
possible isospin-symmetry breaking is discussed.
Then, the difference between the proton and neutron structure functions
is given by 
\begin{equation}
F_2^p(x,Q^2) - F_2^n(x,Q^2) =  
                \frac{1}{3} \, x \, [ u_v(x,Q^2) - d_v(x,Q^2) ]
              + \frac{2}{3} \, x \, [ \bar u(x,Q^2) - \bar d(x,Q^2) ]
\ .
\label{eqn: F2P-M}
\end{equation}
The valence-quark distributions should satisfy
\begin{equation}
\int_0^1 dx \, u_v(x,Q^2) = 2 \ \ , \ \ 
\int_0^1 dx \, d_v(x,Q^2) = 1 
\ \ \ ,
\label{eqn: VSUM}
\end{equation}
due to the proton and neutron charges, $\int dx (2u_v-d_v)/3=1$ and
$\int dx (2d_v-u_v)/3=0$ where elastic scattering amplitudes are
expressed in the parton model by considering an infinite momentum frame.
Substituting Eq. (\ref{eqn: VSUM}) into Eq. (\ref{eqn: F2P-M})
and integrating over the variable $x$, we obtain
\begin{equation}
\int_0^1 \frac{dx}{x} \, 
[ F_2^p(x,Q^2) - F_2^n(x,Q^2) ] 
                   = \frac{1}{3} + 
      \frac{2}{3} \int_0^1 dx \, [ \bar u(x,Q^2) - \bar d(x,Q^2) ]
\ \ \ .
\label{eqn:GINT}
\end{equation}
If the sea is flavor symmetric $\bar u = \bar d$,
the second term vanishes and it becomes the Gottfried sum rule
\cite{GOTT}:
\begin{equation}
\int_0^1 \frac{dx}{x} \, 
 [ F_2^p(x,Q^2) - F_2^n(x,Q^2) ] = \frac{1}{3} 
\ \ \ .
\label{eqn: GOTTFRIED}
\end{equation}

As it is obvious in the above derivation in a naive parton model,
there is a serious assumption of the flavor symmetry in
the light antiquark distributions. Therefore, it is not a rigorous 
one like the Bjorken sum rule. Even if violation of
the sum rule is found in experiments, there is virtually no danger in
the fundamental theory of strong interactions, quantum chromodynamics.
It is nevertheless interesting to test it because its violation
could suggest an SU(2)-flavor asymmetric sea in the nucleon 
as it was found in the neutrino-induced dilepton production
in the case of SU(3). 
Because of small u and d quark masses, large $\bar u/\bar d$
asymmetry cannot be expected in perturbative QCD.
Therefore, a possible sum-rule violation gives an opportunity
for learning more details on internal structure of the nucleon.

\subsection{Early experimental results}\label{EXP}

Because the small $x$ region could have a significant contribution
to the sum rule, it was not possible to test it until recently.
The minimum $x$ is restricted by the lepton-beam energy $E$
as $min(x)=Q^2/2 M E$, where
$Q^2$ should not be smaller than a few GeV$^2$
in order to be deep inelastic scattering.
The first test of the sum rule was studied at SLAC in the 1970's.
The electron-beam energy is 4.5--20 GeV so that 
the smallest $x$ is about 0.02.
Targets are hydrogen, deuterium, and heavier ones.
The data are taken in the $x$ range from 0.02 to 0.82
for the hydrogen and deuterium targets. 
The $Q^2$ varies depending on the $x$ region, but it is from 0.1 GeV$^2$
to 20 GeV$^2$. In the 1975 analysis \cite{SLAC75},
the data with 0.02$\le x\le$0.28 are combined
with previous data in the extended range $x\le 0.82$.
The neutron structure function is extracted by taking
into account Fermi smearing effects:
$F_2^n/F_2^p=(S F_2^d-F_2^p)/F_2^p$, where $S$ is 
the Fermi smearing factor.
The difference becomes $F_2^p-F_2^n=2F_2^p-SF_2^d$.
We define a Gottfried integral by
\begin{equation}
\int_{x_{min}}^{x_{max}} \frac{dx}{x} \, 
        [ \, F_2^p(x,Q^2) - F_2^n(x,Q^2) \, ]
\equiv I_G (x_{min}, x_{max})
\ \ \ .
\end{equation}
According to the SLAC data in 1975 \cite{SLAC75}, it is
\begin{equation}
I_G (0.02,0.82) = 0.200 \pm 0.040 \ \ \ \ \ \text{(in \ 1975)}
\ \ \ .
\end{equation}
It should be noted that the integral contains various $Q^2$ data
ranging from small $Q^2$, where perturbative QCD may not be valid.
In any case, it is interesting to find a significantly smaller value
than the Gottfried sum 1/3.
Therefore, there was earlier indication of the sum-rule violation
in the SLAC data. In fact, the Pauli-blocking and diquark models
were proposed, just after the SLAC finding, 
for explaining the possible deficit in the sum. 
However, it was not conclusive enough to state that
the sum rule is violated experimentally due to a possible large contribution
from the smaller-$x$ region.

Next experimental data came from EMC measurements
at CERN by deep inelastic muon scattering on 
the hydrogen and deuterium \cite{EMC80S}.
The muon-beam energy is 280 GeV, and the measured kinematical
range is 0.03$\le x\le$0.65 and 7$\le Q^2\le$170 GeV$^2$.
The neutron structure function is extracted from the
deuteron data by taking into account the smearing effects
due to the nucleon Fermi motion.
The Hulthen and Paris wave functions are used in
the 1983 and 1987 analyses to estimate the smearing correction. 
The mean $Q^2$ in the data depends on $x$, and it ranges from
10 GeV$^2$ at $x$=0.03 to 90 (80 in 1983) GeV$^2$ at $x$=0.65.
Using the $Q^2$ averaged data at each $x$, they obtained
\begin{equation}
I_G(0.03,0.65) = 0.18 \pm 0.01 \, (stat.) 
                      \pm 0.07 \, (syst.)
\ \ \ ,
\end{equation}
in 1983.
The distribution $F_2^p -F_2^n$ is extrapolated into the unmeasured regions
by using a function $F_2^p -F_2^n = A x^{0.5} (1-x)^\alpha (1+\beta x)$,
where the constants $\alpha$ and $\beta$ are obtained from the data.
The 1983 EMC result in the whole $x$ is then given by 

\vfill\eject
\begin{equation}
I_G(0,1) = 0.24   \pm 0.02 \, (stat.) 
                  \pm 0.13 \, (syst.)
\ \ \ \ \ \text{(in \ 1983)}
\ \ \ .
\end{equation}
In the 1987 report, these values became
\begin{equation}
I_G(0.02,0.8) = 0.197 \pm 0.011 \, (stat.) 
                      \pm 0.083 \, (syst.)
\ \ \ ,
\end{equation}
and
\begin{equation}
I_G(0,1) = 0.235 \, _{-0.099}^{+0.110} \ \ \ \ \ \text{(in \ 1987)}
\ \ \ .
\end{equation}
It should be noted that different $Q^2$ data are collected
to get the integral, whereas the sum rule is valid at
certain $Q^2$.
The above result could be consistent with the sum 1/3
within the experimental error; however, it is also smaller
as the SLAC data indicated.

Another muon group at CERN, BCDMS, also obtained the sum
by analyzing muon scattering on the hydrogen and deuterium
\cite{BCDMS}.
The muon-beam energies are 120, 200, and 280 GeV. 
The kinematical range is 0.06$\le x\le$0.80 and 8$\le Q^2\le$260 GeV$^2$.
The structure function ratio is obtained 
with the smearing factor calculated with
the Paris wave function for the deuteron.
Integrating the distribution $F_2^p-F_2^n$, the BCDMS obtained
\begin{equation}
I_G(0.06,0.8) = 0.197 \pm 0.006 \, (stat.) 
                      \pm 0.036 \, (syst.) \ \ \ \ \ \text{(in\ 1990)}
\ \ \ ,
\end{equation}
at $Q^2$=20 GeV$^2$. The larger-$x$($>$0.8) contribution
is negligible, and the smaller-$x$($<$0.06) one varies from
0.07 to 0.22 by considering the behavior $F_2^p-F_2^n \propto x^\alpha$
with 0.3$\le \alpha\le$0.7.
Because of the large uncertainty from the small $x$ region, 
they did not quote the integral value in the whole range of $x$.
Due to the possible small-$x$ contribution, they concluded that
it could be consistent with the sum 1/3.

\subsection{NMC finding and recent progress}\label{NMC}

\begin{figure}[h]
\parbox[b]{0.46\textwidth}{
   \begin{center}
       \epsfig{file=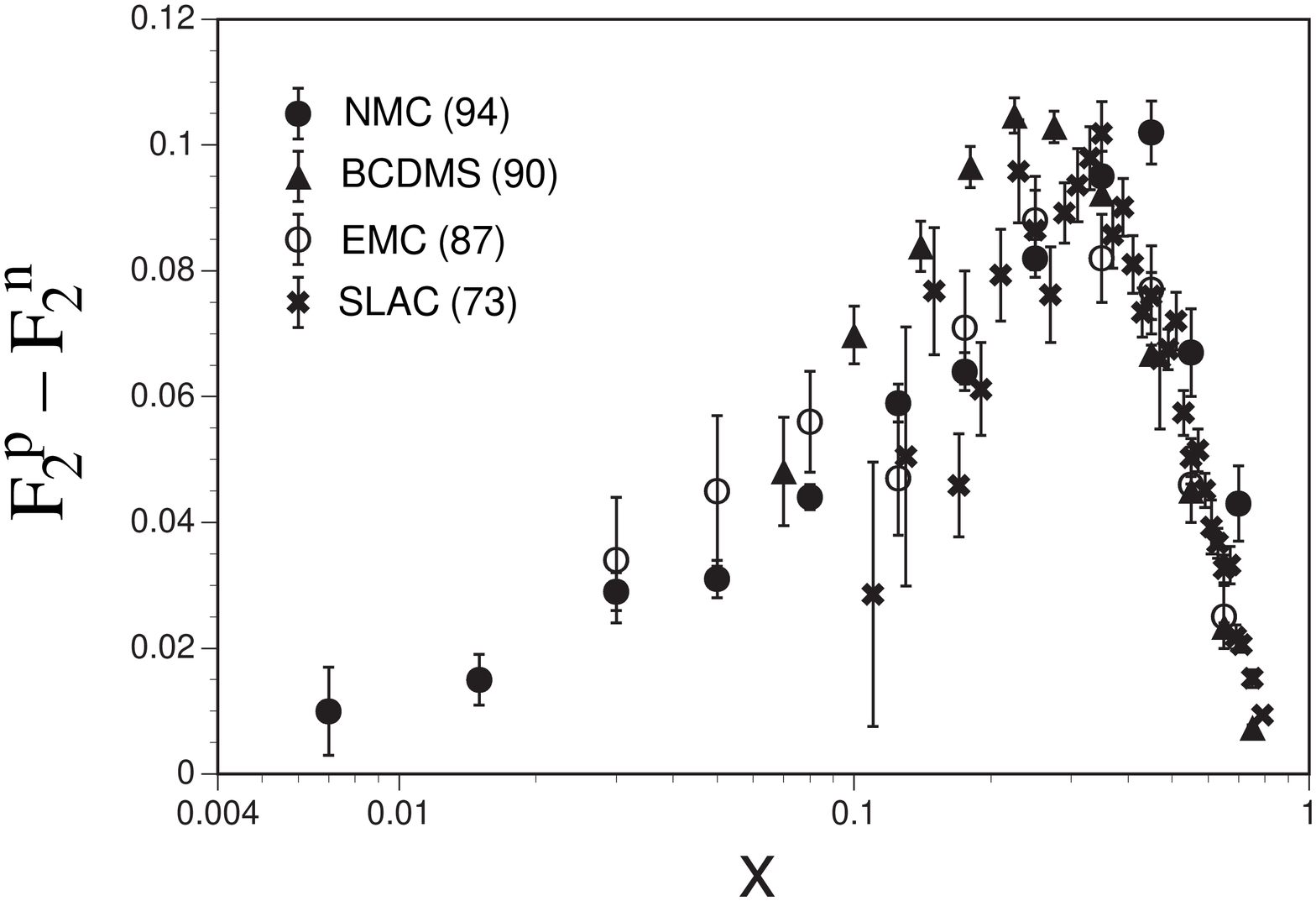,width=7.0cm}
   \end{center}
       \vspace{-0.8cm}
       \caption{\footnotesize
          $F_2^p-F_2^n$ data by SLAC, EMC, BCDMS, and NMC.}
       \label{fig:f2pn}
}\hfill
\parbox[b]{0.46\textwidth}{
   \begin{center}
       \epsfig{file=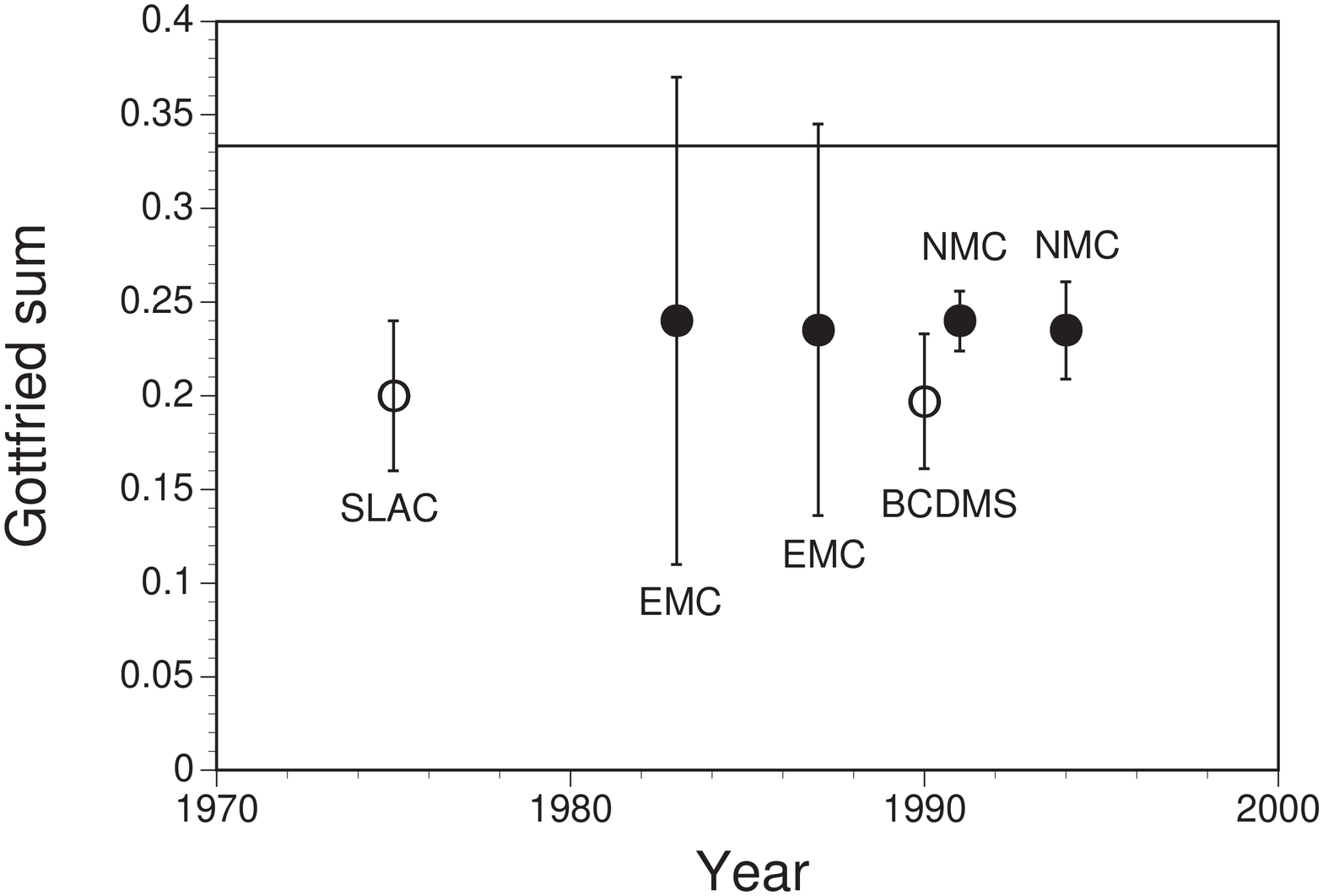,width=6.0cm}
   \end{center}
       \vspace{-0.8cm}
   \caption{\footnotesize 
          Experimental history of the Gottfried sum rule.
          The SLAC and BCDMS integrals are evaluated in the region of
          0.02$\le x\le$0.82 and 0.06$\le x\le$0.8 respectively. 
          Because unmeasured regions are not included, 
          they are shown by the open circles.}
   \label{fig:igyear}
}
\end{figure}

Although the earlier data suggested violation of the Gottfried sum,
it was not conclusive enough because of the large errors and a possible 
large contribution from the small $x$ region. 
In the NMC experiment, the kinematical range was extended
to the small $x$ region.
The NMC obtained 90 and 280 GeV muon scattering data on hydrogen and
deuterium targets at CERN \cite{NMC9194}.
The kinematical range is 0.004$\le x\le$0.8 and 0.4$\le Q^2\le$190 GeV$^2$.
The difference of the structure functions is calculated by
\begin{equation}
F_2^p -F_2^n= 2 \, F_2^d \, \frac{1-F_2^n/F_2^p}{1+F_2^n/F_2^p}
\ \ \ ,
\label{eqn:NMCINTEG}
\end{equation}
where the ratio $F_2^n/F_2^p$$=2F_2^d/F_2^p-1$ is determined 
by the NMC experiment, and the absolute value of the deuteron
structure function $F_2^d$ is given by a fit to various experimental data. 
Nuclear corrections such as the Fermi motion
in section \ref{EXP} are not taken into account.
The $F_2^d$ and $F_2^n/F_2^p$ are determined at $Q^2$=4 GeV$^2$
by interpolation or extrapolation. 
The obtained $F_2^p-F_2^n$ data \cite{NMC9194} are shown
in Fig. \ref{fig:f2pn} together with the previous data by SLAC \cite{SLAC75}, 
EMC \cite{EMC80S}, and BCDMS \cite{BCDMS}.
The NMC result in 1991 is
\begin{equation}
I_G(0.004,0.8) = 0.227 \pm 0.007 \, (stat.) 
                       \pm 0.014 \, (syst.)
\end{equation}
at $Q^2$=4 GeV$^2$.

The contribution from the larger $x$ region is estimated
by extrapolation, and it is a rather small value
$I_G(0.8,1.0)=0.002 \pm 0.001$. 
The extrapolation into the smaller-$x$ region indicates
a behavior $F_2^p-F_2^n=ax^b$ with $a=0.21 \pm 0.03$
and $b=0.62 \pm 0.05$. Then its contribution becomes
$I_G(0,0.004)=0.011 \pm 0.003$. Combing all these results,
they obtained
\begin{equation}
I_G(0,1) = 0.240 \pm 0.016 \ \ \ \ \ \text{(in \ 1991)}
\ \ \ .
\label{eqn:NMC91}
\end{equation}
This time, it clearly indicates the failure of the sum rule
because the value is significantly smaller than 1/3 even
if the experimental error is taken into account.
The violation is obvious in Fig. \ref{fig:igyear},
where the history of the experimental measurements is shown.
However, the small $x$ estimate by the NMC is not unique. A small variation
in the small $x$ data could result in a very different contribution
and may modify Eq. (\ref{eqn:NMC91}) significantly. 
This issue is discussed in section \ref{SMALLX}.
From Eqs. (\ref{eqn:GINT}) and (\ref{eqn:NMC91}), the deficit
could be explained if there is a flavor asymmetry
\begin{equation}
\int_0^1 dx (\bar u-\bar d)=-0.140\pm 0.024
\ \ \ .
\label{eqn:UBMDB}
\end{equation}

\vspace{-0.7cm}
\noindent

\vfill\eject
The NMC reanalyzed the integral by using a new parametrization
for $F_2^d$ including their own data and revised $F_2^n/F_2^p$ ratios.
Their result in 1994 is
\begin{equation}
I_G(0.004,0.8) = 0.221 \pm 0.008 \, (stat.) 
                       \pm 0.019 \, (syst.)
\end{equation}
at $Q^2$=4 GeV$^2$. 
The larger $x$ contribution becomes $I_G(0.8,1.0)=0.001\pm 0.001$.
The smaller $x$ one is $I_G(0,0.004)=0.013\pm 0.005$ by the extrapolation
$F_2^p-F_2^n=ax^b$ with $a=0.20 \pm 0.03$ and $b=0.59 \pm 0.06$.
Then, the overall integral is
\begin{equation}
I_G(0,1) = 0.235 \pm 0.026 \ \ \ \ \ \text{(in \ 1994)}
\ \ \ .
\label{eqn:NMC94}
\end{equation}
The sum is consistent with the previous NMC result; however,
the error is slightly larger due to more extensive examination
of the systematic uncertainties.

In the HERMES experiment \cite{HERMES}, 
the positron beam energy is 27.5 GeV
and hydrogen, deuterium, and $^3 He$ gas targets are used.
The ratio $F_2^p/F_2^n$ is extracted from the unpolarized
hydrogen and deuterium data. 
The measured kinematical range is $0.015 \le x \le 0.55$ 
(averaged in each bin) and $0.4 \le Q^2 \le 11$ GeV$^2$.
Because the obtained ratios agree with the NMC results, 
the HERMES experiment seems to support the sum-rule violation.
However, the sum $I_G$ is not reported yet.
On the other hand, a clearer indication of the flavor asymmetry is
given in semi-inclusive data.
As it is discussed in section \ref{CHARGED},
the charged-hadron production ratio
$r(x,z)=( N^{p\pi^-} - N^{n\pi^-} ) / ( N^{p\pi^+} - N^{n\pi^+} )$
is also related to the $\bar u/\bar d$ asymmetry.
The data analysis \cite{HERMES} clearly favors
the NMC expectation rather than the flavor symmetric one.

Because of the small errors, the NMC 1991 result is the first one
which made us realize that the Gottfried sum rule is actually violated.
It strongly suggests the flavor asymmetry in the light antiquark 
distributions, namely a $\bar d$ excess over $\bar u$ in the proton.
After the NMC finding, many theoretical papers are written on
this topic and independent Drell-Yan experiments are proposed
at CERN and Fermilab. 
Some Drell-Yan experimental results were already taken
at Fermilab and CERN. They indicate also the NMC type flavor
asymmetry. These Drell-Yan results are discussed in section
\ref{DRELLYAN}.

\subsection{Small $x$ contribution}\label{SMALLX}

\begin{wrapfigure}{r}{7.0cm}
   \begin{center}
      \mbox{\epsfig{file=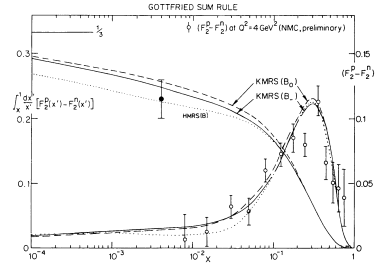,width=5.0cm}}
   \end{center}
 \vspace{-0.4cm}
\caption{\footnotesize Small $x$ contributions 
                       (taken from Ref. {\normalsize\cite{MRS90}}).
}
\label{fig:small}
\end{wrapfigure}

One of the reasons why the Gottfried sum rule was not
investigated in detail in the 1970's and 1980's
is the lack of small $x$ data,
which may contribute significantly.
The smallest $x$ point of the NMC data is 0.004.
Their analysis indicates that the small $x$ contribution
is $I_G(0,0.004)=0.013\pm 0.005$, which is merely 4\% of the sum 1/3.
In evaluating the integral, they extrapolate the data by 
using the fitting $F_2^p-F_2^n=0.20 x^{0.59}$
to the experimental data.
However, it is not very obvious whether the small $x$ contribution
is so small. Slight variations of the NMC small $x$ data
could make a significant change in the integral as it is obvious
in Fig. \ref{fig:small}.

\vfill\eject
\quad
The small $x$ contribution was investigated in Refs. \cite{MRS90} 
and \cite{INDIANA}. 
Three MRS-group (Martin-Roberts-Stirling) parametrizations, 
which were available in 1990, were studied \cite{MRS90}. 
They are HMRS-B, KMRS-B0, and KMRS-B\_
which are fit to various experimental data without any small $x$
constraint for the HMRS-B, $x^0$ type sea-quark and gluon distributions
in the limit $x\rightarrow 0$ for the KMRS-B0, and $x^{-0.5}$ for the KMRS-B\_.
Comparison of these parametrizations with the NMC experimental data
is shown in Fig. \ref{fig:small} \cite{MRS90}.
It indicates that the parametrization
curves are consistent with the data points, and yet they satisfy
the Gottfried sum rule without any flavor asymmetry for the $\bar u$
and $\bar d$ quarks. 
The NMC raw data $I_G(0.004,0.8)$ is shown by the filled circle 
with an error bar. It is also consistent with the three predictions.
The only difference comes from small $x$ behavior of $F_2^p-F_2^n$.
According to the KMRS-B0, valence-quark distributions at small $x$
behave like $x(u_v+d_v)\sim x^{0.27}$ and
            $xd_v\sim x^{0.61}$.
The small $x$ fall-off $x^{0.27}$ is much slower than
the NMC one $x^{0.59}$, which makes a significant
contribution from the small $x$ region. In fact, 
three parametrizations have $I_G(0,0.004)$=0.07$-$0.11 so that
the missing 10\% strength could come from the smaller-$x$ region.
More recent parametrizations are discussed in 
section \ref{PARAMET}.

Therefore, it is not definite whether the small $x$
contribution is relatively small as suggested by the NMC. 
In the HMRS-E case, the sum 1/3 can be reached if
the integral region is extended to very small $x\approx 10^{-10}$.
This is an unrealistic number for experimental measurement.
However, as it is obvious from Fig. 2 of the KL paper \cite{INDIANA},
the small-$x$ contribution should become obvious at $x\approx 10^{-5}$.
There is an experimental possibility of measuring $F_2^d$ at such
small $x$ by accelerating the deuteron at 
the Hadron-Electron Ring Accelerator (HERA) in Hamburg.
However, it is not clear whether such experiment
could be realized at HERA.
Therefore, the best way of testing it, at least at this stage,
is to use other experimental processes. 
The NA51 experimental data \cite{NA51} support the NMC conclusion,
a $\bar d$ excess over $\bar u$ in the nucleon. More complete
information will come from the Fermilab-E866 experiment
in the near future \cite{E866}.

\subsection{Nuclear correction: shadowing in the deuteron}\label{SHADOW}

Because there is no fixed target for the neutron,
the deuteron is usually used for measuring the neutron structure 
function $F_2^n$.
In the NMC analyses, the deuteron and proton structure-function
ratios are measured and they are related to the proton-neutron
ratio by $F_2^n/F_2^p=2F_{2D}/F_2^p-1$. Together with world-averaged
deuteron structure functions, the difference $F_2^p-F_2^n$ is calculated
by Eq. (\ref{eqn:NMCINTEG}).
To be precise, the NMC result can be compared with the Gottfried sum only
if there is no nuclear modification in the deuteron: $F_{2D}=F_2^p+F_2^n$.
Of course, there is a famous Fermi-motion correction at large $x$
and the EMC effect at medium $x$. However, these do not change the sum
significantly because the major contribution comes from
the small $x$ region. 

It is well known that nuclear structure functions
are modified at small $x$, and the phenomena are called shadowing.
It means literally that internal constituents are shadowed due to 
the existence of nuclear surface ones, so that the cross section is smaller
than the each nucleon contribution:
$\sigma_A=A^\alpha \sigma_N$ with $\alpha<1$.
Such phenomena occur at small $x$ in the following way
according to the vector-meson-dominance (VMD) model.
A virtual photon transforms into vector meson states, 
which then interact with a target nucleus.
The propagation length of the hadronic ($v$) fluctuation:
$\lambda \approx 1 /| E_v - E_\gamma | \approx 0.2/x$ fm,
exceeds the average nucleon separation (2 fm) in nuclei for $x<0.1$.
Then, the shadowing takes place due to multiple scattering.
For example, the vector meson interacts
elastically with a surface nucleon and then
interacts inelastically with a central nucleon.
Because this amplitude is opposite in phase to 
the one-step amplitude for an inelastic interaction with the central nucleon,
the nucleon sees a reduced hadronic flux (namely the shadowing).
This multiple scattering picture is valid only in the laboratory frame.
In terms of the terminology in an infinite momentum frame, the phenomena
are explained by parton recombinations, which mean parton
interactions from different nucleons. Such interactions occur
because the localization size of a parton with momentum 
fraction $x$ exceeds 2 fm at $x<0.1$.
Whatever the description is, the shadowing in $F_2$
is a well studied topic, so that we should be able to 
estimate deuteron shadowing effects on the Gottfried sum.

Nuclear corrections in the deuteron to the Gottfried sum rule,
in particular the shadowing effects,
were calculated in various models 
\cite{BK, ZOLLER, KU, EFGS, MST, BGNPZ, MST, PRW}.
So far, VMD, Pomeron, and meson-exchange mechanisms
have been studied, and the results are nicely presented in Ref. \cite{MST}. 
A significant part of the following discussions is based on this paper.
We discuss first a popular description, the VMD model.
Then, its results are compared with other model results.

\begin{wrapfigure}{l}{7.0cm}
   \begin{center}
      \mbox{\epsfig{file=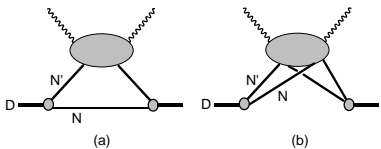,width=5.0cm}}
   \end{center}
 \vspace{-0.4cm}
\caption{\footnotesize  Virtual photon interaction with the deuteron
               (a) in the impulse approximation and 
               (b) in the double scattering case. } 
\label{fig:double}
\end{wrapfigure}
\quad
The shadowing is traditionally described by
the VMD model in particular at small $Q^2$.
Estimates of its effects on the Gottfried sum are found in Refs. 
\cite{MST, PRW}.
The virtual photon transforms into vector-meson states ($v$),
which then interact with the deuteron.
The hadron-deuteron cross section is given by 
an individual nucleon term in Fig. \ref{fig:double}(a) and 
a double scattering term in Fig. \ref{fig:double}(b)
in the Glauber theory: $\sigma_{vD}=2\sigma_{vN}+\delta\sigma_{vD}$, where

\vfill\eject
\begin{equation}
\delta\sigma_{vD} = - \frac{\sigma_{vN}^2}{8\pi^2}
                    \int d^2 \vec k_T \, S_D (\vec k^2)
\ \ \ .
\end{equation}
The $S_D (\vec k^2)$ is the deuteron form factor given 
by the S and D state wave functions:
$S_D(\vec k^{\, 2})=\int dr [u^2(r)+w^2(r)]j_0(kr)$.
Then, the virtual-photon cross section is written as
$\delta \sigma_{\gamma^* D} = \sum_v
  (e^2 /f_v^2)  \, \delta \sigma_{vD} / (1+Q^2/M_v^2)^2$.
This equation is expressed in the $F_2$ form:
\begin{equation}
\delta F_{2D} (x) = \frac{Q^2}{\pi} \sum_v
                    \frac{\delta\sigma_{vD}}
                     {f_v^2 \, (1+Q^2/M_v^2)^2} 
\ \ \ .
\end{equation}
The $\rho$, $\omega$, and $\phi$ mesons are included
as the vector mesons.
The most contribution comes from the $\rho$ meson and it is about 80\%.
With this shadowing correction, the deuteron $F_2$ becomes
$F_{2D}=F_2^p+F_2^n+\delta F_{2D}$.
Because no nuclear correction is assumed in the NMC analysis, namely
$[F_2^p-F_2^n]_{NMC}=2F_2^p-F_{2D}$, the Gottfried sum becomes
\begin{equation}
I_G=I_G^{NMC} + \int \frac{dx}{x} \, \delta F_{2D} (x)
\ \ \ .
\end{equation}
In Ref. \cite{PRW}, the VMD model is investigated further by
including $q\bar q$ continuum in addition to the vector mesons,
$\rho$, $\omega$, and $\phi$. 
The model can explain the NMC shadowing data for various nuclei.
Applying the same model to the deuteron, they find 
the shadowing correction from $\delta I_G(0.004,1)$=$-$0.039 to
$-$0.017 depending on different nuclear potentials.
The results qualitatively agree with those in Ref. \cite{MST},
where the Pomeron and meson exchange contributions are added
to the VMD one.

Because the shadowing effects on $I_G$ are more or less same
in all realistic models, other descriptions are not explained in detail. 
There are other studies in the Pomeron and meson exchange models.
We briefly discuss these ideas in the following.
For the details of formalism, the reader may read the original papers.
Historically, the first estimate of shadowing contribution 
to $I_G$ is discussed by the Pomeron exchange model \cite{BK, ZOLLER}.
A possible way of describing the high-energy scattering in the
diffractive region is in terms of Pomeron exchange.
The virtual photon transforms into a $q\bar q$
pair which then interacts with the deuteron.
In the diffractive case, the target is remain intact
and only vacuum quantum number, namely the Pomeron, could be exchanged
between the $q\bar q$ pair and the nucleons.
In the earlier works, the shadowing correction in this model
was rather large $\delta I_G\approx - 0.08$ \cite{ZOLLER,BGNPZ}.
However, the Pomeron contribution is reduced if more realistic
deuteron wave functions are used according to Ref. \cite{MST}.
Next, meson-exchange corrections were investigated in Refs. \cite{KU,MST}.
The studied mesons are $\pi$, $\omega$, and $\sigma$ in Ref. \cite{KU},
and $\rho$ is also included in Ref. \cite{MST}.
The formalism is essentially the same with the one in subsection 
\ref{MESON-1}.
If the corrections due to the $\pi$, $\omega$, and $\sigma$ mesons were
taken into account, the NMC result became $I_G=0.29\pm 0.03$ \cite{KU}. 
Therefore, meson-exchange contributions reduce the discrepancy
between the NMC data and the Gottfried sum.

\begin{figure}[h]
\parbox[b]{0.46\textwidth}{
   \begin{center}
      \epsfig{file=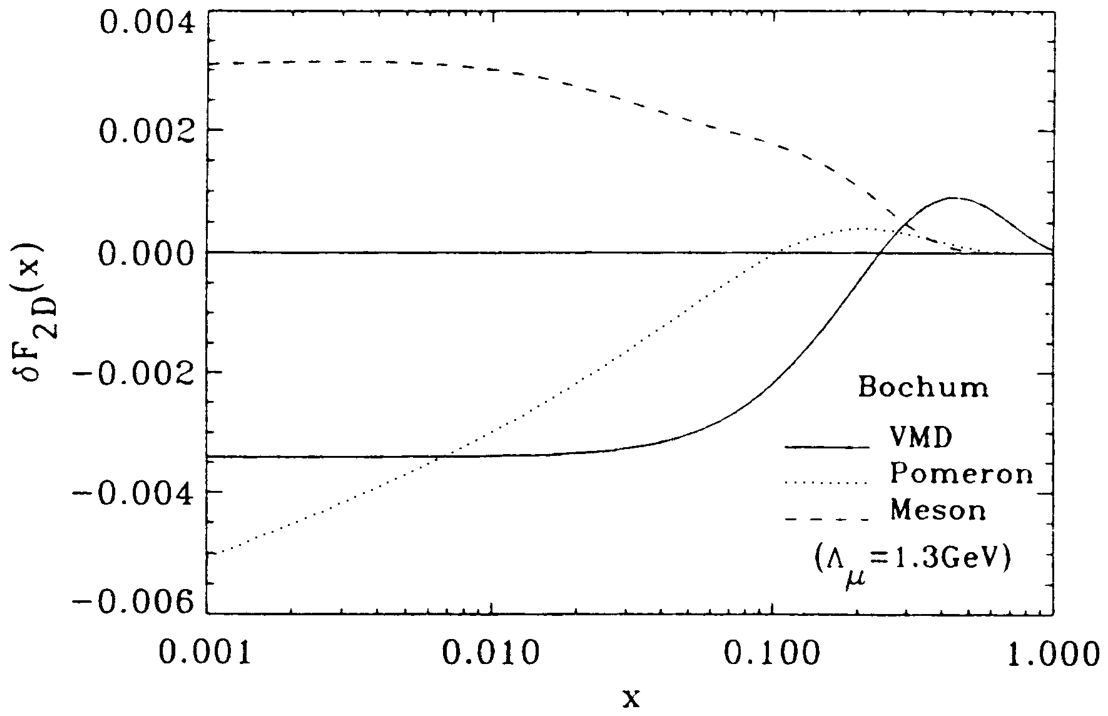,width=6.0cm}
   \end{center}
 \vspace{-0.5cm}
   \caption{\footnotesize Vector-meson-dominance, Pomeron, and
                       meson-exchange contributions at $Q^2$=4 GeV$^2$
                       (taken from Ref. {\normalsize\cite{MST}}).}
   \label{fig:each}
}\hfill
\parbox[b]{0.46\textwidth}{
   \begin{center}
      \epsfig{file=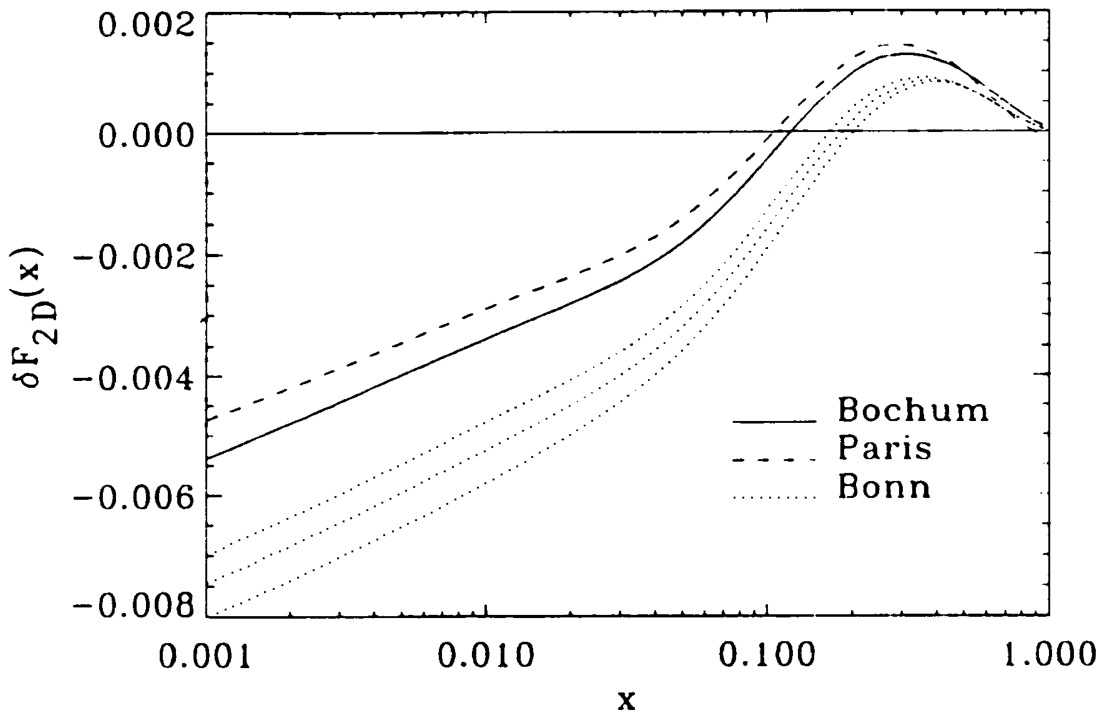,width=6.0cm}
   \end{center}
 \vspace{-0.5cm}
\caption{\footnotesize 
          Total shadowing in the deuteron. 
          The Bochum, Bonn, and Paris wave functions are used
                       (taken from Ref. {\normalsize\cite{MST}}).}
\label{fig:all}
}
\end{figure}

The VMD contributions are compared with the Pomeron and meson
exchange results in Fig. \ref{fig:each}.
The Pomeron contribution is of the same order of magnitude with
the VMD effect at $Q^2$=4 GeV$^2$.
Because the meson exchange produces extra sea-quark distributions,
its effects show antishadowing. This fairly large antishadowing
cancels much of the shadowing produced by the vector-meson dominance
and the Pomeron exchange. 
The shadowing due to the Pomeron is rather small at $x>0.05$ compared
with other contributions, and it becomes comparable only 
in the small $x$ region, $x<0.01$.
Adding these contributions, we show the total deuteron shadowing
in Fig. \ref{fig:all}. From these results, we obtain
the correction to the NMC analysis at $Q^2$=4 GeV$^2$:
$F_2^p-F_2^n=(F_2^p-F_2^n)_{NMC}+\delta F_{2D}$.
The correction to the sum ranges
from $\delta I_G$=$-$0.026 to $-$0.010 depending on
the nuclear potential \cite{MST}.
The description in Ref. \cite{MST} is consistent with
the Fermilab-E665 data \cite{E665} on the deuteron shadowing.
This fact suggests the above result of $\delta I_G$ is a correct estimate.

\vspace{-0.5cm}
\noindent

In the beginning, the estimated shadowing effects on the Gottfried sum
were fairly large, $\delta I_G=-0.08$.
However, the recent numerical values seem to converge into
about $\delta I_G=-0.02$ although there are still
uncertain factors due to the nuclear potential.
In comparison with the NMC value $I_G=0.235\pm 0.026$,
it is about 10\% effect. 
The shadowing studies do not alter the NMC conclusion.
However, it has to be taken into account carefully because
the shadowing magnifies the deviation from the Gottfried sum.

\subsection{Parametrization of antiquark distributions}\label{PARAMET}

There are various factors which affect the NMC finding.
Even the failure of the Gottfried sum is not undoubtedly 
confirmed. So present parametrizations of the flavor asymmetry
$\bar u-\bar d$ is subject to change depending on future
experimental results. We introduce several parametrizations
in the following, but these should be considered as 
preliminary versions. If independent Fermilab Drell-Yan experiments
confirm the NMC result and the NA51, they should be taken seriously.

Early version of the parametrization was proposed in Ref. \cite{FF}
for explaining the SLAC data.
The Fermilab-E288 collaboration analyzed its Drell-Yan data 
and obtained a parametrization in 1981 \cite{E288}. 
After the NMC measurement, several parametrizations have been proposed.
The first one is Ref. \cite{PRS}. 
They find that the global parametrization MRS-B (1988)
overestimates the NMC $F_2^p-F_2^n$ data at small $x$
even though it works for the neutrino structure function $xF_3$
and the Gross-Llewellyn Smith sum rule. The differences between
the MRS-B and the NMC data are used for finding the flavor asymmetric
distribution. In the similar way, the difference from
the parametrization EHLQ1 is used for finding $\bar u-\bar d$
in Ref. \cite{ES}.

New global MRS parametrizations were proposed by including the NMC data.
The total sea-quark distribution at $Q_0^2$ is parametrized as
\begin{equation}
xS = 2x(\bar u + \bar d + \bar s + \bar c)
   = A_S x^{-\lambda} (1-x)^{\eta_S}
         (1+\epsilon_S x^{1/2}+\gamma_S x)
\ \ \ ,
\end{equation}
and each distribution is
\begin{align}
\bar u &= 0.2 \, S \, (1 - \delta) - \Delta/2 \ \ , \nonumber \\
\bar d &= 0.2 \, S \, (1 - \delta) + \Delta/2 \ \ , \nonumber \\
\bar s &= 0.1 \, S \, (1 - \delta)            \ \ ,           \\
\bar c &= \delta \, S /2                   \ \ , \nonumber \\
\text{with} \ \ \ \ \ 
x \Delta &= A_\Delta x^{\eta_\Delta} (1-x)^{\eta_S} (1+\gamma_\Delta x)
\ \ .
\end{align}
The parameters are determined by fitting many experimental data.

\begin{wrapfigure}{r}{6.5cm}
   \begin{center}
      \mbox{\epsfig{file=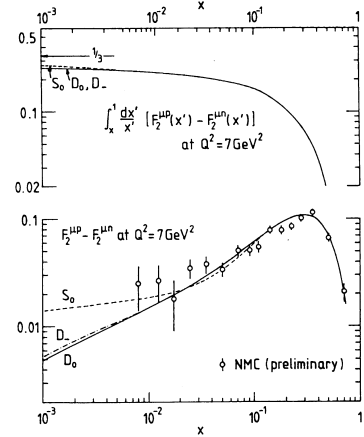,width=5.0cm}}
   \end{center}
 \vspace{-0.6cm}
\caption{\footnotesize MRS-1993 parametrizations are compared with
                       NMC data $F_2^p-F_2^n$ 
                       (taken from Ref. {\normalsize\cite{MRS}}).  
}
\label{fig:mrs}
\end{wrapfigure}
\quad
The 1993 version is a good example in showing flavor symmetric
and asymmetric distributions in comparison with the NMC $F_2$ data \cite{MRS}.
Therefore, we explain them in the following.
Three possibilities are studied in the parametrization:
(1) S (same),      flavor symmetric sea $\bar u=\bar d$,
(2) D$_0$ (different),   asymmetric sea $\bar u\ne \bar d$,
(3) D$_-$,               asymmetric sea $\bar u\ne \bar d$ with a singular
gluon distribution. 
The parametrizations are compared with the NMC data,
and they explain the data fairly well as shown in Fig. \ref{fig:mrs}.
The flavor symmetric distribution ($S_0$) deviates from 
the asymmetric ones ($D_0$, $D_-$) only at small $x(<0.01)$,
where the data do not exist.
In the NMC kinematical region, it is not possible to detect
the differences between these parametrizations.
Because the $S_0$ distribution recovers the Gottfried sum 1/3,
there is significant contribution from the very small $x$ region.
However, the sum 1/3 can be reached only at very small $x\approx 10^{-10}$.
The situation should be clarified by the Fermilab Drell-Yan
experiments.
The MRS group published new ones after the MRS-1993 version,
in particular by including the NA51 asymmetry, HERA data, and
the single jet cross sections at the Fermilab $p\bar p$ collider.
Obtained parameters of the recent 1996 version 
are listed in Table \ref{tab:MRS} together with those of the 1993 version. 
The comparison of the recent one with other parametrizations
is discussed in the end of this subsection.

\vspace{+0.0cm}
\begin{table}[h]
\hspace{0.0cm}
\footnotesize
\begin{center}
\begin{tabular}{||c||c|c|c||c|c|c|c||} \hline
Year           & \multicolumn{3}{c||}{1993} &
                 \multicolumn{4}{c||}{1996}  \\
\hline
Name           &    $S_0$   &    $D_0$   &   $D_-$   &
                     $R_1$  &    $R_2$   &   
                     $R_3$  &    $R_4$   \\
\hline
\ $Q_0^2$      &     4.0    &    4.0     &     4.0   & 
                     1.0    &    1.0     & 
                     1.0    &    1.0     \\
\hline
$\Lambda_{\overline{MS}}^{n_f=4}$
                  &  215     &  215      &  215      & 
                     241     &  344      & 
                     241     &  344      \\
\hline
$A_S$           &    1.87  &    1.93   &    0.054  & 
                     0.42  &    0.37   &
                     0.92  &    0.92   \\
$\lambda$       &    0     &    0      &    0.5    &
                     0.14  &    0.15   &
                     0.04  &    0.04   \\ 
$\eta_S$        &   10.0   &   10.0    &    6.5    & 
                     9.04  &    8.27   & 
                     9.38  &    8.93   \ \\
$\epsilon_S$    & $-$2.21  & $-$2.68   &   19.5    &
                     1.11  &    1.13   &
                  $-$1.65  & $-$2.34   \ \\
$\gamma_S$      &   6.22   &    7.38   & $-$3.28   &
                    15.5   &   14.4    &
                    11.8   &   12.0    \ \\
$A_\Delta$      &    0     &    0.163  &    0.144  &
                     0.039 &    0.036  & 
                     0.040 &    0.038  \ \\
$\eta_\Delta$    &    /    &   0.45    &    0.46   &
                     0.3   &   0.3     &
                     0.3   &   0.3     \ \\
$\gamma_\Delta$ &     /    &    0      &    0      &
                    64.9   &  64.9     & 
                    64.9   &  64.9     \ \\
$\delta$        &    0     &   0       &    0      &
                     0     &   0       &
                     0     &   0       \ \\
\hline
\end{tabular}
\end{center}
\vspace{-0.3cm}
\caption{\footnotesize Parameters in the MRS 1993 and 1996 versions.
            The $Q_0^2$ and $\Lambda_{\overline{MS}}^{n_f=4}$ are listed in the
            units of GeV$^2$ and MeV respectively.}
\label{tab:MRS}
\end{table}
\normalsize

\vspace{-0.2cm}

There are other parametrizations, for example by 
the CTEQ (Coordinated Theoretical/Experimental Project on QCD 
Phenomenology and Tests of the Standard Model) group \cite{CTEQ}.
The recent CTEQ parametrizations included the HERA data,
which provided information on
the small $x$ behavior of the parton distributions.
The HERA data and others from the CCFR, 
the Collider Detector at Fermilab (CDF), the NA51 on $\bar u/\bar d$
are included in the new parametrization analyses.
The functional forms of the parton distributions are
similar to the MRS ones, and they are provided at $Q^2$=2.56 GeV$^2$.
According to the $\overline{MS}$ version of the CTEQ4 parametrization,
the light antiquark distributions are obtained as \cite{CTEQ}
\begin{align}
x(\bar d+\bar u)/2 &= 0.255 x^{-0.143} (1-x)^{8.041} 
                       (1+6.112 \sqrt{x}+ x) 
                                                     \ \ , \nonumber \\
x(\bar d-\bar u)   &= 0.071 x^{0.501} (1-x)^{8.041}
                       (1+30.0x) 
                                                     \ \ ,      
\end{align}
at $Q^2$=2.56 GeV$^2$. 
The scale parameter is $\Lambda_{\overline{MS}}^{n_f=5}$=202 MeV.

In contrast to the above parametrizations, the GRV (Gl\"uck, Reya, and Vogt)
model supplies input distributions at very small $Q^2$ ($\approx$0.3 GeV$^2$).
The original motivation was to set $\bar q(x)=g(x)=0$ at certain small
$Q^2$ ($\equiv \mu^2$) by allowing only the valence-quark distributions. 
Then, the sea-quark and gluon distributions are considered to be produced 
perturbatively through the evolution from $\mu^2$. This attempt is slightly
modified to the form including valence-like sea-quark and gluon distributions
even at $\mu^2$ so as to explain the HERA data. 
Although it would be dubious that the perturbative QCD 
can be used in such a small $Q^2$ region, the model seems successful
in explaining various experimental data.
The light-antiquark distributions are given at $Q^2$=0.23 GeV$^2$ \cite{GRV}:
\begin{align}
x(\bar d+\bar u) &= 1.09 \, x^{0.30} \, (1+2.65 \, x)  \, (1-x)^{8.33}
                                                 \ \ , \nonumber \\
x(\bar d-\bar u) &= 0.0525 \, x^{0.381} \, (1+15.2 \, x+132 \, x^{3/2}) 
                           \, (1-x)^{8.65}
                                                 \ \ , 
\end{align}
with $\Lambda_{NLO,\overline{MS}}^{(n_f=3,4,5)}$=248, 200, and 131 MeV.

\begin{floatingfigure}{7.0cm}
   \begin{center}
      \mbox{\epsfig{file=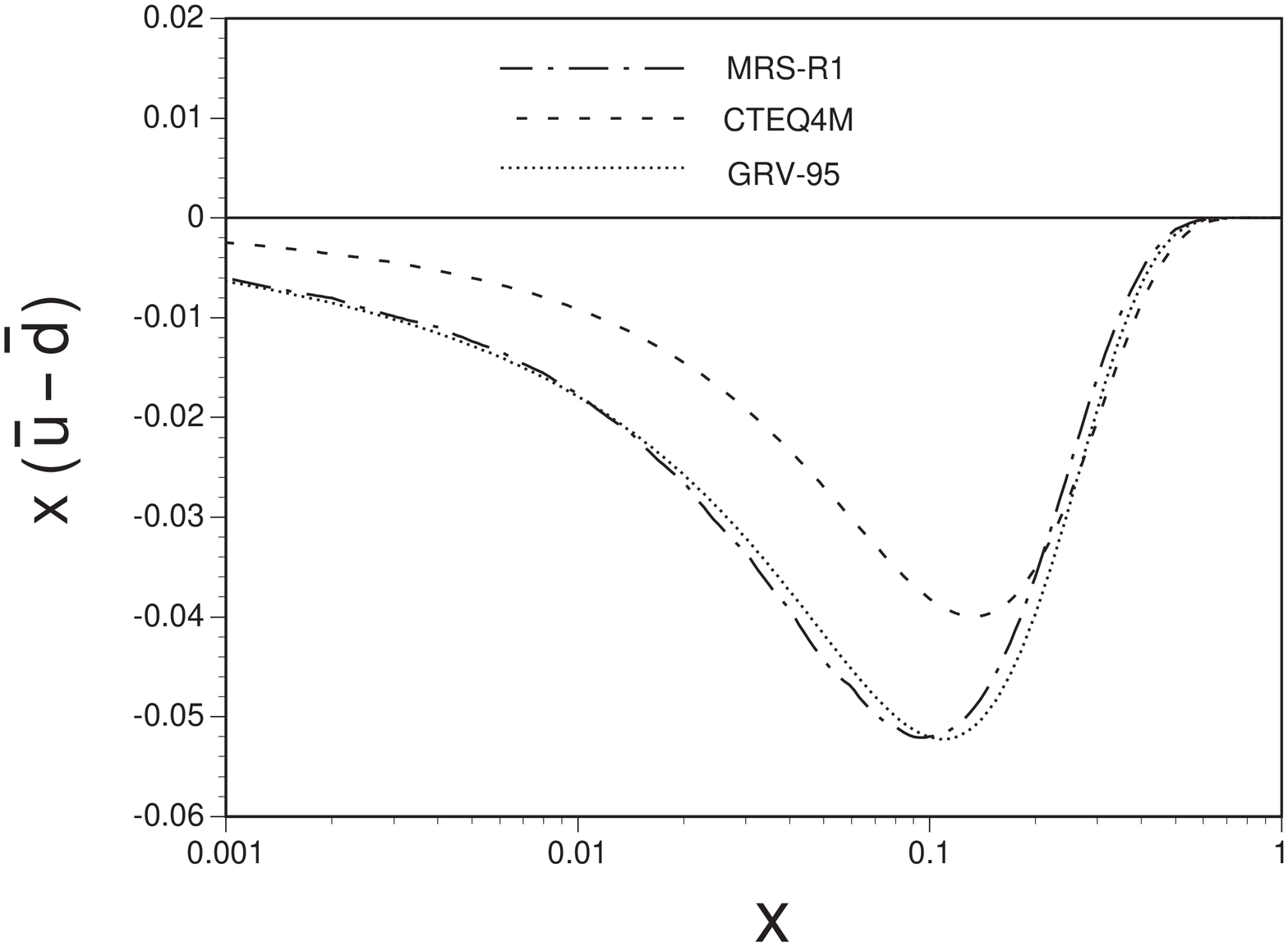,width=6.2cm}}
   \end{center}
 \vspace{-0.5cm}
\caption{\footnotesize $x(\bar u-\bar d)$ distributions at $Q^2$=4 GeV$^2$
                       in the MRS-R1, CTEQ4M, and GRV-95 parametrizations.} 
\label{fig:para}
\end{floatingfigure}
\quad
The recent $x(\bar u-\bar d)$ distributions in the
$\overline{MS}$ scheme are plotted in Fig. \ref{fig:para}.
They are the MRS-R1 \cite{MRS}, CTEQ4M \cite{CTEQ}, and GRV-95 \cite{GRV}
distributions at $Q^2$=4 GeV$^2$. Because the GRV-95 was made so as to agree
with the older version (MRS-A) of MRS at $Q^2$=4 GeV$^2$, 
the MRS-R1 and GRV-95 distributions are almost
the same. Even though the CTEQ4M agrees with the others in the region $x>0.2$,
it is very different in the small $x$ region.
The CTEQ4M distribution is much smaller than the others.
Because the NA51 Drell-Yan result is taken into account in 
the parametrizations, these distributions are almost the same at $x=0.18$. 
However, we do not have enough constraint in the small $x$ region.
The only one is the NMC data on the Gottfried sum. 
The small $x$ region should be clarified by future experiments,
for example by the Fermilab-E866.

Updated information on the various parametrizations of 
the parton distributions is given at http://durpdg.dur.ac.uk/HEPDATA/PDF.

\vfill\eject
\section{{\bf Expectations in perturbative QCD}}\label{PQCD}
\setcounter{equation}{0}
\setcounter{figure}{0}
\setcounter{table}{0}

According to the NMC conclusion, the Gottfried sum rule
should be violated. In this section, we discuss how much
corrections are expected in perturbative QCD.
First, a general treatment of operator product expansion is
discussed. Then possible perturbative QCD corrections
to the sum rule are discussed.

\subsection{Operator product expansion}\label{OPE}

In order to discuss QCD corrections to the Gottfried sum rule,
we introduce operator-product expansion which is used
in applying perturbative QCD methods to the structure functions.
The hadron tensor $W_{\mu\nu}$ is expressed as the current product
in Eq. (\ref{eqn:wmunu}). It is known in the light-cone limit
$\xi^2\rightarrow 0$ that the product
is expressed in terms of local operators and their coefficients. 
For example if the current is given by
$J_\mu (\xi)=\bar\psi(\xi)\gamma_\mu {\cal Q} \psi(\xi)$ with
the charge matrix $\cal Q$ in the free massless Dirac theory, 
it becomes \cite{IZ}
\begin{multline}
\frac{1}{2} \left ( \, [J_\mu (\xi/2),J_\nu (-\xi/2)] 
                   +[J_\nu (\xi/2),J_\mu (-\xi/2)] \, \right )
    \\
\begin{CD}
@>>{\xi^2\rightarrow 0}>
\end{CD}
      \sum_{odd\ n} \, \frac{i}{2^{n} n! \pi} \, 
             \xi^{\mu_1} \cdot\cdot\cdot \xi^{\mu_n} \, 
             S_{\mu\alpha\nu\beta} \, 
        \left [ \, \partial_\xi^\alpha \, \delta(\xi^2) \,
                  \varepsilon(\xi^0) \, \right ] \, 
             O_{\mu_1 \cdot\cdot\cdot \mu_n}^\beta
\ \ ,
\label{eqn:JJ}
\end{multline}
where $\delta(\xi^2)$ is the $\delta$ function, 
$\varepsilon(\xi^0)$ is a step function: 
$\varepsilon(\xi^0)=+1$ for $\xi^0>0$
               and $-1$ for $\xi^0<0$,
$S_{\mu\alpha\nu\beta}$ is given by
$S_{\mu\alpha\nu\beta}=$$g_{\mu\alpha}g_{\nu\beta}
                      +g_{\mu\beta}g_{\nu\alpha}$$
                      -g_{\mu\nu}g_{\alpha\beta}$,
and the operator is defined by 
\begin{equation}
O_{\mu_1 \cdot\cdot\cdot \mu_n}^\beta = i \bar\psi (0) {\cal Q}^2
 \gamma^\beta \stackrel{\leftrightarrow}{\partial}_{\mu_1}\cdot\cdot\cdot
\stackrel{\leftrightarrow}{\partial}_{\mu_n} \psi (0)
\ \ \ .
\end{equation}

Virtual forward Compton amplitude is usually analyzed 
instead of the hadron tensor $W_{\mu\nu}$,
because it is more convenient to use time-ordered product
and to treat interference terms.
The hadron tensor is related to the imaginary part of
the Compton amplitude by the optical theorem
$2M \, W_{\mu\nu}=(4/\pi) \, Im T_{\mu\nu}$, where
$T_{\mu\nu}$ is given by the time-ordered product of currents:
\begin{equation}
T_{\mu\nu}(q^2,\nu)=\frac{i}{4} \, \, \overline{\sum_\sigma} 
          \int d^4 \xi \,  e^{iq\cdot \xi}
          <p, \sigma \, | \, T(J_\mu(\xi) J_\nu(0)) \, | \, p, \sigma >
\ \ \ .
\label{eqn:COMPTON}
\end{equation}
Here, only the unpolarized case is considered.
The amplitude is decomposed into three invariant ones \cite{BBDM}:
\begin{equation}
T_{\mu\nu}(q^2,\nu)=e_{\mu\nu} T_L(q^2,\nu)+d_{\mu\nu} T_2(q^2,\nu)
                   -i\epsilon_{\mu\nu\alpha\beta}
                  \, \frac{p^\alpha p^\beta}{\nu} \, T_3(q^2,\nu)
\ \ \ ,
\label{eqn:TK}
\end{equation}
where the tensors are defined by
$e_{\mu\nu}=g_{\mu\nu}-q_\mu q_\nu/q^2$ and
$d_{\mu\nu}=-p_\mu p_\nu q^2 /\nu^2 +(p_\mu q_\nu +p_\nu q_\mu)/\nu
            -g_{\mu\nu}$.
The amplitude $T_L$ is the longitudinal one, $T_2$ is the longitudinal
plus transverse one, and $T_3$ appears only in the weak current case.

As it is shown in Eq. (\ref{eqn:JJ}), a product of current operators
could be written by local operators and their coefficients.
The singular behavior at $\xi^2\rightarrow 0$ can be absorbed into
the coefficients. Therefore,
the Compton amplitude is expanded in terms of possible operators.
However, infinite number of operators contribute to the amplitude
in the expansion near the light cone.
A convenient way to classify the contributions is to introduce
twist $\tau$, which is defined by the mass dimension of the operator 
minus its spin: $\tau=d_O-n$.
For example, the twist for the operator 
$\bar\psi \gamma^\beta {\partial}_{\mu_1} 
\cdot\cdot\cdot {\partial}_{\mu_n} \psi$
is two because the mass dimension of $\psi$ is 3/2,
the dimension of the derivatives are n, and the spin is n+1.
In this way, the current product is expanded near the light cone,
and the amplitude becomes \cite{MUTA,RGRBOOK}
\begin{equation}
i \,  T(J (\xi) J (0)) \longrightarrow
\sum_{\tau=2}^\infty \sum_{n=0}^\infty C_{n}^\tau (\xi^2,\mu^2) 
\, \xi^{\mu_1} \cdot\cdot\cdot \xi^{\mu_n} \, 
O_{\mu_1 \cdot\cdot\cdot \mu_n}^\tau (\mu^2) 
\ \ \ ,
\label{eqn:OPE}
\end{equation}
where $C_n^\tau (\xi^2,\mu^2)$ are called coefficient functions
and $O_{\mu_1 \cdot\cdot\cdot \mu_n}^\tau (\mu^2)$ are operators.
For simplicity, the Lorentz indices $\mu$ and $\nu$ are dropped
in the above equation.
In the case of interacting fields, it is necessary to introduce
a scale $\mu^2$ in renormalizing the operators. This is the reason
why explicit dependence on the renormalization point $\mu^2$
is written in the above equation. In this way,
the Compton amplitude is factorized into the long distance part
and the light-cone part which could be handled in perturbative QCD.
As it is given in Eq. (\ref{eqn:JJ}), 
the product of the currents has a singular
behavior in the limit $\xi^2 \rightarrow 0$, so that the coefficients
could be written in a singular form
$C_n^\tau (\xi^2) \sim (1/\xi^2)^{d_C/2}$.
Counting dimensions in Eq. (\ref{eqn:OPE}), we obtain
$d_C=n-d_O+2d_J= -\tau +2d_J$ where $d_O$ and $d_J$ are mass dimensions
of the operator and the current.  
From the dimensional counting, we find that the lowest-twist contribution, 
namely the twist-two, 
is most singular in the operator product expansion. 
From Eqs. (\ref{eqn:COMPTON}), (\ref{eqn:TK}), and (\ref{eqn:OPE}), 
the Compton amplitude becomes
\begin{equation}
T (q^2,\nu)  \longrightarrow
                   \sum_{\tau,n} \overline{C}_{n}^{\, \tau}(Q^2,\mu^2) 
                       \, \overline{O} _{n}^{\, \tau} (\mu^2) \, 
                  \frac{1}{x^n}
\ \ \ ,
\label{eqn:TOPE}
\end{equation}
where $T (q^2,\nu)$ represents $T_L$, $T_2$, or $T_3$.
The above $\overline{C}_{n}^{\, \tau}(Q^2,\mu^2)$ and  
$\overline{O}_n^{\, \tau} (\mu^2)$
are defined by 
\begin{align}
 \int d^4 \xi \, e^{iq\cdot\xi} \, C_n^\tau (\xi^2,\mu^2) \,  
    \xi^{\mu_1} \cdot\cdot\cdot \xi^{\mu_n}
&= \frac{q^{\mu_1} \cdot\cdot\cdot q^{\mu_n}}{(Q^2/2)^n}
     \, \overline{C}_{n}^{\, \tau} (Q^2,\mu^2) \, 
\ \ \ , \\
\frac{1}{4} \, \, \overline{\sum_\sigma} <p ,\sigma \, 
              | \, O_{\mu_1 \cdot\cdot\cdot \mu_n}^\tau (\mu^2) \, 
              | \, p, \sigma >
&=\overline{O}_n^{\, \tau} (\mu^2) \, p_{\mu_1} \cdot\cdot\cdot p_{\mu_n}
\ \ \ .
\end{align}
In relating the Compton amplitudes to the structure functions,
the following dispersion relation is used:
\begin{align}
T(q^2,\nu) = \frac{2}{\pi} \int _{-q^2/2M}^\infty 
             \frac{\nu' d\nu'}{{\nu'}^2-\nu^2} \, Im T(q^2,\nu)
          &= \int _{-q^2/2M}^\infty 
             \frac{\nu' d\nu'}{{\nu'}^2-\nu^2} \, M \, W (q^2,\nu)
\nonumber \\
          &= \sum_n \frac{1}{x^n} \int _{0}^1 dx'
                \, {x'}^{n-1} \, M \, W (q^2,x')
\ .
\label{eqn:disp}
\end{align}
Comparing Eq. (\ref{eqn:TOPE}) with Eq. (\ref{eqn:disp}),
we obtain moments of the corresponding structure function.
They are then expressed by the scaling functions:
\begin{equation}
\int_{0}^{1} dx x^{n-1} F_1(x,Q^2) = 
           \sum_{\tau} \overline{C}_{1,n}^\tau (Q^2,\mu^2) \, 
                       \overline{O}_{1,n}^{\, \tau} (\mu^2)
\ \ , \ \ \ n=2, 4, 6 \cdot \cdot \, \cdot  
\ \ \ ,
\label{eqn:FKN}
\end{equation}
and similar equations for $F_2$ and $F_3$, except that
the moments are given by $\int dx x^{n-2} F_2$ in the
$F_2$ case. 
Because of the crossing properties of the structure function
under $\mu\leftrightarrow\nu$ and $x\leftrightarrow -x$,
the only even-spin operators contribute in Eq. (\ref{eqn:FKN}).
The moments of the structure functions are thus given by
the long-range part, which cannot be calculated 
without resorting to nonperturbative methods such as lattice QCD,
and the light-cone part which can be evaluated in perturbative QCD.

There exist only even twists in the expansion Eq. (\ref{eqn:FKN})
in the massless quark case. Therefore, higher-twist contributions
are suppressed by the factor of $1/Q^2$ compared with
the twist-two.
The Gottfried sum rule is a flavor nonsinglet one.
A twist-two nonsinglet operator is given by
\begin{equation}
O_{\mu_1 \cdot\cdot\cdot \mu_n}^{\tau=2, NS}
= \frac{i^{n-1}}{n!}
\left [ \bar \psi \frac{\lambda^a}{2} \gamma_{\mu_1} 
               D_{\mu_2} \cdot\cdot \cdot D_{\mu_n} \psi
         + permutations \right ]
\ \ \ ,
\end{equation}
where $D^\mu$ is the covariant derivative 
$D_\mu =\partial_\mu -ig T^a A_\mu^a$ with
eight generators $T^a$ of the color SU(3) group. 

The renormalization point $\mu^2$ is an arbitrary constant,
so that physical observable should not depend on its scale.
This fact leads to a renormalization group equation.
It can be applied to the coefficients $C_{k,n}^\tau (Q^2,\mu^2)$
by comparing a renormalization group equation for a
Green's function with the one for the local operator. 
In the nonsinglet case, it is given by
\begin{equation}
   \left [ \mu \frac{\partial}{\partial \mu} 
    + \beta(g) \frac{\partial}{\partial g}
    - \gamma^{n,NS}(g) \right ]
\overline{C}_{k,n}^{\, NS} \left ( \frac{Q^2}{\mu^2},g^2\right ) = 0
\ \ \ ,
\label{eqn:RGE}
\end{equation}
where $k$ indicates the structure-function type ($k$=1, 2, or 3)
and $\tau$ is omitted for simplicity.
The $\gamma^{n,NS}$ is anomalous dimension of the operator
which is related to the renormalization factor of the operator
($Z_n^{NS}=O_n^{0,NS}/O_n^{NS}$)
by $\gamma^{n,NS}(g)=\mu (\partial / \partial \mu) \ln Z_n^{NS}$.
The $\beta$ function is given by 
$\beta(g)=\mu (\partial / \partial \mu) g(\mu)$.
The solution of Eq. (\ref{eqn:RGE}) is
\begin{equation}
\overline{C}_{k,n}^{\, NS} \left (\frac{Q^2}{\mu^2},g^2\right ) =
\overline{C}_{k,n}^{\, NS} \left (1,\bar g ^2 \right )
exp \left [ -\int_{\bar g(\mu^2)}^{\bar g(Q^2)} dg' 
                 \frac{\gamma^{n, NS}(g')}{\beta(g')} \right ]
\ \ \ .
\end{equation}
The anomalous dimension, coefficient function,
and $\beta$ function are expanded
in $\alpha_s$:
$\gamma^{n,NS}(g)= \gamma_0^{n,NS} (g^2/16\pi^2) 
                 + \gamma_1^{n,NS} (g^2/16\pi^2)^2
                   + \cdot\cdot\cdot$,
$\overline{C}_{k,n}^{NS} (1,\bar g^2)
       = 1 + B_{k,n}^{NS} (\bar g^2/16\pi^2) 
           + \cdot\cdot\cdot$,
and 
$\beta(g)= - g [  \beta_0 (g^2/16\pi^2) 
                + \beta_1 (g^2/16\pi^2)^2 +\cdot\cdot\cdot ]$.
Then the moments of the structure function become
\begin{equation}
M_{k,n}^{NS}(Q^2) = M_{k,n}^{NS}(Q_0^2) 
       \left [ \frac{\alpha_s(Q^2)}{\alpha_s(Q_0^2)} \right ]^{d_n}
       \left [ 1 + C_{k,n}^{NS} \left ( 
              \frac{\alpha_s(Q^2)-\alpha_s(Q_0^2)}{4\pi} \right ) \right ]
\ \ \ ,
\label{eqn:MNNS}
\end{equation}
where 
\begin{equation}
d_n= \frac{\gamma_0^{n,NS}}{2 \beta_0}
\ \ , \ \ \ 
C_{k,n}^{NS}=B_{k,n}^{NS} 
       + \frac{\gamma_1^{n,NS}}{2\beta_0}
       - \frac{\beta_1 \, \gamma_0^{n,NS}}{2\beta_0^2}
\ \ \ .
\end{equation}
Because $Q_0^2$ is an arbitrary scale, it is often convenient to 
express the above equation without $Q_0^2$:
\begin{equation}
M_{k,n}^{NS}(Q^2) = A_{k,n}^{NS}  \left [ \alpha_s(Q^2) \right ]^{d_n}
                   \left [ 1 + C_{k,n}^{NS}  
                         \frac{\alpha_s(Q^2)}{4\pi}
                \right ]
\ \ \ ,
\label{eqn:MNNS2}
\end{equation}
where $A_{k,n}^{NS}$ is a constant given 
by $M_{k,n}^{NS}(Q_0^2)=A_{k,n}^{NS}
[1+C_{k,n}^{NS}\alpha_s(Q_0^2)/(4\pi)]
[\alpha_s(Q_0^2)]^{d_n}$.
In getting various sum rules, 
$A_{k,n=1}^{NS}$ may be evaluated in the parton model.
Then LO and NLO anomalous dimensions 
$\gamma_{0}^{1,NS}$ and $\gamma_{1}^{1,NS}$ are calculated by
studying renormalization of the nonsinglet operator.
In order to obtain $B_{k,n=1}^{NS}$, we calculate first
perturbative correction to the Compton amplitude and
then $O_{k,n}^{\tau=2, NS}(p^2/\mu^2,g^2)$ by considering
a matrix element of the nonsinglet operator between
quark states. From these results, the NLO correction
to the coefficient function $B_{k,1}^{NS}$ is obtained \cite{BBDM}.
Combining these anomalous dimensions and the coefficient,
we obtain the NLO correction $C_{k,n=1}^{NS}$ in Eq. (\ref{eqn:MNNS2}).

\subsection{Perturbative correction to the Gottfried sum}\label{QCDIG}

In the previous subsection, it is derived how the moments
of a structure function at certain $Q^2$ can be calculated 
with given moments at $Q_0^2$ by using the prescriptions
of the operator product expansion and the 
renormalization-group equation. Before discussing the Gottfried sum rule,
we first check NLO corrections to another nonsinglet quantity, for example
the Gross-Llewellyn Smith sum rule.
It is related to the $F_3$ structure functions in
neutrino scattering: 
$\int dx (F_3^{\nu N}+F_3^{\bar\nu N})/2$,
where $F_3^N=(F_3^p+F_3^n)/2$. 
In the parton model without NLO effects, 
$(F_3^{\nu N}+F_3^{\bar\nu N})/2$ is given by
$u_v(x,Q^2)+d_v(x,Q^2)$ so that it's integration over $x$ is 
three ($A_1^{NS}=3$). Because the first LO nonsinglet
anomalous dimension vanishes ($\gamma_0^{n=1,\, NS}$=0),
the coefficient $d_1$ becomes $d_{1}$=0.
The NLO corrections are given by $B_{3,\, 1}^{NS}=-4$ 
and $\gamma_{1}^{n=1,\, NS(-)}$=0 \cite{BURA},
so that we obtain $C_{3,1}^{NS(-)}=-4$. 
The notation $NS(-)$ indicates a $q-\bar q$ type nonsinglet distribution.
Including the NLO correction, we have the sum rule:
\begin{equation}
\int_0^1 dx \left [ F_3^{\nu N}(x,Q^2)+F_3^{\bar\nu N}(x,Q^2) \right ] 
              \, / \, 2 \, = 
         3 \left [ 1 - \frac{\alpha_s(Q^2)}{\pi} \right ]
\ \ \ .
\label{eqn:glssum}
\end{equation}
It is evaluated as 2.66$\pm$0.04 with the QCD scale parameter
$\Lambda$=210$\pm$50 MeV. 
The Columbia-Chicago-Fermilab-Rochester (CCFR) neutrino data
[$2.50\pm 0.018(stat.) \pm 0.078(syst.)$ \cite{CCFR}]
confirmed the QCD correction within the experimental errors.

Odd-spin operators contribute to $F_3^{\nu+\bar\nu}$, so that
there is no problem in deriving Eq. (\ref{eqn:glssum}).
On the other hand, the $F_2$ moments are given only for even n 
(see in Eq. (\ref{eqn:FKN})), and
the Gottfried sum is the n=1 moment of $F_2^p-F_2^n$.
Strictly speaking, the only even-n anomalous dimensions and coefficient
functions have meaning. However, the QCD parton model is successful
in reproducing the OPE results, and it provides all the moments. 
Therefore, the perturbative corrections are studied by analytically
continuing the even-n results to the odd-n values.

\begin{wrapfigure}{r}{6.0cm}
   \begin{center}
      \mbox{\epsfig{file=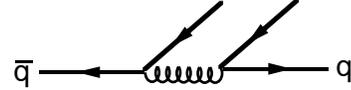,width=5.0cm}}
   \end{center}
 \vspace{-0.3cm}
\caption{\footnotesize NLO contribution to the splitting 
                       $\bar q \rightarrow q$.}
\label{fig:nlo}
\end{wrapfigure}
\quad
The NLO correction to the Gross-Llewellyn Smith sum rule
is about 11\%; however, 
the correction to the Gottfried sum is very different.
Because the NLO term in the coefficient function vanishes
($B_{2,1}^{NS}=0$) for the nonsinglet structure function $F_2^{NS}$,
the only contribution is from the NLO anomalous dimension 
$\gamma_{1}^{NS(+)}$. 
Because the structure-function combination in Eq. (\ref{eqn:glssum})
is given by $F_3^{\nu N}+F_3^{\bar\nu N}=(u-\bar u)+(d-\bar d)$
in the leading order, it is a $q-\bar q$ type distribution. 
On the other hand, the Gottfried integrand is given
in the parton model by
$(F_2^p-F_2^n)/x=(1/3)[(u+\bar u)-(d+\bar d)]$, which is a
$q+\bar q$ type.
This difference makes the anomalous dimension
$\gamma_1^{n=1,\, NS(+)}$ finite.
Even though the LO anomalous dimension vanishes in both cases,
there is a finite contribution from the NLO process in 
Fig. \ref{fig:nlo}.
Namely, the $\bar q\rightarrow q$ splitting becomes possible.
Because evolution of the $q\pm \bar q$ distributions is controlled
by the splitting functions $P_{qq}\pm P_{q\bar q}$, the $q+\bar q$
evolution is different from the $q-\bar q$ one \cite{KKKM}.
Because of baryon number conservation, the first anomalous dimension
in the $NS(-)$ case has to vanish. However,
there is an extra contribution from the $P_{q\bar q}$
(note: $P_{qq}+P_{q\bar q}=[P_{qq}-P_{q\bar q}]+2P_{q\bar q}$)
in the $NS(+)$ case.
The anomalous dimension is calculated as \cite{CFP}
\begin{align}
\gamma_1^{n=1,\, NS(+)} &= - 8 \, P_{NS(+)}^{(1)} (n=1) \nonumber \\
                        &= - 16 \, (C_F^2-C_F C_A/2) P_A(n=1) \, \nonumber \\
                        &= - 4 \, (C_F^2-C_F C_A/2) \, [13+8\zeta(3)-2\pi^2]
\ \ \ .
\end{align}
With the numerical values
$\zeta(3)$=1.2020569..., $C_F=(N_c^2-1)/2N_c$, $C_A=N_c$, and $N_c$=3,
we obtain $\gamma_1^{n=1,\, NS(+)}$=+2.5576.
In this way, the NLO term becomes 
$C_{1}^{NS(+)}=\gamma_{1}^{n=1, NS(+)}/(2\beta_0)=
  -6(C_F^2-C_F C_A/2)[13+8\zeta (3)-2\pi^2]/(33-2n_f)$.
Including the NLO correction, we obtain
the Gottfried sum \cite{RS,HK}:
\begin{align}
I_G &= \frac{1}{3} \left [ 1 + 
      \frac{3(C_F C_A/2 - C_F^2)}{2(33-2n_f)} \, 
       ( 13+8 \zeta (3)-2 \pi^2 ) \,   
      \frac{\alpha_s (Q^2)}{\pi} 
     \right ]
\nonumber \\
&= \frac{1}{3} \left [ 1 +    
\begin{pmatrix} 0.03552 \, (n_f=3) \\
                0.03836 \, (n_f=4) 
\end{pmatrix}
              \frac{\alpha_s (Q^2)}{\pi} \right ]
\ \ \ .
\label{eqn:ignlo}
\end{align}
The NLO contribution is merely 0.3\% at $Q^2$=4 GeV$^2$.
It obviously cannot explain the large violation found by the NMC.
The NNLO $\alpha_s$ correction is estimated recently 
in Ref. \cite{KKPS}:
\begin{equation}
I_G= \frac{1}{3}  \left [ 1 +
\begin{pmatrix} 0.036 \, (n_f=3) \\
                0.038 \, (n_f=4) 
\end{pmatrix}
                                        \frac{\alpha_s(Q^2)}{\pi}
                            + 
\begin{pmatrix} 0.72 \, (n_f=3) \\
                0.55 \, (n_f=4)  
\end{pmatrix}
                                    \left \{ \frac{\alpha_s(Q^2)}{\pi}
                                    \right \} ^2 \right ]
\ \ \ .
\label{eqn:ignnlo}
\end{equation}
The NNLO correction is about 0.4\% at $Q^2$=4 GeV$^2$.
We find from these higher-order analyses that
the perturbative corrections are too small to account the
NMC deficit.

The tiny scaling violation is understood in the following way.
The $Q^2$ dependence comes from the difference between
the flavor-diagonal and nondiagonal splitting processes.
Because there are two identical particles in the 
flavor-diagonal case, they should be antisymmetrized.
If it could be neglected, the $Q^2$ evolution is flavor
symmetric and there is no scaling violation in the Gottfried sum.
However, the above-mentioned antisymmetrization provides the very small
scaling violation \cite{FBBG}. 

We comment on experimental information about
possible $Q^2$ dependence in Ref. \cite{ST8461}.
The neutron $F_2(x,Q^2)$ is obtained from various proton and deuteron
measurements with nuclear corrections.
With the $F_2^p$ parametrization for explaining
the NMC, H1, or ZEUS data, the $Q^2$ variation 
\begin{equation}
I_G(Q^2) = S_0 \, \left [ \, 1 + c_1 \, (\alpha_s/\pi)
                               + c_2 \, (\alpha_s/\pi)^2 \, \right ]
\ \ \ 
\end{equation}
is investigated. The obtained parameters averaged over the NMC92, NMC95,
and H1 are $S_0=0.242\pm 0.21$, $c_1=-6.00\pm 0.74$, and $c_2=40.4\pm 11.1$.
The result indicates large $Q^2$ dependence which cannot be accounted
by the perturbative QCD.
However, the analysis with the ZUES $F_2^p$ shows rather different
values: $S_0=0.383$, $c_1=-12.9$, and $c_2=76.2$.
Therefore, accurate information cannot be obtained at this stage. 
Future HERA measurement of $F_2^D$ at small $x$ is necessary
to find the precise $Q^2$ variation.

The perturbative QCD studies show that perturbative mechanisms cannot account 
for the large violation of the Gottfried sum rule.
If the violation is confirmed by further experiments, 
the deficit should come from another source, 
namely a nonperturbative mechanism.

\vfill\eject
\section{{\bf Theoretical ideas for the sum-rule violation}}\label{THEORY}
\setcounter{equation}{0}
\setcounter{figure}{0}
\setcounter{table}{0}

The NMC results in 1991 and in 1994 indicate a significant deviation
from the Gottfried sum.
We showed that the perturbative mechanisms cannot account for
the possible violation of the sum rule. It is even not clear
whether or not the sum rule is in fact violated by considering
the small $x$ part.
A possible way to answer these problems theoretically is
to use a nonperturbative approach. 
Various theoretical ideas have been proposed
for explaining the deficit in terms of explicit flavor asymmetry
$\bar u-\bar d\ne 0$. 
These ideas are discussed in the following.

\subsection{Lattice QCD}\label{LATTICE}

The most fundamental way to treat nonperturbative physics
is to use lattice QCD. 
The following discussions are based on Ref. \cite{LD}.
The forward Compton amplitude Eq. (\ref{eqn:COMPTON}) could be computed
by taking the ratio of a four-point function and
a two-point function:
\begin{align}
\widetilde{W}_{\mu\nu}(\vec{q}^{\, \, 2},\tau) 
  &= \left. \frac{\frac{1}{2M_N}< O_N(t) \int \frac{d^3\xi}
                    {2\pi} e^{-i \vec{q}\cdot\vec{\xi}}
  J_{\mu}(\vec{\xi},t_1)J_{\nu}(0,t_2) O_N(0)>}{<O_N(t- \tau) O_N(0)>} \,
    \right | _{t -t_1, t_2 >> 1/\Delta M_N}
\ \ ,
\end{align}
where $\tau$ is given by $\tau=t_1-t_2$, $\Delta M_N$ is the mass
difference between the nucleon and the first excitation state,
and $O_N(t)$ is the interpolation field for the nucleon.
The hadron tensor $W_{\mu\nu}$ is then calculated 
by the inverse Laplace transformation.

\begin{floatingfigure}{7.0cm}
   \begin{center}
      \mbox{\epsfig{file=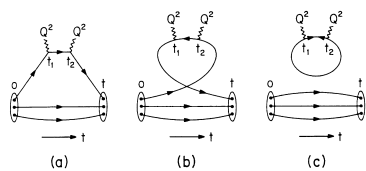,width=5.0cm}}
   \end{center}
 \vspace{-0.4cm}
\caption{\footnotesize Twist-two contributions 
                        (taken from Ref. {\normalsize\cite{LD}}).
}
\label{fig:ld1}
\end{floatingfigure}
\quad
Euclidean path-integral formalism can be used for evaluating 
the four-point function.
The leading-twist contributions come from the diagrams
in Fig. \ref{fig:ld1}.
Quark propagators are involved in the diagrams of 
Figs. \ref{fig:ld1}(a) and (c), and antiquark ones are in 
Figs. \ref{fig:ld1}(b) and (c).
Therefore, antiquark contributions come from either
the connected insertion in Fig. \ref{fig:ld1}(b)
or the disconnected one in Fig. \ref{fig:ld1}(c).
We may call the contribution in Fig. \ref{fig:ld1}(b)
from ``cloud'' antiquarks and the one in Fig. \ref{fig:ld1}(c)
from ``sea'' antiquarks, so that an antiquark distribution
could be written as $\bar q_i(x)=\bar q_i^{\, c}(x) +\bar q_i^{\, s}(x)$.
In the same way, a quark distribution is expressed as 
$q_i (x) = q_i^V (x) + q_i^c (x) + q_i^s (x)$.
If the light-quark masses are equal $m_u=m_d$, 
there is no contribution to the flavor asymmetry $\bar u-\bar d$
from the sea graphs in Fig. \ref{fig:ld1}(c).
Then, the contributions become
$I_G=1/3+(2/3)\int_0^1 dx [\bar u_c(x)-\bar d_c(x)]$.

The hadron tensor has not been calculated directly
due to a huge numerical task,
so that three-point function with one current may be investigated. 
However, the Gottfried sum cannot be calculated because
the first moment of $F_2^p-F_2^n$ cannot be expressed in terms of 
the matrix element of a twist-two operator.
Therefore, real lattice QCD estimate of the Gottfried sum
is not available at this stage. 
Instead, scalar matrix elements were studied in Ref. \cite{LD}
in order to learn about the cloud contributions to
the $\bar u-\bar d$ number.
The scalar charge
$\int d^3 x \bar \Psi \Psi = \int d^3 k (m/E) 
 \sum_s [b_{k,s}^\dagger b_{k,s}+d_{k,s}^\dagger d_{k,s}]$
is the sum of quark and antiquark numbers with the weight factor m/E,
so that they could be a measure of the difference $\bar u-\bar d$.
Because we are interested in the cloud antiquarks,
only the connected insertion (CI) is discussed in the following.
In order to reduce the artificial lattice effects, it is better
to investigate ratios of matrix elements.
The ratio of isoscalar and isovector matrix elements 
for the CI is then approximated as
\begin{equation}
R_s= \frac{\langle p|\bar{u}u|p\rangle  - \langle p|\bar{d}d|p\rangle}
{\langle p|\bar{u}u|p\rangle  + \langle p|\bar{d}d|p\rangle}
\, \begin{array}{|l}  \\ _{CI} \end{array} \!\!
=\,\frac{1 + 2\int dx [\bar{u}_c(x) - \bar{d}_c(x)]}
{3 + 2\int dx [\bar{u}_c(x) + \bar{d}_c(x)]}
\ \ \ .
\end{equation}

\begin{floatingfigure}{7.0cm}
   \begin{center}
      \mbox{\epsfig{file=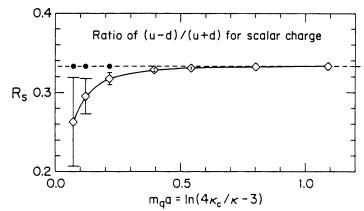,width=5.0cm}}
   \end{center}
 \vspace{-0.4cm}
\caption{\footnotesize Ratio of the isoscalar to isovector scalar charge
                       (taken from Ref. {\normalsize\cite{LD}}).
}
\label{fig:ld2}
\end{floatingfigure}
\quad
Numerical results are obtained by using 16$^3\times$24 lattices with
$\beta$=6 and the hopping parameter $\kappa$=0.105$-$0.154.
In Fig. \ref{fig:ld2}, the obtained ratios $R_s$ are plotted
as a function of the quark mass $m_q a$. 
Because the antiquark number is positive,
the ratio has to be smaller than 1/3.
The cloud antiquarks are suppressed in the heavy-quark case,
so that the ratio agrees with the valence-quark expectation 1/3
in Fig. \ref{fig:ld2}.
As the quark mass decreases, the ratio becomes smaller than 1/3.
This decrease should be interpreted as the cloud effects.
In order to verify this interpretation, we consider a valence approximation, 
which means amputating the backward time hopping.
The ratios with this approximation are shown by the filled circles
and they are 1/3 as expected.
Next, we discuss comparison with the NMC result.
The state with the $q\bar q$ clouds has higher
energy than one of the valence-quark state, which means that 
the $m/E$ factor is smaller. Therefore, 
we could estimate the upper bound for the $\bar u-\bar d$ number by
$n_{\bar u}-n_{\bar d} \le 
[<N|\bar uu-\bar dd|N>_{cloud}/<N|\bar uu-\bar dd|N>_{valence}-1 ]/2$.
Their result extrapolated to the chiral limit
is $n_{\bar u}-n_{\bar d} \le -0.12\pm 0.05$.
It indicates a $\bar d$ number excess over the $\bar u$.
The obtained result could be used as a measure
of the flavor asymmetry although the quantitative comparison with
the Gottfried sum is not obvious.
It is, however, interesting to find that
the obtained value is consistent with the NMC asymmetry
in Eq. (\ref{eqn:UBMDB}).

Even though there is no direct estimate of the Gottfried sum
in the lattice QCD right now, there is an indication of the large 
light-antiquark flavor asymmetry. 
It is shown that the difference comes from
the connected insertion involving quarks propagating
backward in time by studying the isovector-isoscalar charge ratio.
Because the flavor asymmetry comes from the cloud antiquarks,
physics mechanism behind the above results is considered as
the Pauli blocking and/or the mesonic effects.

\subsection{Pauli exclusion principle}\label{PAULI}

Pauli exclusion model was investigated in Refs. \cite{FF,OTHEREXC} 
for explaining the SLAC data.
Because the proton has two valence u quarks and one valence d quark, 
the $u\bar u$ pair creation receives more Pauli exclusion effect
than the $d\bar d$ pair creation does. 
This results in the difference between $\bar u$ and $\bar d$ in the nucleon.
No qualitative calculation is done in Ref. \cite{FF} except for
a parametrization based on the above intuition. In order to explain
the SLAC data ($I_G$=0.27 according to the analysis in Ref. \cite{FF})
the following parametrization was proposed
\begin{equation}
x\bar u=0.17(1-x)^{10} \ \ \ \ \ , \ \ \ \ \ 
x\bar d=0.17(1-x)^7
\ \ \ .
\label{eqn:FF}
\end{equation}

First, we explain how the flavor asymmetry can be calculated
in a model even though a realistic four dimensional
calculation is not available at this stage.
A qualitative calculation on the Pauli blocking effects
is discussed in Refs. \cite{ST} and \cite{ADELAIDE}.
A parton distribution in the nucleon is calculated by \cite{JAFFE85}
\begin{align}
\bar q_i(x) &= \frac{\sqrt{2}}{4\pi} \int d\xi^- e^{-ixp^+ \xi^-}
          <p \, | \, \psi_{i,+}(\xi^-) \, 
                     \psi_{i,+}^\dagger (0) \, | \, p>_c |_{\xi^+=0}
\nonumber \\
            &= \frac{1}{\sqrt{2}} \sum_n
            \int \frac{dp_n'}{4\pi p_n^0}
            \delta(p_n^+-(1-x)p^+) \, \, 
            |<n \, |\psi_{i,+}^\dagger (0)| \, p>|^2
\ \ \ ,
\label{eqn:BARQ1}
\end{align}
where the subscript $c$ indicates a connected matrix element,
$\psi_+$ is defined by $\psi_+=\gamma^-\gamma^+\psi/2$,
and $p_n'$ is the momentum of the intermediate state.
It is the probability of removing an antiquark $q_i$
with momentum $xp^+$, leaving behind a state $| \, n>$.
The 1+1 dimensional MIT bag model is used for evaluating the antiquark 
distribution.
However, a realistic 3+1 dimensional calculation has not been
done yet. We estimate the effect by a simple counting estimate.

In the 1+1 dimension, there are three color states for each flavor.
Two of the three u-quark ground states and one of the three d-quark states
are occupied. It is possible to have only one more u-quark in the
ground state, but two more d-quarks can be accommodated.
Therefore, expected sea-quark asymmetry is fairly large: $\bar d=2\bar u$ 
in the 1+1 dimensional bag picture.
In the four dimensional case, there are six states (three-color times
two-spin states) in the ground state.
There are four available ground states for u-quarks
and five states for d-quarks, so that 
the asymmetry becomes 
\begin{equation}
\frac{\bar d}{\bar u}= \frac{5}{4}
\ \ \ \ \ \text{(in a naive counting estimate)}
\ \ \ .
\label{eqn:count}
\end{equation}
Because there is no valence antiquarks in the bag,
Eq. (\ref{eqn:BARQ1}) indicates that the contribution comes
from a quark being inserted, interacting in the bag, and then being
removed. Therefore, the $\bar d$ excess is related to the distribution
associated with a four-quark intermediate state $f_4(x)$
\begin{equation}
\int_0^1 dx [d_{sea}(x)-u_{sea}(x)]=\int_0^1 dx f_4(x)=1-P_2
\ \ \ ,
\end{equation}
where $P_2$ is the integral of a distribution associated with
a two-quark intermediate state.
Because the $\bar u$ and $\bar d$ distributions are not calculated
in the four dimensional model, the Pauli contributions are 
given by a simple function $x^\alpha(1-x)^\beta$ in Ref. \cite{ADELAIDE}.
The constant $\alpha$ is chosen to match the small $x$ behavior
of used valence distributions, and $\beta$=7 is taken so that
it contributes only at small $x$. The overall normalization
is not determined theoretically at this stage. 
Roughly speaking, we expect to have $1-P_2$
in the 25\% range because of 
the naive counting estimate in Eq. (\ref{eqn:count}).
The obtained $x$ dependent results are studied together 
with pionic effects in the following subsection.

It is shown that the Pauli blocking effects could produce
the excess of $\bar d$ quark over $\bar u$.
The naive counting in four dimension indicates
$\bar d/\bar d$=5/4.  Unfortunately, qualitative $x$ dependence
is not calculated except for the 1+1 dimensional model. 
It is indicated that 10\% Pauli effects together with
pionic contributions can explain the NMC data
$F_2^p-F_2^n$ fairly well \cite{ADELAIDE}.
However, it was found recently that
the conclusion should be changed drastically if the antisymmetrization
between quarks is considered in addition \cite{ST}.
The same $u$-valence excess, which suppresses the $u\bar u$ pair creation,
also produces extra diagrams involved in the $u\bar u$ creation
because of the antisymmetrization with the extra $u$.
These extra diagrams contribute to a $\bar u$ excess over $\bar d$.
According to Ref. \cite{ST}, if the Pauli-principle and antisymmetrization
effects are combined, the $\bar u$ could be larger than $\bar d$,
which is in contradiction to the NMC conclusion.

\subsection{Mesonic models}\label{MESON}

The Pauli exclusion mechanism produces the flavor
asymmetry; however, its effects on the sum rule do not seem to be
large enough to explain the NMC result according to 
the counting estimate. Furthermore, if they are combined with
the antisymmetrization contributions, we could have a $\bar u$
excess over $\bar d$.
On the other hand, the meson-cloud mechanism is the most
successful one in explaining the major part of the NMC flavor asymmetry.
Because a significant amount of papers are written on 
this idea, we explain the model in detail.
We first discuss conventional virtual-meson contributions
in section \ref{MESON-1}. Second, chiral models in a similar spirit
are explained in \ref{CHIRAL}.  
Third, possible modification of the $Q^2$ evolution due to 
the meson emission is discussed in \ref{MESON-Q2}.

\subsubsection{Meson-cloud contribution}\label{MESON-1}

It is well known that virtual mesons play a very important dynamical
role in nucleon structure, as they have been studied in the context
of cloudy bag or other chiral models. 
The proton decays into $\pi^+$ and neutron,
$\pi^0$ and proton, and other states within the time
allowed by the uncertainty principle. 
The virtual pion state is essential for explaining many dynamical
properties, for example the large $\Delta$ decay width
and negative square charge radius of the neutron. 
Therefore, it is important to study whether or not
the mechanism could produce a flavor asymmetry.
It should be noted that perturbative contributions to 
the antiquark distributions through the gluon splitting into $q\bar q$
should be much larger than the mesonic ones at reasonably large $Q^2$.
However, these contributions are supposed to be flavor symmetric,
and the asymmetric distributions $(\bar u+\bar d)/2-\bar s$
and $\bar u-\bar d$ could be used for testing the meson mechanism.

\begin{wrapfigure}{l}{7.0cm}
   \begin{center}
      \mbox{\epsfig{file=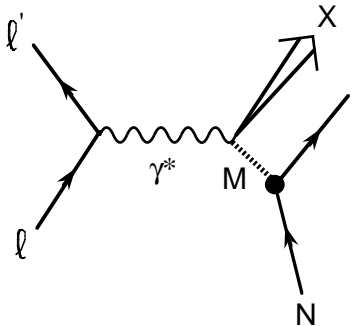,width=5.0cm}}
   \end{center}
 \vspace{-0.4cm}
\caption{\footnotesize Mesonic contribution to $\bar q$.}
\label{fig:sull}
\end{wrapfigure}
\quad
The original idea stems from Ref. \cite{SULL} in 1972,
so that the process in Fig. \ref{fig:sull} is sometimes called
``Sullivan process''.
The proton splits into a pion and a nucleon, and
the virtual photon interacts with the pion.
Antiquark distributions in the pion contribute to
the corresponding antiquark distributions in the proton. 
Although the idea is interesting, it had not been a very popular topic
until recently. It is partly because experimental data are not accurate
enough to shed light on the mechanism. 
After the NMC discovery, it is shown that the pion-cloud 
mechanism could explain a significant part of the NMC finding
\cite{HM,INDIANA,ADELAIDE}. This idea is developed further by
including many meson and baryon states \cite{JULICH}
and by considering different form factors at the meson-baryon vertices
\cite{KFS} so as to explain the whole NMC asymmetry.
 
The formalism in Ref. \cite{SULL} is the following.
The cross section of Fig. \ref{fig:sull} with $M=\pi$
is derived by replacing the $\gamma^* +p\rightarrow X$ vertex 
in the $e+p\rightarrow e'+X$ formalism in section \ref{GOTT}
by
\begin{equation}
{\mathcal M}_\mu = \, <X|e J_\mu (0)|\pi> \, 
                   \frac{1}{p_\pi^2 - m_\pi^2} \, 
                   F_{\pi NN} (t) \, 
         \bar u(p') \, i \, g_{_{\pi NN}} \, \widetilde \phi_\pi^{\, *} 
                             \cdot \widetilde\tau \, 
                             \gamma_5 \, u(p)   
\ \ \ ,
\end{equation}
where $F_{\pi NN}(t)$ is the $\pi NN$ form factor,
$g_{_{\pi NN}}$ is the $\pi$NN coupling constant,
and $\widetilde \phi_\pi^{\, *} \cdot \widetilde\tau$ is the isospin factor.
The $W_2$ structure function for the pion is defined in the same way with
Eq. (\ref{eqn:HADRON}) by the replacements $p\rightarrow p_\pi$ and
$M\rightarrow m_\pi$. Then, projecting out the $F_2$ part,
we obtain
\begin{equation}
F_2^{pionic}(x,Q^2) = | \widetilde \phi_\pi^{\, *} 
                      \cdot \widetilde\tau  |^2
         \, \frac{g_{_{\pi NN}}^2}{16 \pi^2} \,
                  \int _x^1 dy \, y  \, F_2^\pi  (x/y,Q^2) 
                 \int_{-\infty}^{t_m} dt\, \frac{-t}{(t-m_\pi^2)^2}\, 
         \left | F_{\pi NN} (t) \right |^2
\ \ ,
\label{eqn:f2pi}
\end{equation}
where $t_m$ is the maximum energy transfer: $t_m=-m_N^2y^2/(1-y)$.
This equation is understood by the convolution of the pion
structure function with a light-cone momentum distribution
of the pion. The formalism is used for antiquark distributions
in the same manner.

The studies of the pionic mechanism used to be somewhat confusing.
The direct interaction of the photon with mesons being
present in the cloud of a nucleon does not contribute to the
Gottfried sum. This does not mean that the mesons do not
contribute to the Gottfried sum and the $\bar u-\bar d$
distribution as explained in this subsection.
In dealing with this issue, there are two types
of descriptions. One is to calculate only mesonic contributions
to the $\bar u-\bar d$ distribution \cite{HM,INDIANA,KFS} 
and another is to include recoiling baryon interaction 
with the virtual photon in addition \cite{ADELAIDE,JULICH}.
Both are essentially the same. 
The details of the compatibility are discussed in the following. 

\vspace{0.5cm}
\noindent
{\bf [I. Models with only meson contributions]}
\vspace{0.3cm}

First, we discuss the former approach with the only meson
contributions. The Seattle \cite{HM} and Indiana \cite{INDIANA}
papers proposed pionic ideas for the sum-rule violation
in this description.
The only major difference is the inclusion of $p\rightarrow\pi\Delta$
processes in Ref. \cite{INDIANA} in addition to the $\pi NN$ ones.
Relative magnitude and sign of the $\pi NN$ and $\pi N\Delta$ contributions 
can be understood in the following way. 
We consider the processes $p \to \pi^+ +n$, $\pi^0 +p$,
$\pi^+ +\Delta^0$, $\pi^0 +\Delta^+$, and $\pi^- +\Delta^{++}$,
where the virtual photon interacts with the pions.
Assuming the flavor symmetry in the pion sea,
we have the $\bar u -\bar d$ distributions in the pions:
\begin{equation}
(\bar u-\bar d)_{\pi^+} = - V_\pi \ \ , \ \ \ \ 
(\bar u-\bar d)_{\pi^0} = 0       \ \ , \ \ \ \
(\bar u-\bar d)_{\pi^-} = + V_\pi \ \ , 
\end{equation}
where $V_\pi$ is the valence-quark distribution in the pions.
The flavor symmetry assumption in the pions does not alter
our conclusion unless at very small $x$ with the following reason.
For example, let us consider the $\bar u-\bar d$ distribution at 
$x$=0.1. The light-cone momentum distribution of the pion
is peaked at $y\sim$0.25; therefore, the most important
kinematical region for $(\bar u-\bar d)_\pi$ is at
$x/y\sim$0.4. The valence distribution still dominates 
in this region, so that the sea asymmetry in the pion does not matter.
Including isospin coefficients at the
$\pi$NN and $\pi$N$\Delta$ vertices,
\begin{alignat}{3}
| \widetilde \phi^{\, *}_{\pi^+} \cdot \widetilde \tau |^2 &= \, 2  
\ \ , \ \ \ \ 
&| \widetilde \phi^{\, *}_{\pi^0} \cdot \widetilde \tau |^2 &= \, 1
\ \ , \ \ & \ & 
\nonumber \\  
| \widetilde \phi^{\, *}_{\pi^+} \cdot \widetilde  T   |^2 &= \frac{1}{3}
 \ \ , \ \ \ \ 
&| \widetilde \phi^{\, *}_{\pi^0} \cdot \widetilde  T   |^2 &= \frac{2}{3}  
\ \ , \ \ \ \ 
&| \widetilde \phi^{\, *}_{\pi^-} \cdot \widetilde  T   |^2 &= \, 1   \ \ ,
\label{eqn:isospin}
\end{alignat}
we have the isospin times the $(\bar u-\bar d)_\pi$ factors as
\begin{alignat}{2}
\sum_\pi \, | \widetilde \phi^*_\pi \cdot \widetilde \tau | ^2
\, (\bar u-\bar d)_\pi &=\, -2 \,  V_\pi & \qquad
&\text{for the $\pi$NN process} \ \ ,
\nonumber \\
\sum_\pi \, | \widetilde \phi^*_\pi \cdot \widetilde T | ^2
\, (\bar u-\bar d)_\pi &=\, +\, \frac{2}{3} \, V_\pi & \qquad   
&\text{for the $\pi$N$\Delta$} \ \ . 
\label{eqn:ndelta}
\end{alignat}
In this way, we find that
the $\pi$NN contribution to $\bar u-\bar d$ is negative 
and is partly canceled by a positive contribution from the $\pi$N$\Delta$.
The other important factor is the meson-baryon vertex form factor.
Because the exact functional form is not known, the following
monopole, dipole, and exponential forms are usually used:

\vfill\eject
\begin{alignat}{2}
F_{MNB}(t) &= \frac{\Lambda_m^2-m_M^2}{\Lambda_m^2-t}
    & \qquad &\text{monopole}
\nonumber \\
    &= \left(\frac{\Lambda_d^2-m_M^2}{\Lambda_d^2-t}\right)^{\!\!2}
    & \qquad &\text{dipole} 
\nonumber \\
    &= e^{(t-m_M^2)/\Lambda_e^2}
    & \qquad &\text{exponential} \ ,
\label{eqn:dipolef}
\end{alignat} 
at the $MNB$ vertex.
The different parameters could be related, for example, by
$\Lambda_m=0.62\Lambda_d=0.78\Lambda_e$ \cite{INDIANA}.

\begin{wrapfigure}{l}{7.0cm}
   \begin{center}
      \mbox{\epsfig{file=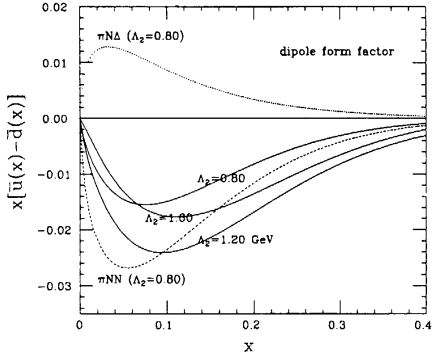,width=6.0cm}}
   \end{center}
 \vspace{-0.4cm}
\caption{\footnotesize $\pi NN$ and $\pi N\Delta$ contributions 
                       to $\bar u-\bar d$
         (taken from Ref. {\normalsize\cite{INDIANA}})
}
\label{fig:sk}
\end{wrapfigure}
\quad
Detailed numerical results are shown in Fig. \ref{fig:sk},
where pionic contributions from the $\pi NN$ and $\pi N\Delta$
processes are shown \cite{INDIANA}. 
The dipole cutoff parameter 
$\Lambda_2$=0.8 GeV [$=\Lambda_d$ in Eq. (\ref{eqn:dipolef})]
is fixed by fitting the $(\bar u+\bar d)/2-\bar s$
experimental data. The dotted curves are $\pi NN$ and $\pi N\Delta$
contributions. As it is shown in the naive discussion, the $\pi NN$
effect is negative and it is canceled by the positive $\pi N\Delta$
contribution. The total contribution with $\Lambda_2$=0.8 GeV
is shown by a solid curve together with those at $\Lambda_2$=1.0
and 1.2 GeV. It is noteworthy that the total $\bar u-\bar d$ curve
is not very sensitive to the cutoff although the distribution
$(\bar u+\bar d)/2-\bar s$ does depend much on it.
Integrating the pionic contribution over $x$,
we obtain $\Delta I_G =2/3 \int dx (\bar u -\bar d)_\pi =-0.04$,
which accounts for about a half the discrepancy found by the NMC.
It is encouraging that the mesonic model gives 
a reasonable value for the magnitude obtained by the NMC.
Although we discussed only the $\pi$NN and $\pi$N$\Delta$ processes, 
other processes should be investigated.
For example, kaon, $\Lambda$, $\Sigma$, and $\Sigma^*$
are added to $\pi$, $N$, and $\Delta$ in Ref. \cite{KFS}.
The first method is well summarized in Ref. \cite{KFS}, so that
we quote its results in the following.

Mesonic contributions to an antiquark distribution in the nucleon
are given by the convolution of the corresponding antiquark
distribution in a meson with the light-cone momentum distribution
of the meson.
The contributions are given by the equation 
\begin{equation}
x \, \overline q_N(x,Q^2) = \sum_{MB} \alpha_{MB}^q 
                               \int_x^1 dy \, f_{MB}(y) \, \frac{x}{y} \, 
                               \overline q_M (x/y,Q^2)
\ \ \ ,
\label{eqn:CONV1}
\end{equation}
where the summations are taken over combinations of
meson states $M=(\pi,\, K)$ and baryon states
$B=(N,\, \Delta,\, \Lambda,\, \Sigma,\, \Sigma^*)$,
and $\alpha_{MB}^q$ is the spin-flavor SU(6) Clebsch-Gordan
factors. 
This equation corresponds to Eq. (\ref{eqn:f2pi}) in the $F_2$ case.
The light-cone momentum distribution of the virtual meson is 
\begin{equation}
f_M(y) \, = \, \sum_B f_{MB}(y)
\ \ \ ,
\end{equation}
with 
\begin{equation}
f_{MB} (y) = \frac{g_{MNB}^2}{16\pi^2} \, y
                      \int_{-\infty}^{t_m} dt \,
                      \frac{{\mathcal I}(t,m_N,m_B)}{(-t+m_M^2)^2}
                      \, [F_{MNB} (t)]^2
\ \ \ .
\label{eqn:fmb}
\end{equation}
The integrand factor ${\mathcal I}(t,m_N,m_B)$ is given by
\begin{alignat}{2}
{\mathcal I}(t,m_N,m_B) &= - \, t+(m_B-m_N)^2 
            & \qquad & \text{for $B \in$ {\bf 8}} 
\nonumber \\
&= \frac{\big[ \, (m_B+m_N)^2-t \, \big]^2 \,
         \big[ \, (m_B-m_N)^2-t \, \big]}
         {12 \, m_N^2 \, m_B^2 }
            & \qquad & \text{for $B \in$ {\bf 10}} \ ,
\end{alignat}
depending whether the baryon $B$ is in the baryon octet
or in the decuplet.
The upper limit of the integral is given by
\begin{equation}
t_{m} = m_N^2 \, y - \frac{m_B^2 \, y}{1-y}
\ .
\end{equation}

Because the coupling constants are relatively well known, 
the only factor to be paid attention to is the $MNB$ form factors.
There are recent studies on whether the form factor is hard or soft. 
Instead of stepping into the details, we summarize 
briefly the historical background and the present situation. 
A hard form factor with the typical monopole cutoff
$1.0 < \Lambda_m < 1.4$ GeV is essential for 
explaining the deuteron D-state admixture and nucleon-nucleon 
scattering experiments. 
On the other hand, softer ones are obtained in quark models:
for example $\Lambda_m\approx 0.6$ GeV in the cloudy-bag model
\cite{CLOUDY} $0.7 < \Lambda_m < 1.0$ GeV 
in a flux-tube model \cite{FLUX}.
However, a conflicting result came from the studies
of the flavor asymmetric distribution $(\bar u+\bar d)/2-\bar s$.
It was originally announced that the cutoff should be much softer,
$\Lambda_m < 0.5$ GeV \cite{FMS}, 
which contradicts awfully to the OBEP one.
Later analysis with renewed experimental data show a slightly
larger cutoff $\Lambda_m\approx 0.6$ GeV \cite{INDIANA}, 
which could be consistent with those in the quark models. 
In the J\"ulich approach, which is discussed in the following paragraphs,
the obtained cutoff becomes larger $\Lambda_m\approx 0.74$ GeV
(note: monopole cutoff is estimated by $\Lambda_m=0.62\Lambda_d$ 
       with $\Lambda_d=1.2$ GeV \cite{JULICH})
because more meson and baryons are added to 
$\pi$, $N$, $\Delta$ and
because the following normalization factor $Z$ is taken into account.
The probability to find the bare nucleon is reduced by the factor $Z$
due to the presence of the meson clouds as explained in detail
in the model II.
The recent one in Ref. \cite{KFS} without explicit baryon 
contributions indicates a similar value $\Lambda_m\approx 0.8$ GeV.
Furthermore, it is discussed that kinematical regions, which
contribute to the low-energy NN scattering and deep inelastic
processes, are very different in the form factor. The discrepancy
between the hard OBEP form factor and the soft one should not be taken
seriously. 
Whatever the outcome may be, it does not change 
our $\bar u-\bar d$ studies significantly if the parameter is
fixed so as to explain the $(\bar u+\bar d)/2-\bar s$
distribution.

\vspace{0.5cm}
\noindent
{\bf [II. Models with meson and baryon contributions]}
\vspace{0.3cm}

\begin{figure}[h]
\parbox[c]{0.46\textwidth}{
   \begin{center}
       \epsfig{file=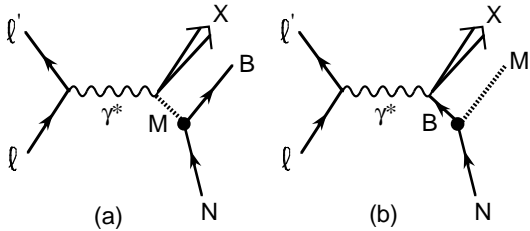,width=7.0cm}
   \end{center}
       \vspace{-0.5cm}
\caption{\footnotesize Baryon terms in (b) are included
         in addition to the meson contributions in (a).}
       \label{fig:ade}
}\hfill
\parbox[c]{0.46\textwidth}{
   \begin{center}
\footnotesize
\bottomcaption{\footnotesize 
                       Coefficient $A_i$ and $B_i$ in two different 
                       descriptions {\normalsize\cite{JULICH}.}}
\label{tab:julich}
\begin{supertabular}{|l|cccc|}
\hline
Process      & $A_i^M$ & $A_i^B$ & $B_i^M$ & $B_i^B$ \\
\hline
$\pi N$      &    0    &  $-$1/3 &  $-$2/3 &    0    \\
$\rho N$     &    0    &  $-$1/3 &  $-$2/3 &    0    \\
$\omega N$   &    0    &     1   &     0   &    0    \\
$\sigma N$   &    0    &     1   &     0   &    0    \\
$\eta N$     &    0    &     1   &     0   &    0    \\
$\pi\Delta$  &    0    &     5/3 &     1/3 &    0    \\
$\rho\Delta$ &    0    &     5/3 &     1/3 &    0    \\
$K\Lambda$   &    1    &     0   &     0   &    0    \\
$K^*\Lambda$ &    1    &     0   &     0   &    0    \\
$K\Sigma$    & $-$1/3  &     4/3 &     0   &    0    \\
$K^*\Sigma$  & $-$1/3  &     4/3 &     0   &    0    \\
$KY^*$       & $-$1/3  &     4/3 &     0   &    0    \\
$K^*Y^*$     & $-$1/3  &     4/3 &     0   &    0    \\
\hline
\end{supertabular}
\end{center}
}
\end{figure}

Next, we discuss the second approach, which includes
recoil-baryon interactions with the virtual photon
in addition to the meson interactions as shown
in Fig. \ref{fig:ade}.
This type of description is studied in the Adelaide paper,
\cite{ADELAIDE}, the J\"ulich \cite{JULICH},
and Ref. \cite{ZPION}.
In particular, the J\"ulich group developed this model
by including many meson and baryon states. So far
$\pi N$, $\rho N$, $\omega N$, $\sigma N$, $\eta N$, $\pi\Delta$,
$\rho\Delta$, $K\Lambda$, $K^*\Lambda$, $K\Sigma$, $K^*\Sigma$,
$KY^*$, and $K^* Y^*$ states are included.
The pions do not contribute to $F_2^p-F_2^n$, so that
the $\pi NN$ and $\pi N\Delta$ contributions are given
by \cite{ADELAIDE}
\begin{multline}
F_2^p(x) - F_2^n(x) = Z \left \{ \frac{x}{3} [u_v(x)-d_v(x)]
         -\frac{1}{3} \int_0^{1-x} dy \frac{f_N (y)}{1-y}
          \frac{x}{3} \left [ u_v \left ( \frac{x}{1-y} \right ) 
                            - d_v \left ( \frac{x}{1-y} \right )
                      \right ] 
\right. \\
\left.
         +\frac{1}{6} \int_0^{1-x} dy \frac{f_\Delta (y)}{1-y}
          \frac{10 x}{3} d_v \left ( \frac{x}{1-y} \right )  
     \right \}
\ \ \ .
\end{multline}
The functions $f_N(y)$ and $f_\Delta (y)$ are pion light-cone momentum
distributions, and $Z$ is the valence normalization factor
$Z=1/(1+N_\pi+\Delta_\pi)$ with probability of finding a pion
$N(\Delta)_\pi=\int_0^1 dy f_{N(\Delta)}(y)$.
From these equations, the sum becomes
$I_G=(Z/3)(1-N_\pi/3+5\Delta_\pi /3)$.
According to this equation, the failure of the sum rule is not due to 
the photon interaction with the virtual pion but it is due to
the interaction with the recoil baryons.
This may seem contradictory to the conclusion in the first approach.
However, it is not a paradox as explained in Refs. \cite{JULICH, KFS}.

\vspace{-0.4cm}
\noindent
\normalsize

Adding other contributions from light meson and baryon states, 
we write the previous equation as
\begin{equation}
I_G= \frac{1}{3} \, (Z+\sum_i A_i) \ , \ \ \ 
\text{with} \ \ 
A_i=\int_0^1 dx \, (u_i + \bar u_i -d_i - \bar d_i)_{Sull}
\label{eqn:pi2nd}
\end{equation}
where the meson and baryon contributions $A_i$ are given in
Table \ref{tab:julich}. It should be noted that all the coefficients
in the table should be multiplied by the probabilities of finding
the meson-baryon states in the nucleon.
The nucleon ``core" satisfies the valence sum
$\int dx (u_v-d_v)_{core}=1/3$ but its probability
is reduced by the normalization factor $Z$ due to
the virtual $MB$ states.
On the other hand, the sum could be written in a different form.
Whatever the normalization mechanism is, the valence sum should
be exactly satisfied. 
Therefore, a part of the meson and baryon contributions can be included
into the valence sum 1/3. Then, the deviation from 1/3 is identified
with the flavor asymmetry due to the Sullivan processes
$\int dx(\bar u-\bar d)=\int dx(\bar u-\bar d)_{Sull}$
\cite{INDIANA,JULICH,KFS}:
\begin{equation}
I_G= \frac{1}{3} \, (1+\sum_i B_i) \ , \ \ \ 
\text{with} \ \ 
B_i=\int_0^1 dx \, (u_i - \bar d_i)_{Sull}
\ \ .
\label{eqn:pi1st}
\end{equation}
Equation (\ref{eqn:pi1st}) corresponds to the first approach
without the baryon contributions
and Eq. (\ref{eqn:pi2nd}) to the second method by the Adelaide and
J\"ulich. 
As the coefficients are listed in Table \ref{tab:julich},
there is no contribution from the pion and rho mesons
to the sum in the second approach. Therefore, the violation comes
from the normalization factor $Z$ and the baryon contributions.
Because the distribution $u+\bar u-d-\bar d$ vanishes
for example in the pion, this is a natural consequence.
However, the virtual $\pi B$ contributes to the renormalization
of the valence-quark distributions. Therefore, we may take out
the pionic renormalization contributions and put them
into the obvious valence-sum factor 1/3 in Eq. (\ref{eqn:pi1st})
\cite{INDIANA}.
Then, it becomes apparent that
the pion contributes to the deviation from the Gottfried
sum as indicated in Eq. (\ref{eqn:pi1st}).
Because of the flavor symmetry assumption in the $MB$,
the pion and rho are the only light hadrons which contribute to the 
violation. 
In this way, the two different mesonic descriptions are 
equivalent, and both numerical results have to be the same.

In the beginning, it was shown that a significant part,
approximately a half, of the NMC deficit could be explained
by the virtual pions. 
The Adelaide group tried to interpret
the whole deficit by adding the Pauli exclusion effects.
In the J\"ulich model, it is explained only by the meson
model with the additional meson and baryon states \cite{JULICH}.
The other possibility for explaining the whole result 
is to consider different form factors
in the $\pi NN$ and $\pi N\Delta$ vertices \cite{KFS}.
It is also discussed in Ref. \cite{KFS} that 
the normalization factor does not affect
the meson part because the bare coupling $g_{_{MNB}}^0$ has to
be used in the wave function of a physical particle in terms
of its constituents. Although there are slight differences
among the above models, the meson-cloud approach is successful
in explaining at least the major part of the NMC flavor asymmetry.
   
\begin{wrapfigure}{r}{7.0cm}
   \begin{center}
      \mbox{\epsfig{file=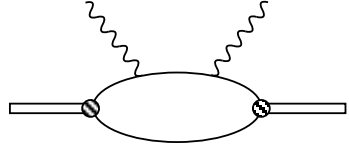,width=4.5cm}}
   \end{center}
 \vspace{-0.5cm}
\caption{\footnotesize Off-shell pion structure function in the NJL model.}
\label{fig:njl}
\end{wrapfigure}
\quad
According to the above conclusion, we should have 
the antiquark distributions at least in the virtual pion and rho
for a realistic evaluation of the violation.
So far, we have been using the distributions in the pion 
measured in the Drell-Yan processes, namely those in the on-shell pion.
If the distributions in the off-shell pion are significantly
modified, numerical results in this subsection should be reanalyzed.
There is an experimental proposal for measuring virtual pion
structure functions by detecting a recoil nucleon at HERA \cite{OFFSHELL}. 
Here, we discuss model estimates of the off-shell effects \cite{SST,SS}. 
Tokyo-Metropolitan-University (TMU) \cite{SST} and Brooklyn \cite{SS} groups
studied this topic within the Nambu-Jona-Lasinio (NJL) model.
While the TMU group calculated distributions at a hadronic scale
($Q^2$=0.25 GeV$^2$) and they were evolved to larger $Q^2$,
the Brooklyn group calculated them at large $Q^2$ directly.
The Compton amplitude in Fig. \ref{fig:njl} is calculated
with the $\pi qq$ vertex given by the NJL model.
Then the pion structure function is projected out from the 
amplitude, and the sea-quark distributions in the proton are calculated
in the pion model.
Although the $\pi NN$ cutoff depends much on the off-shell nature
of the pion, the obtained contribution to $I_G$ is not very significant.
According to Ref. \cite{SST}, the deviation $\Delta I_G=-$0.0557 with
the on-shell pion becomes $\Delta I_G=-$0.0586 with the NJL off-shell
pion structure function. 
The difference is merely 5 \% effect 
(0.9\% in the Gottfried sum) because of
the cancellation between the $\pi NN$ and $\pi N\Delta$
in Eq. (\ref{eqn:ndelta}).
In any case, if model parameters are fixed by fitting other
distributions such as $(\bar u+\bar d)/2-\bar s$, present mesonic
contributions are not significantly changed because of
the off-shell nature.

\subsubsection{Chiral models}\label{CHIRAL}

In the  previous subsection, 
we find that the mesonic contributions could explain 
the major part of the Gottfried-sum-rule violation.
The difference between $\pi^+$ and $\pi^-$ production 
in the process $p\rightarrow B\pi$ gives rise to the antiquark
flavor asymmetry. The pion production ratio $\pi^+$:$\pi^0$:$\pi^-$
is  2:1:0 in the processes $p\rightarrow N\pi$ [Eq. (\ref{eqn:isospin})]. 
However, as it is obvious from Eq. (\ref{eqn:ndelta}), the contribution
is partly canceled by the $p\rightarrow\Delta \pi$ process.
In order to have a better estimate, other resonances have to be included.
Their contributions could be included in a more microscopic approach with
effective chiral models. In such models, the pion ratio
$\pi^+$:$\pi^0$:$\pi^-$ becomes 4:3:2 if they are produced 
in the process $q\rightarrow q\pi$.
We explain this kind of approaches in this subsection.
Although chiral quark-meson models were studied slightly
earlier \cite{SC,WAKA,BPG}, we discuss first a chiral-field-theory
approach in Ref. \cite{EHQ,KRE,SBF,CL} because of similarity in its formalism
to those in the previous subsection.
Later, the chiral-meson models are discussed.

In describing hadron properties at low energies, it is important to explain
spontaneous chiral symmetry breaking. As an effective model for describing
such a property, we have the chiral field theory.
This model is used for evaluating the Gottfried-sum deficit \cite{EHQ}. 
Appropriate degrees of freedom in describing
low-energy hadron structure are quarks, gluons, and Goldstone bosons.
The effective interaction Lagrangian is
\begin{equation}
{\mathcal L}= \overline\psi (iD_\mu+V_\mu) \gamma^\mu \psi
              +i g_{_A} \overline\psi A_\mu \gamma^\mu \gamma_5 \psi 
              +\cdot\cdot\cdot
\ \ \ ,
\end{equation}
where $\psi$ is the quark field and 
$D_\mu$ is the covariant derivative.
The vector and axial-vector currents are expressed by Goldstone-boson fields
\begin{equation}
\begin{pmatrix} V_\mu \\ A_\mu \end{pmatrix}
        = \frac{1}{2} (\xi^\dagger \partial_\mu \xi \pm
                       \xi \, \partial_\mu \xi^\dagger)
\ \ \ ,
\end{equation}
\begin{equation}
\xi= exp(i\Pi/f) \ \ \ , \ \ \ 
\Pi=\frac{1}{\sqrt{2}}
\begin{pmatrix}
    \pi^0/\sqrt{2}+\eta/\sqrt{6}  & \pi^+ & K^+ \\
    \pi^- & -\pi^0/\sqrt{2}+\eta/\sqrt{6} & K^0 \\
    K^-   & \overline K^0     & -2\eta/\sqrt{6}
\end{pmatrix}
\ \ \ .
\end{equation}
Expanding the currents in power of $\Pi/f$, we have
$V_\mu=O(\Pi/f)^2$ and $A_\mu=i\partial_\mu\Pi/f+O(\Pi/f)^2$.
Then the quark-boson interaction becomes
${\mathcal L}_{\Pi q} = - (g_{_A}/f)
         \overline\psi\partial_\mu \Pi\gamma^\mu\gamma_5\psi$.

\begin{floatingfigure}{7.0cm}
   \begin{center}
      \mbox{\epsfig{file=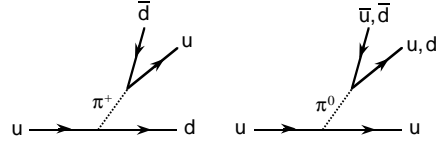,width=6.0cm}}
   \end{center}
 \vspace{-0.5cm}
\caption{\footnotesize Valence $u$ quark splitting.}
\label{fig:usplit}
\end{floatingfigure}
\quad
We give an idea how the deficit arises in this model
by considering
the splitting processes in Fig. \ref{fig:usplit}.
A valence $u$ quark splits into $\pi^+$ and $d$, and subsequently into 
$d$, $u$, and $\bar d$
in the left figure. It also splits into  $\pi^0$ and $u$, then into
$u$, $u$, and $\bar u$ or into $u$, $d$, and $\bar d$.
Noting isospin factors and assigning the factor $a$ for the 
splitting probability $u\rightarrow d\pi^+$, we have
the final state
\begin{equation}
u \rightarrow a\pi^+ +ad +\frac{a}{2}\pi^0 +\frac{a}{2} u 
            = \frac{7a}{4} u +\frac{5a}{4} d
                               +\frac{a}{4}\bar u +\frac{5a}{4}\bar d 
\ \ \ .
\end{equation}
In the same way, the d quark splitting becomes
\begin{equation}
d \rightarrow  a\pi^- +au +\frac{a}{2}\pi^0 +\frac{a}{2} d 
            =  \frac{5a}{4} u +\frac{7a}{4} d
                               +\frac{5a}{4}\bar u +\frac{a}{4}\bar d 
\ \ \ .
\end{equation}
In this simple picture, the sum deficit is estimated by taking
the difference between $\bar u$ and $\bar d$ in the above equations:
\begin{equation}
\Delta I_G = \frac{2}{3} (\bar u-\bar d) = - \frac{2a}{3}
\ \ \ .
\end{equation}
The probability $a$ is calculated in the chiral field theory.
With the interaction Lagrangian ${\mathcal L}_{\Pi q}$, 
the splitting function is given by
\begin{equation}
P_{\Pi q'\leftarrow q} (z) = \frac{g_{_A}^2}{f^2} \, 
                             \frac{(m_q+m_{q'})^2}{32\pi^2} \, 
            z \int_{-\Lambda^2}^{t_{m}} dt \, 
                     \frac{(m_q-m_{q'})^2-t}{(t-M_\Pi^2)^2}
\ \ \ ,
\end{equation}
where $t_{m}=m_q^2 z-m_{q'}^2z/(1-z)$ and the ultraviolet cutoff
is taken a chiral-symmetry-breaking scale:
$\Lambda\approx$1169 MeV. This equation is analogous to
Eq. (\ref{eqn:fmb}) in the previous meson model.
Integrating the function over $z$, we obtain the probability
$a$ for the $u\rightarrow\pi^+ d$ process:
\begin{equation}
a = \frac{g_{_A}^2 \, m_u^2}{8\pi^2 f^2} 
    \int_0^1 dz \, \theta (\Lambda^2-\tau(z)) \, z  
     \left \{ \ln \left [ \frac{\Lambda^2+M_\pi^2}{\tau(z)+M_\pi^2} \right ]
              +M_\pi^2 \left [ \frac{1}{\Lambda^2+M_\pi^2} -
                               \frac{1}{\tau(z)+M_\pi^2} \right ] \right \}
\ ,
\end{equation}
where $\tau(z)=m_u^2 z^2/(1-z)$ and 
$\theta (x)$ is a cutoff function defined by
$\theta(x)=1$ for $x>0$ and 0 for $x<0$.
With the cutoff $\Lambda=$1169 MeV, the probability becomes
$a=0.083$ which leads to $I_G=(1-2a)/3=0.278$.
If a larger cutoff is taken, for example $\Lambda=$1800 MeV, the sum
becomes smaller ($I_G$=0.252). 
The obtained deficit is qualitatively in agreement with
the meson-cloud models in the previous subsection.
The $x$ distribution is calculated by the convolution \cite{EHQ,KRE}
\begin{equation}
\bar q_i (x) = \sum_{j=u,d} \sum_{k,l}
               \left (  \delta_{jl} \delta_{ik} 
                 - \frac{1}{n_f} \delta_{jk} \delta_{il} \right )^2
               \int_x^1 \frac{dy}{y} \int_{x/y}^1 \frac{dz}{z}
               \, \bar q_i^{\, (\Pi)}(x/(yz)) 
               \, P_{\Pi k\leftarrow j}(z)
               \, q_{v \, j}^{(N)}(y)
\ ,
\label{eqn:cpt}
\end{equation}
where the flavor summation is taken for the indices $k$ and $l$. 
However, the calculated $\bar u-\bar d$ distribution is
concentrated in the small $x$ region at the NA51 scale
$Q^2=27$ GeV$^2$, so that the model only with the pion
has difficulty in explaining the large NA51 flavor asymmetry
at $x=0.18$ \cite{KRE}.
The model is compared with the meson-cloud results and
it is extended to study the strange quark distribution \cite{SBF}.
On the other hand, model consistency is studied 
among different quantities:
the asymmetry $\bar u-\bar d$, the $\bar s$ distribution, 
and quark polarizations \cite{CL}.

As another effective model to describe low-energy properties of hadrons,
a chiral quark-meson model was proposed.
This is an extension of the linear-sigma model with replacement of the nucleon
field by the quark field. The Lagrangian density is given by
\begin{align}
\mathcal{L} (x) &= \bar\psi (x) \{ i\gamma\cdot\partial 
             +g[\sigma (x) +i{\tilde\tau}\cdot{\tilde\phi}(x)\gamma_5]\}\psi(x)
\\ \nonumber
       & \ \ \ \ \ \ 
             + \frac{1}{2} \partial_\mu \sigma (x) \partial^\mu \sigma (x)
             + \frac{1}{2} \partial_\mu {\tilde\phi} (x) \cdot
                             \partial^\mu {\tilde\phi} (x)
\\ \nonumber
       & \ \ \ \ \ \ 
             - \frac{\lambda^2}{4} [\sigma(x)^2+{\tilde\phi}(x)^2 -\nu^2]^2
             - F_\pi m_\pi^2 \sigma(x)
\ \ \ .
\end{align}
The meson fields are treated as classical
mean fields, in which the quarks form bound states.
In this model, the nucleon consists of valence
quarks and a coherent superposition of mesons, and
it is generated from mean-field hedgehog solution.
However, it is known that the hedgehog states are not
eigenstates of spin nor isospin. 
The nucleon with definite spin and isospin
should be obtained by a semi-classical cranking method.
For slow rotations, cranked meson spin and isospin are
linear in the angular velocity.  The moment of inertia is given by
valence-quark and pion contributions: 
${\mathcal I}={\mathcal I}_q+{\mathcal I}_\pi$.
For discussing the flavor asymmetry in the pion model,
an important factor is the number difference between
$\pi^+$ and $\pi^-$ in the proton.
The difference is equal to the fraction of the proton
electric charge carried by the pions:
$N_{\pi^+}-N_{\pi^-} 
 = \frac{\mathcal{I}_\pi}{\mathcal{I}} \left < I_3 \right > _p$.
Then, the Gottfried sum becomes \cite{SC}
\begin{equation}
I_G=\frac{1}{3} \left ( 1 - \frac{\mathcal{I}_\pi}{\mathcal{I}} \right )
\ \ \ .
\label{eqn:SCGOTT}
\end{equation}
The deviation from 1/3 is related to
the fact that a fraction of the nucleon isospin is carried by the pions.
The fraction is expressed by the moments of inertia for the nucleon and
pion. 

More rigorous derivation of the sum-rule violation, which is similar to
Eq. (\ref{eqn:SCGOTT}), is given in Ref. \cite{WAKA} where
a similar chiral-quark-soliton model is used.
The model consists of quark and pion fields with the following functional 
\begin{equation}
Z=\int \mathcal{D} \pi \, \mathcal{D} \psi \, \mathcal{D} \psi^\dagger
     exp \left [ \, i \int d^4x \, \bar \psi \, (i \partial \hspace{-6.0pt} /
                 -M U^{\gamma_5} -m) \, \psi \, \right ] 
\ \ \ ,
\end{equation}
with $U^{\gamma_5}(x)= e^{i\gamma_5 {\tilde\tau} \cdot {\tilde\pi} (x)/f_\pi}$.
The nucleon is treated in the same way with the previous chiral 
quark-meson model. However, the matrix element
$I_G= (1/3) <p\, | \, \hat O \, |\, p>$ is calculated with a plausible operator
$\hat O = \int d^3x [\psi_+^\dagger (x)\tau_3 \psi_+(x)-
                     \psi_-^\dagger (x)\tau_3 \psi_-(x)]$,
where $\psi_+$ and $\psi_-$ are positive and negative energy parts of $\psi$.
As a result, the integral becomes a similar equation to Eq. (\ref{eqn:SCGOTT}):
\begin{equation}
I_G=\frac{1}{3} \left ( 1 - 
           \frac{{\mathcal I}_{val}+{\mathcal I}_{vp}}{{\mathcal I}} \right )
\ \ \ ,
\label{eqn:WAKA}
\end{equation}
with the moments of inertia
\begin{align}
{\mathcal I}_{val} &= \frac{N_c}{2} \sum_{m\ne 0} 
           \frac{ <0\, |\tau_3|\, m><m\, |\{ \tau_3, P_- \}_+|\, 0>}{E_m-E_0} 
\ \ \ ,
\nonumber \\
{\mathcal I}_{vp}  &= \frac{N_c}{8} \sum_{m,n} f(E_m,E_n;\Lambda)
             <n\, |\tau_3|\, m><m\, |\{ \tau_3, P_- \}_+|\, n>
\ \ \ .
\end{align}
The state $|\, m>$ is an eigenstate of the single-quark Dirac equation and
$f(E_m,E_n;\Lambda)$ is a cutoff function. 
The ${\mathcal I}_{val}$ and ${\mathcal I}_{vp}$
are valence quark and vacuum polarization parts respectively, and 
the $P_-$ is the projection operator of the negative-energy eigenstates.
Numerical results depend on the choice of the dynamical quark mass $M$;
the sum ranges from $I_G$=0.235 for $M$=450 MeV
                 to $I_G$=0.288 for $M$=350 MeV.
Similar calculation is discussed in Ref. \cite{BPG} by using
the NJL model.
From these results, we find that the chiral models give similar qualitative
results to those of the mesonic models in section \ref{MESON-1}.
However, we have to be careful in comparing the effective-model
results with the NMC value. 
The effective models are considered to be valid at small $Q^2$,
typically $Q^2\sim \Lambda_{QCD}^2$. 
On the other hand, the NMC sum is given
at $Q^2$=4 GeV$^2$ where the perturbative QCD is usually applied.
It is not obvious whether these different $Q^2$ results could be compared
directly.

We also comment on other studies in the
chiral models. In Ref. \cite{WH}, the Gottfried sum is assumed 
to be related to the matrix element: $I_G=(1/3)<p|u\bar u-d\bar d|p>$. 
Then it is estimated in a soliton model. The obtained result is a simple relation:
$I_G\approx (1/3)(M_n-M_p)/(m_d-m_u)$ \cite{WH,SF}.
In Ref. \cite{LI}, the $\bar u-\bar d$ is related to the $\sigma$ term by
$\sigma=23(m_u+m_d)(1/2+\bar u-\bar d)$.
Choosing the mass $m_u+m_d$ to fit the pion mass and the $\sigma$
term, we find $\bar u-\bar d=-0.134$.
There is also an attempt to relate the Gottfried sum to 
kaon-nucleon scattering cross sections \cite{KORE}.
On the other hand, there is an instanton model approach
\cite{KOCH}. The instanton induced quark-nucleon interaction
is described by a Lagrangian with terms which do not vanish
only for different quark flavors. This feature could be 
related to the observed $\bar u/\bar d$ asymmetry.
Because we do not discuss the details of these works,
the interested reader should look at the original papers.

\subsubsection{Anomalous $Q^2$ evolution}\label{MESON-Q2}

We have learned that the scaling violation phenomena
in the structure functions are successfully described by
the perturbative QCD (pQCD). We believe that the pQCD
can be used in the $Q^2$ region, $Q^2>\, \, $a few GeV$^2$.
In the recent years, the experimental data improved much
and they provide us an opportunity to investigate
beyond the pQCD in the small $Q^2$ region. 
We explained that the lattice QCD cannot handle 
the structure functions themselves at this stage.
Studies of small $Q^2$ physics are difficult and 
they are inevitably model dependent. 
It is nevertheless important to study such a $Q^2$ region
by some kind of model in order to learn about underlying physics.
For this reason, we have discussed
the low-energy effective models in the previous subsection.
In the following, we discuss possible meson effects
on the scaling violation at small $Q^2$.

\begin{floatingfigure}{7.0cm}
   \begin{center}
      \mbox{\epsfig{file=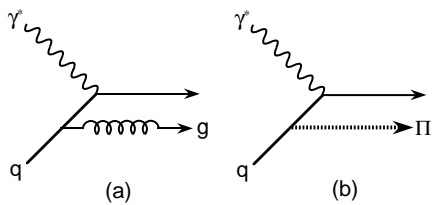,width=6.0cm}}
   \end{center}
 \vspace{-0.4cm}
\caption{\footnotesize
          $Q^2$ evolution due to (a) gluon and (b) meson $\Pi$ emissions.}
\label{fig:mq2}
\end{floatingfigure}
\quad
We found that the mesonic mechanism is a strong candidate
in explaining the failure of the Gottfried sum. 
Because of similarity between the quark splitting $q\rightarrow gq$
in Fig. \ref{fig:mq2}(a) and the one into a quark and a meson
($q\rightarrow \Pi q$) in Fig. \ref{fig:mq2}(b), 
the meson emission is expected to modify the $Q^2$ evolution.
The process (a) together with other splitting processes
gives rise to the standard Dokshitzer-Gribov-Lipatov-Altarelli-Parisi
(DGLAP) evolution equations. 
Because the appropriate degrees of freedom at large $Q^2$
are the quarks and gluons, the evolution has to be described by
the interactions among these fundamental particles, namely by
the DGLAP equations.
However, meson degrees of freedom may become important at
relatively small $Q^2$. If the meson effects are included in the
evolution equations, special care should be taken for double counting.
We discuss possible mesonic effects
on the $Q^2$ evolution as discussed in Ref. \cite{FBBG}.

If the meson interactions are taken into account in addition,
the DGLAP evolution equations are modified as
\begin{align}
\frac{\partial}{\partial t} \, q_i &= \sum_j P_{q_i q_j} \otimes q_j
                              +\sum_j P_{q_i \bar q _j} \otimes \bar q_j
                              +\sum_a P_{q_i \Pi_a} \otimes \Pi_a 
\ \ \ , \nonumber \\
\frac{\partial}{\partial t} \, \bar q_i &= \sum_j P_{\bar q_i \bar q_j} 
\otimes \bar q_j
                              +\sum_j P_{\bar q_i q _j} \otimes q_j
                              +\sum_a P_{\bar q_i \Pi_a} \otimes \Pi_a 
\ \ \ , \nonumber \\
\frac{\partial}{\partial t} \, \Pi_a &= \sum_j P_{\Pi_a q_j} \otimes q_j
                              +\sum_j P_{\Pi_a \bar q _j} \otimes \bar q_j
                              +\sum_b P_{\Pi_a \Pi_b} \otimes \Pi_b 
\ \ \ ,
\end{align}
where $\Pi_a$ is a distribution function of a meson,
and $t$ is defined by $t=\ln(Q^2/\mu^2)$.
Since the current interest is the nonsinglet evolution,
obvious gluon terms are omitted for simplicity.
The notation $\otimes$ indicates a convolution integral:
$f\otimes g=\int_x^1 dy f(x/y) g(y)$.
The function $P_{q\Pi}$ is the splitting probability
of the meson $\Pi$ into a quark and an antiquark,
and $P_{\Pi q}$ ($P_{\Pi \Pi}$) represents 
the $\Pi$ emission probability from a quark 
(from another meson $\Pi$).
Light pseudoscalar mesons are taken into account in the above equations.
Isospin and charge-conjugation invariance suggests
$P_{u\Pi^0}=P_{d\Pi^0}=P_{\bar u \Pi^0}=P_{\bar d \Pi^0}$
for neutral pseudoscalar mesons, 
$P_{u\pi^+}=P_{d\pi^-}=P_{\bar d \pi^+}=P_{\bar u \pi^-}\equiv P_{q\pi}$ and
$P_{d\pi^+}=P_{u\pi^-}=P_{\bar u \pi^+}=P_{\bar d \pi^-}=0$
for charged pions.
Defining the distribution $q^+=(u+\bar u)-(d+\bar d)$, we obtain
a nonsinglet evolution equation
\begin{equation}
\frac{\partial}{\partial t} \, q^+ = ( Q_{qq} + Q_{q\bar q} ) \otimes q^+
\ \ \ ,
\label{eqn:q+}
\end{equation}
where $Q$ is defined by the difference between flavor-diagonal
and nondiagonal splitting functions $Q=P^D-P^{ND}$.
They are given by $P_{q_i q_j}^D=P_{q_i q_j\, (i=j)}$
and $P_{q_i q_j}^{ND}=P_{q_i q_j \, (i \ne j)}$.
The meson terms cancel out in Eq. (\ref{eqn:q+}); however,
there are mesonic contributions to the splitting functions.

A contribution of the $\Pi$ emission to the splitting
function $P_{q_i q_j}$ is given by 
$[P_{q_i q_j}]^{\Pi} = \partial \sigma_{q_i q_j}^{\gamma^* \Pi}/\partial t$.
Here, $\sigma_{q_i q_j}^{\gamma^* \Pi}$ is the total cross section
integrated over $\vec k_\perp$ for absorption of a virtual photon
and emission the meson $\Pi$.
Its anomalous dimension is then calculated by the Mellin transformation:
$\gamma^{ud}_{_N} =\gamma^{du}_{_N} =
 \partial \sigma^{\gamma^*\pi^+}_{_N}/\partial t =
 \partial \sigma^{\gamma^*\pi^-}_{_N}/\partial t$, 
$\gamma^{uu}_{_N} =\gamma^{dd}_{_N} =(\partial /\partial t)
\left({\sigma^{\gamma^*\pi^0}_{_N}/2
+\sigma^{\gamma^*\eta}_{_N}/6
+\sigma^{\gamma^*\eta '}_{_N}}/3\right)$,
where $\sigma_{_N}$ is the Nth moment of the cross section.
In discussing the evolution of the Gottfried sum, 
we calculate the anomalous dimension
$\gamma_1^+$, which is the first moment of the splitting function for
the flavor diagonal minus the one for the nondiagonal. 
The mesonic contribution is
\begin{align}
\gamma_{_N}^+ = \gamma_{_N}^{uu}-\gamma_{_N}^{ud} &=
\frac{\partial}{\partial t}
\left( \frac{1}{2} \sigma^{\gamma^*\pi^0}_{_N}+
\frac{1}{6}\sigma^{\gamma^*\eta}_{_N} +
\frac{1}{3}\sigma^{\gamma^*\eta^\prime}_{_N}-
\sigma^{\gamma^*\pi^+}_{_N}\right)
\nonumber \\
            &\simeq \frac{\partial}{\partial t}
\left(\frac{1}{6}\sigma^{\gamma^*\eta}_{_N} +
\frac{1}{3}\sigma^{\gamma^*\eta^\prime}_{_N}-
\frac{1}{2}\sigma^{\gamma^*\pi}_{_N}\right)
\ \ \ .
\end{align}
In this way, evolution of the Gottfried sum due to
the meson emissions becomes
\begin{multline}
I_G(Q^2)= \Delta_1^+ (t,t_0) \, I_G(Q_0^2) \  , \ \ \  
\Delta_1^+ (t,t_0) = exp \left \{ 
               \left [ \frac{1}{6} \sigma_1^{\gamma^* \eta} (t)
                      +\frac{1}{3} \sigma_1^{\gamma^* \eta '} (t)
                      -\frac{1}{2} \sigma_1^{\gamma^* \pi} (t) \right ] 
\right. \\
\left.
              -\left [ \frac{1}{6} \sigma_1^{\gamma^* \eta} (t_0)
                      +\frac{1}{3} \sigma_1^{\gamma^* \eta '} (t_0)
                      -\frac{1}{2} \sigma_1^{\gamma^* \pi} (t_0) \right ] 
   \right \}
\ \ .
\end{multline} 
Explicit expressions for the above cross sections are presented in
the Appendix of Ref. \cite{FBBG}.
In the large mass limit $M\rightarrow\infty$, the cross section
falls off like $\sigma_1^{\gamma^* \Pi} \sim 1/M^2$, so that
massive meson contributions are smaller than those of the light mesons
($\pi$, $\eta$, and $\eta '$).

We comment on the consistency with the DGLAP equations at large $Q^2$
and on double counting. First, the cross section has 
$\sigma_1^{\gamma^* \Pi} \sim 1/Q^2$ behavior at large $Q^2$.
It means that the anomalous dimension has higher-twist
behavior. As $Q^2$ becomes sufficiently large, the meson
contribution becomes eventually smaller than the logarithmic pQCD effect.
Therefore, the above formalism is consistent with the DGLAP evolution
at large $Q^2$.
Second, the meson contribution to the anomalous dimension
should not be added to the pQCD one in order to avoid the double
counting. We may just use the meson picture below a certain boundary
$Q^2$ and the pQCD evolution above it.

\begin{wrapfigure}{l}{7.0cm}
   \begin{center}
      \mbox{\epsfig{file=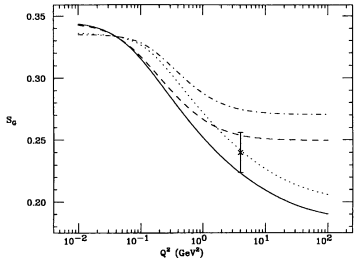,width=5.0cm}}
   \end{center}
 \vspace{-0.4cm}
\caption{\footnotesize $Q^2$ dependence of the Gottfried sum
           (taken from Ref. {\normalsize\cite{FBBG}})
}
\label{fig:igmq2}
\end{wrapfigure}
\quad
Numerical results are shown in Fig. \ref{fig:igmq2},
where the $Q^2$ dependence of the Gottfried sum is shown.
Parameters in the model are dynamical quark mass $m_d$, 
derivative and pseudoscalar coupling ratio $g_\pi$, 
and vertex cutoff parameters $\Lambda$ and $\tilde\Lambda$
for two different couplings.
The mass $m_d$ is determined by the cutoff $\Lambda$ (=$\tilde\Lambda$) with
a normalization condition of the vertex function.
In the range of $\Lambda$=0.4$-$0.8 GeV, it is consistent
with the constituent quark mass $m_d\approx M_\rho/2\approx M_p/3$.
Therefore, the cutoff is chosen 0.5 GeV.
The solid and dashed curves are the results for
$g_\pi$=0.0 and 1.0 respectively, and 
the dotted and dot-dashed ones are those for $g_\pi$=0.0 and 1.0
including dynamical quark-mass corrections.
As it is discussed in section \ref{PQCD}, the perturbative QCD
predicts a very small $Q^2$ variation. Therefore, the scaling
violation is mostly controlled by the mesonic contributions.
Although the results depend much on unknown parameters,
it is evident from the figure that the sum rule is violated
due to the scaling violation caused by the quark-meson interactions.
This is an interesting result, which suggests another possibility
of explaining the sum-rule failure. 
Furthermore, $x$ distribution of $F_2^p-F_2^n$ is calculated at
$Q^2$=4 GeV$^2$ by using input distributions at small $Q^2$
without sea-quark distributions \cite{FBBG}. 
The results agree reasonably well with the NMC $F_2^p-F_2^n$ data.
At this stage, these results should be considered as naive ones.
We need further refinement of the formalism, including discussions on
applicability of the modified evolution equations at small $Q^2$.
In any case, the model predicts very strong $Q^2$ dependence
of the Gottfried sum, so that it can be checked 
in principle by future experiments.

\subsection{Diquark model}\label{DIQUARK}

Flavor asymmetry $\bar u-\bar d$ in a diquark model was already 
noticed in 1976 \cite{PAV76}. Later it was elaborated to compare with
the NMC result in Ref. \cite{AP,ABCP}.
According to the ordinary quark model,
baryons consists of three pointlike quarks 
with spin-parity (1/2)$^+$ and charges 2/3, $-$1/3, $-$1/3. 
It is successful in explaining gross properties of hadrons. 
However, it was rather difficult to understand
the small ratio $F_2^n/F_2^p \sim 0.29$ at large $x$ and missing
SU(6) baryon multiplets $\underline{20}$ in 1970's,
although these problems could be explained within
the usual quark-model framework \cite{RECENT}.
These difficulties are understood in a diquark model.
This model is based on the observation that diquark degrees
of freedom are most relevant for some observables.

In the SU(6) model \cite{FEC}, 
two quarks form twenty-one symmetric states and fifteen 
antisymmetric ones:
$\underline{6}  \times \underline{6}=
 \underline{21} +  \underline{15}$.
In combination with the remaining quark, the antisymmetric
part $\underline{15}$ becomes
$\underline{15} \times \underline{6}=\underline{20} + \underline{70}$.
If the antisymmetric part $\underline{15}$ does not
couple to the quark, it is possible to explain
the missing states $\underline{20}$.
The SU(3) content of the representation $\underline{21}$
is expressed as
$\underline{21}=\{6\} \times 3 +\{\bar 3\}\times 1$,
where the brackets indicate irreducible representation
of SU(3) and the factors 3 and 1 are spin degrees of
freedom of the diquark.
There are SU(3)-sextet axial-vector diquarks and
SU(3)-triplet scalar diquarks.
We introduce a mixing angle $\Gamma$ between 
the vector and scalar diquark states.
The usual SU(6) model is recovered in the limit $\Gamma=\pi/4$.
The proton state in this diquark model is given by
\begin{multline}
\left | \ p, s_z=\pm \frac{1}{2} \ \right > =
    \pm \frac{1}{\sqrt{18}} 
    \left \{ \, {\text{\LARGE [}} \, 
              \sqrt{2} \, V_{\pm 1} (ud) \, u_{\mp} 
                       - 2 \, V_{\pm 1} (uu) \, d_{\mp}
                + \sqrt{2} \, V_{0} (uu) \, d_{\pm}   \right .
    \\
      \left.  -   V_{0} (ud) \, u_{\pm} \, {\text{\LARGE ]}} \, 
          \sqrt{2} \, \sin\Gamma
          \mp S(ud) \, u_{\pm} \, \sqrt{2} \, \cos\Gamma \right \}
\ \ \ ,
\label{eqn:DIQUARK}
\end{multline}
where $V_m(q_1 q_2)$ and $S (q_1 q_2)$ denote the vector and scalar
diquark states consist of $q_1$ and $q_2$ quarks, and the subscript m
is the spin state.

Quark and diquark contributions to the structure function $F_2$
are given by \cite{ACLS}

\vfill\eject
\begin{align}
F_2^{(q)}  &= x \sum_q e_q^2 \, q(x)   \ \ \ ,       \nonumber  \\
F_2^{(S)}  &= x \, e_{_S}^2 \, S(x) \, D_S (Q^2)   \ \ \ ,   \nonumber      \\
F_2^{(V)}  &= x \sum_V \, e_{_V}^2 \, \frac{1}{3} \, V(x)
      \left \{ \, \left [ \, \left ( 1 + \frac{\nu}{m_N x} \right ) D_1 (Q^2)
                                       - \frac{\nu}{m_N x} \, D_2 (Q^2)
                \right. \right. \nonumber \\
  & \ \ \ \ \ \ \   \left. \left.
  + 2 \, m_N \, \nu \, x \left ( 1+ \frac{\nu}{2m_N x} \right ) 
                                               D_3 (Q^2) \right  ]^2
  + 2 \left [ D_1^2 (Q^2) +\frac{\nu}{2m_N x} \, D_2^2 (Q^2) \right ] \, \right \}
,
\label{eqn:F2QSV}
\end{align}
where $D_S(Q^2)$ and $D_{1,2,3}(Q^2)$ are scalar and vector diquark
form factors. The $D_1$, $D_2$, and $D_3$ are defined by tensor structure
of the virtual-photon coupling to a spin-one particle:
\begin{multline}
V^\alpha = i e_{_V} \left \{ (2k+q)^\alpha g^{\mu\nu} D_1(Q^2)
                  -[(k+q)^\nu g^{\mu\alpha} +k^\mu g^{\nu\alpha} ] D_2(Q^2)
         \right.
   \\
         \left.
   + k^\mu (k+q)^\nu (2k+q)^\alpha D_3(Q^2) \right \} 
       \epsilon_{1,\nu}(\lambda_1) \epsilon_{2,\mu}^* (\lambda_2) 
\ \ \ ,
\end{multline}
where $\epsilon_{1,\nu}(\lambda_1)$ and $\epsilon_{2,\mu} (\lambda_2)$
are the polarization vectors of initial and final diquarks with helicities
$\lambda_1$ and $\lambda_2$.
In the limit of pointlike diquarks, the form factors are given by
$D_S(0)=1$, $D_1(0)=1$, $D_2(0)=1+\kappa$, and $D_3(0)=0$,
where $\kappa$ is the anomalous magnetic moment. Therefore,
it is natural to choose 
$D_S(Q^2)=D_1(Q^2)=D_2(Q^2)=Q_0^2/(Q_0^2+Q^2)\equiv D(Q^2)$, 
as expected from a dimensional counting rule, and $D_3=0$
for simplicity.
From Eqs. (\ref{eqn:DIQUARK}) and (\ref{eqn:F2QSV}), valence quark and diquark
contributions to the proton $F_2$ are
\begin{multline}
F_2^p = x \left [ \, \left \{ \, \frac{4}{9} \, \frac{1}{3} \, f_u(x)
                              +\frac{1}{9} \, \frac{2}{3} \, f_d(x) \right \}
                              \sin^2 \Gamma
                        + \frac{4}{9} \, f_u(x) \, \cos^2 \Gamma
                        + \frac{1}{9} \, f_{_S}(x) \, \cos^2 \Gamma \, D^2 (Q^2)
        \right.
     \\
        \left.
     + \left \{ \frac{16}{9} \, \frac{2}{3} \, f_{V_{uu}}(x) 
           + \frac{1}{9} \, \frac{1}{3} \, f_{V_{ud}}(x) \right \} \sin^2 \Gamma
            \left ( 1 + \frac{\nu}{3m_N x} \right ) D^2 (Q^2) \, \right ]
\ .
\end{multline}
New distribution functions $f_{q,S,V}(x)$ are introduced 
in the above equation, and they are normalized as
$\int_0^1 dx f_{q,S,V}(x)=1$.
In the same way, the neutron $F_2$ is given by
\begin{multline}
F_2^n = x \left [ \, \left \{ \frac{4}{9} \, \frac{2}{3} \, f_d(x)
                              +\frac{1}{9} \, \frac{1}{3} \, f_u(x) \right \}
                              \sin^2 \Gamma
                        + \frac{1}{9} \, f_u(x) \, \cos^2 \Gamma
                        + \frac{1}{9} \, f_{_S}(x) \, \cos^2 \Gamma \, D^2 (Q^2)
        \right.
     \\
        \left.
     + \left \{ \frac{4}{9} \, \frac{2}{3} \, f_{V_{uu}}(x)
               + \frac{1}{9} \, \frac{1}{3} \, f_{V_{ud}}(x) \right \} 
                     \sin^2 \Gamma
            \left ( 1 + \frac{\nu}{3m_N x} \right ) D^2 (Q^2) \, \right ]
\ .
\end{multline}
In addition, the virtual-photon scattering off a quark
inside the diquark is considered with a diquark-breakup
probability $1-F^2(Q^2)$.
From these equations, the Gottfried sum becomes \cite{ABCP}
\begin{multline}
I_G= \frac{1}{3}  - \frac{4}{9} \, sin^2 \Gamma
     + \frac{8}{9} \, \sin^2 \Gamma
       \int_0^1 dx \, f_{V_{uu}} (x) 
      \left ( 1 + \frac{\nu}{3m_N x} \right ) D^2 (Q^2)
   \\
      + \frac{4}{9} \, [1-F^2(Q^2)] \, sin^2 \Gamma
\ \ .
\end{multline}
It should be noted that sea-quark contributions are not
taken into account. In other words, the sea-quark distributions
are assumed to be flavor symmetric. 
If the nucleon consists of a scalar diquark and a quark
($\Gamma=0$), the integral becomes the Gottfried sum 1/3.

In the earlier investigations \cite{PAV76,AP}
without the breakup term,
the diquark model seemed to account for the deficit of 
the Gottfried sum. For example, $I_G$=1/3$-$0.384$\ \sin^2 \Gamma$
at $Q^2$=4 GeV$^2$ was obtained in Ref. \cite{AP}.
However, it means that the two quarks in a diquark act
as a single extended object, which never breaks apart.
This is certainly not realistic. 
The breakup mechanism plays an important role in the Gottfried sum.
With the distribution $f_{V_{uu}}(x)=12x^2(1-x)$ used
in the study of polarized structure function $g_1$ and
with the assumption $F(Q^2)=D(Q^2)$, 
the sum at $Q^2$=4 GeV$^2$ becomes \cite{ABCP}
\begin{equation}
I_G=\frac{1}{3} + \frac{4}{9} \, (0.12) \, \sin^2 \Gamma
\ \ \ .
\label{eqn:IGDIQUARK}
\end{equation}
On the contrary to the previous results,
the model produces a positive modification to the sum.
The parameter $\Gamma$ could be taken from other observable
such as the ratio of the axial vector to the vector
neutron $\beta$-decay coupling constant $g_{_A}$. 
Comparing $g_{_A}=1+(2/3)sin(2\Gamma)$
in the diquark model with experimental value $g_{_A}$=1.261$\pm$0.004,
we obtain $sin^2\Gamma\approx$0.04. 
Then the sum becomes $I_G=1/3+0.002$, which is a very small positive
correction to the sum 1/3.

There are two factors which changed the early result
in Ref. \cite{AP}. The first one is the addition of 
the breakup term, and it could result in the positive
contribution in Eq. (\ref{eqn:IGDIQUARK}). The second one
is the small mixing angle $sin^2\Gamma\approx\, $0.04, which
is consistent with $g_{_A}$. It should be noted in Ref. \cite{AP}
that the NMC result could be explained if the mixing were
$sin^2\Gamma\approx\, $0.27!
Sensitivity of the obtained $I_G$ on the assumed functions 
is also discussed in Refs. \cite{AP,ABCP}.
If the function $F(Q^2)$ is different from $D(Q^2)$,
the deviation may become negative instead of 
the positive one in Eq. (\ref{eqn:IGDIQUARK}).
The results depend, of course, on the assumed functions
for $f_{V_{uu}} (x)$, $F(Q^2)$, and $D(Q^2)$; however,
the diquark effects are all suppressed by the 
small factor $sin^2\Gamma$. In this way,
the diquark model predicts a very small deviation
from the Gottfried sum, so that it cannot explain the NMC results.

\vfill\eject
\subsection{Isospin symmetry violation}\label{ISOSPIN}

Isospin symmetry is usually taken for granted in discussing
parton distributions in the proton and neutron.
In fact, it is assumed
[$u_n=d_p$, $d_n=u_p$, $\bar u_n=\bar d_p$, $\bar d_n=\bar u_p$, and etc.]
in deriving Eq. (\ref{eqn: F2P-M}).
Electromagnetic interactions are weak compared with strong interactions,
so that typical isospin-violation effects are expected to be of
the order of the fine structure constant $\alpha=1/137$.
This is in general true, for example, the mass difference of the nucleons
is $(m_n-m_p)/m_p$=0.14\%. Therefore, we cannot believe that
the NMC result is explained only by the isospin-symmetry violation
in antiquark distributions. 
However, it is worth investigating its contributions to the Gottfried sum 
and to various high-energy processes because isospin-violation effects
on the parton distributions are not known.
This topic is discussed in Ref. \cite{MSG}.

What would happen to the Gottfried sum if the isospin
symmetry cannot be assumed?
Without using the isospin symmetry, the sum is expressed as
\begin{multline}
I_G= \frac{1}{3} + \frac{2}{9} \int_0^1 dx
            \, \,  {\text{\Large (}} \, \left [ \, 4 \{ \bar u(x)+\bar c(x) \}
                               +\{ \bar d(x)+\bar s(x) \} \, \right ]_p 
\\        
                    - \left [ \, 4 \{ \bar u(x)+\bar c(x) \}
                          +\{ \bar d(x)+\bar s(x) \} \, \right ]_n \, 
              {\text{\Large )}}
\ \ .
\label{eqn:isv0}
\end{multline}
If the antiquark distributions are flavor symmetric
and if the $\bar s$ and $\bar c$ terms vanish: 
$\int dx (s_p-s_n)=0$ and $\int dx (c_p-c_n)=0$, it becomes
\begin{equation}
I_G= \frac{1}{3} +\frac{10}{9} 
        \int dx \left [ \bar q_p (x)-\bar q_n (x) \right ]
\ \ \ ,
\label{eqn:ISV}
\end{equation}
where $\bar q_p(x)$ is the light antiquark distribution in the proton
[$\bar q_p = \bar u_p =\bar d_p$] and $\bar q_n(x)$ 
is the one in the neutron.
If the isospin-symmetry breaking were the only origin of the NMC finding,
Eqs. (\ref{eqn:isv0})  and (\ref{eqn:ISV}) 
could suggest that there are more antiquarks
in the neutron than those in the proton.
If the NMC 1991 data in Eq. (\ref{eqn:NMC91}) is identified
with Eq. (\ref{eqn:ISV}), we get
$\int dx [ \bar q_p (x) - \bar q_n (x) ]=-0.84\pm 0.014$.
Because the Adler sum rule 
$I_A=\int [F_2^{\bar\nu p}(x)-F_2^{\nu p}(x)] dx/2x=1$ and
the Gross-Llewellyn Smith sum rule 
in Eq. (\ref{eqn:glssum}) are independent of
the flavor asymmetry and the isospin symmetry violation,
these mechanisms cannot be distinguished.
We have to find other observables.
Future experimental possibilities 
are discussed in section \ref{ISOSPINEX}.

\subsection{Flavor asymmetry $\bar u-\bar d$ in nuclei}\label{NUCLEI}

The NMC finding of the flavor asymmetry can be tested by
the Drell-Yan experiments. There exist Drell-Yan data for various nuclear
targets, so that some people use, for example, the tungsten data 
in investigating the flavor asymmetry \cite{E772}.
However, we have to be careful in comparing the NMC result
with the tungsten data because of possible nuclear medium effects.
In the analysis of Ref. \cite{E772}, no nuclear correction
is made except for the overall shadowing correction.
If the nuclear modification in the $\bar u-\bar d$ distribution
is very large, the Drell-Yan 
analysis cannot be compared directly with the NMC result.
Therefore, it is worth while estimating the nuclear effect
in order to find whether or not such a modification should
be taken into account.

In discussing antiquark distributions in a nucleus,
it is essential to describe the shadowing phenomena.
In the small $x$ region, it is experimentally observed that
nuclear structure functions per nucleon are smaller than the
deuteron's. 
There are various models in describing the shadowing
as they are discussed in section \ref{SHADOW}.
The interesting point is to find whether
there is any nuclear mechanism to create extra flavor asymmetry
whatever the shadowing model is.
It could be possible according to Ref. \cite{SK95},
in which the nuclear modification is calculated in a parton-recombination
model. The following discussion is based on this investigation.

First, we discuss the nuclear $\bar u-\bar d$ distribution
without the nuclear modification.
If the isospin symmetry could be applied to the parton distributions
in the proton and neutron, the distribution per nucleon is given by 
$[\bar u(x)-\bar d(x)]_A = - \varepsilon [\bar u(x)-\bar d(x)]_{proton}$.
It is simply the summation of proton and neutron contributions.
The neutron-excess parameter $\varepsilon$ is defined by
$\varepsilon =(N-Z)/(N+Z)$. The above equation indicates that
the flavor asymmetry has to vanish if the antiquark distributions 
are flavor symmetric in the nucleon. However, it is not the case
in the recombination model.

In the parton-recombination picture, partons in different nucleons
could interact in a nucleus.
These interactions become important especially at small $x$.
Parton recombination effects on the antiquark distribution
$\bar q_i(x)$ are given by
\begin{equation}
 \Delta \bar q_{i,A} (x)  =  w_{pp} \, \Delta \bar q_{i,pp} (x)
                          +  w_{pn} \, \Delta \bar q_{i,pn} (x)
                          +  w_{np} \, \Delta \bar q_{i,np} (x)
                          +  w_{nn} \, \Delta \bar q_{i,nn} (x)
\ ,
\label{eqn:PREQ}
\end{equation}
where $w_{n_1 n_2}$
is the combination probabilities of the two nucleons $n_1$ and $n_2$.
For example, $w_{pp}$ is the probability of the proton-proton combination
($w_{pp}=Z(Z-1)/[A(A-1)]$).
The distribution
$\Delta \bar q_{i,n1 \hspace{0.05cm} n2} (x)$ is the modification
of the antiquark distribution with flavor $i$ due to a parton interaction
in the nucleon $n1$ with a parton in the nucleon $n2$.
If the isospin symmetry can be used, the flavor asymmetry becomes
\begin{align}
x[\Delta \bar u(x) - \Delta \bar d(x)]_A 
        &= -(w_{nn} - w_{pp}) \, x \, 
            [\Delta \bar u(x) - \Delta \bar d(x)]_{pp}
             \\ \nonumber
        &= - \varepsilon \, x \, [\Delta \bar u(x) - \Delta \bar d(x)]_{pp}
\ \ \ ,
\end{align}
where $[\Delta \bar u(x)- \Delta \bar d(x)]_{pp}$
is the asymmetry produced in the proton-proton combination. 

\begin{floatingfigure}{7.0cm}
   \begin{center}
      \mbox{\epsfig{file=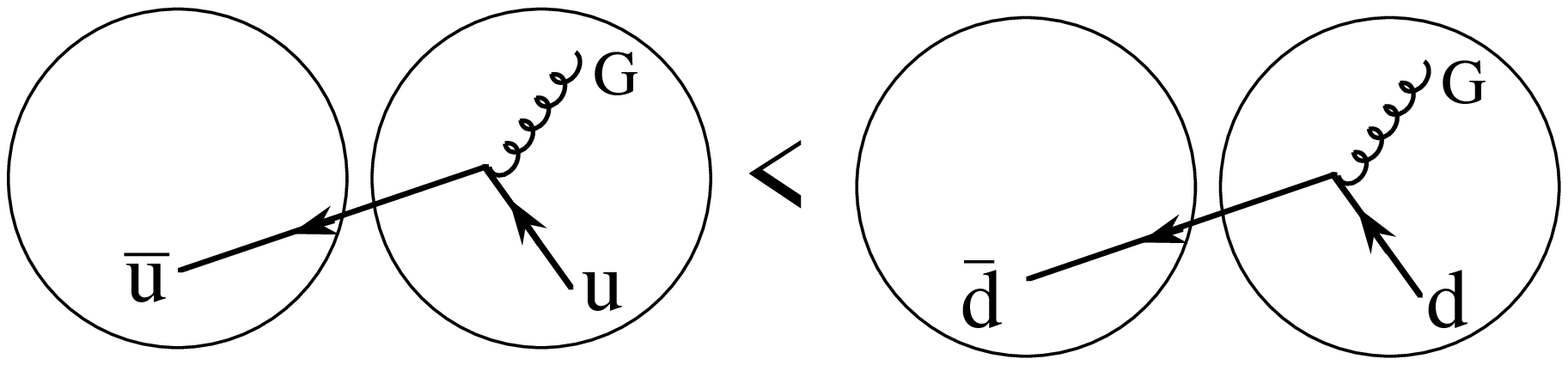,width=6.2cm}}
   \end{center}
 \vspace{-0.6cm}
\caption{\footnotesize Mechanism of creating the flavor asymmetry
                       in a nucleus.} 
\label{fig:recomb}
\end{floatingfigure}
\quad
Next, we discuss how the flavor asymmetric distribution is created
in this model for the simplest situation, $\bar u - \bar d=0$ in the nucleon.
In this case, many recombinations cancel each other, and the only remaining
term is the following:
\begin{equation}
x[\Delta \bar u(x) - \Delta \bar d(x)]_A= 
\varepsilon
\frac{4K_0}{9} x \int_0^1 dx_2  \,  x \, \bar u^* (x) 
        \, x_2 [u_v(x_2)-d_v(x_2)] \,  \frac{x^2+x_2^2}{(x+x_2)^4}
\ \ \ , 
\label{eqn:NUCL}
\end{equation}
where $u_v(x)$ and $d_v(x)$ are the u and d 
valence-quark distributions in the proton,
and the asterisk indicates a leak-out parton in the recombination.
The factor $K_0$ is defined by $K_0=9A^{1/3}\alpha_s(Q^2)/(2R_0^2Q^2)$
with $R_0=1.1$ fm.
The physics mechanism of creating the asymmetry in Eq. (\ref{eqn:NUCL})
is the following.
In a neutron-excess nucleus ($\varepsilon >0$) such
as the tungsten, more $\bar d$ quarks are lost than $\bar u$ quarks
in the parton recombination process 
$\bar q q \rightarrow G$ in Fig. \ref{fig:recomb} because of
the $d$ quark excess over $u$ in the nucleus.
The $\bar q q \rightarrow G$ type
recombination processes produce positive contributions at small $x$.
The details of the recombination formalism are discussed in
Ref. \cite{SK95}.

\begin{wrapfigure}{l}{7.0cm}
   \begin{center}
      \mbox{\epsfig{file=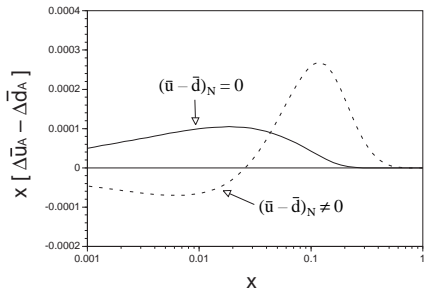,width=6.2cm}}
   \end{center}
 \vspace{-0.4cm}
\caption{\footnotesize Created flavor asymmetry in the tungsten nucleus
                       by a recombination model 
                       {\normalsize\cite{SK95}}.}
\label{fig:w}
\end{wrapfigure}
\quad
The recombination contributions are evaluated 
for the tungsten $_{74}^{184} W_{110}$ nucleus
by using the input parton distributions MRS-D$_0$ at $Q^2$=4 GeV$^2$.
The obtained results are shown in Fig. \ref{fig:w}, 
where the solid (dashed) curve shows the $x[\Delta\bar u-\Delta\bar d]_A$ 
distribution of the tungsten nucleus with the flavor symmetric (asymmetric)
sea in the nucleon.
In the $(\bar u-\bar d)_N = 0$ case, the positive contribution
at small $x$ can be understood by the processes in Fig. \ref{fig:recomb}.
In the $(\bar u-\bar d)_N \ne 0$ case, 
the $\bar q(x) G\rightarrow \bar q$ process
is the dominant one kinematically at small $x$.
Its contribution to $\bar u(x)-\bar d(x)$ becomes negative
due to the neutron excess.
On the other hand, the $\bar q G\rightarrow \bar q(x)$
process becomes kinematically favorable in the medium $x$ region.
Because it produces
$\bar q$ with momentum fraction $x$, its contribution becomes
opposite to the one at small $x$.

The above results are obtained at $Q^2$=4 GeV$^2$. 
Because the factor $K_0$ is proportional to $\alpha_s(Q^2)/Q^2$,
the nuclear flavor asymmetry may seem to be very large at small $Q^2$.
However, the quark distribution $u_v(x)-d_v(x)$  
becomes very small in the small $x$ region, so that the overall
$Q^2$ dependence is not so significant according to Eq. (\ref{eqn:NUCL}).
There are merely factor-of-two differences
between the asymmetric distribution 
at $Q^2$=4 GeV$^2$ and the one at $Q^2 \approx 1$ GeV$^2$.
Considering this factor of two,
we find that the nuclear modification is of the order of 2\%--10\%
compared with the asymmetry $\bar u-\bar d$ 
suggested by the MRS-D0 distribution.
Therefore, special care should be taken in comparing flavor asymmetry
data of the nucleon with the nuclear ones.
On the other hand, because the Drell-Yan experiments on 
various targets are in progress at Fermilab, 
the nuclear modification of $\bar u-\bar d$ could be
tested experimentally. The studies could provide important clues
in describing nuclear dynamics in the high-energy region.

\subsection{Relation to nucleon spin}\label{SPIN}

We explained various mechanisms of creating the flavor asymmetry.
It is natural that there is a certain relationship between
the light-antiquark flavor asymmetry and the nucleon spin issue. 
We discuss possible relations in the exclusion principle \cite{BS}
and in the chiral quark model \cite{EHQ,CL}. 

One of the ideas in explaining the flavor asymmetry is
the Pauli blocking model in section \ref{PAULI}.
This idea could be extended to the proton-spin problem 
according to Ref. \cite{BS}.
In a naive quark model, polarized valence-quark distributions
are related to the matrix elements of axial charges:
\begin{equation}
u_v^\uparrow   =1+F \ \ , \ \ 
u_v^\downarrow =1-F \ \ , \ \
d_v^\uparrow   =\frac{1+F-D}{2} \ \ , \ \ 
d_v^\downarrow =\frac{1-F+D}{2} \ \ ,
\label{eqn:udspin}
\end{equation}
where $F$ and $D$ are axial parameters, and the current values are
$F+D$=1.2573$\pm$0.0028 $F/D$=0.575$\pm$0.016 experimentally.
From these equations, fractions of the proton spin carried by
the valence quarks are
\begin{equation}
\Delta u_v = u_v^\uparrow - u_v^\downarrow = 2F \ \ , \ \ 
\Delta d_v = d_v^\uparrow - d_v^\downarrow = F-D \ \ .
\end{equation}
The Pauli blocking mechanism in the flavor case was the following.
Because there is an extra u-valence quark over d-valence,
$u\bar u$ pair creations suffer more exclusion effects
than $d\bar d$ creations. 
Substituting numerical values of $F$ and $D$ into Eq. (\ref{eqn:udspin}),
we obtain $u_v^\uparrow$=1.46, $u_v^\downarrow$=0.54, 
$d_v^\uparrow$=0.33, and $d_v^\downarrow$=0.67. 
The proton spin is dominated by the
$u_v^\uparrow$ distribution. Because $u_v^\uparrow$ is significantly
larger than $u_v^\downarrow$, the Pauli blocking could be applied to
the spin case in the similar way. As a rough estimate, 
the fraction of the spin asymmetry created in the exclusion principle 
is assumed to be the same with the one for the flavor asymmetry
\begin{equation}
\frac{u_{s}^\downarrow - u_{s}^\uparrow}
 {u_{v}^\uparrow - u_{v}^\downarrow} =
\frac{d_{s} - u_{s}}{u_{v} - d_{v}} 
\ \ \ ,
\end{equation}
where $q_s$ denotes a sea-quark distribution.
The NMC result in 1991 indicates $d_s-u_s$=0.14 for the first 
moments $u_s$ and $d_s$. Therefore, the difference
becomes
\begin{equation}
u_{s}^\uparrow - u_{s}^\downarrow = - 0.14 \, \Delta u_v 
= -0.28 \, F 
\ \ \ .
\end{equation}
We also assume that the exclusion mechanism is applied in
the same way to the d quark:
\begin{equation}
d_{s}^\uparrow - d_{s}^\downarrow = - 0.14 \, \Delta d_v 
= -0.14 \, (F-D)
\ \ \ .
\end{equation}
From these equations, the first moment of $g_1^p(x)$ becomes
\begin{equation}
\int_0^1 g_1^p(x) dx = \frac{1}{18} (9F-D)(1-0.28) =0.14,
\end{equation}
which is in fair agreement with polarized experimental data.
The Pauli blocking interpretation of the proton-spin issue
is summarized in the following way.
Because of the $u_v^\uparrow$ excess over $u_v^\downarrow$ and
the $d_v^\downarrow$ excess over $d_v^\uparrow$, the u-quark (d-quark) 
sea is negatively (positively) polarized. However, 
magnitude of the exclusion effect
is expected to be larger in the u-quark sea because of
$u_v^\uparrow/u_v^\downarrow > d_v^\downarrow/d_v^\uparrow$.
In fact, we have $\Delta\bar u$=$-0.28F=-$0.13
and $\Delta\bar d=-0.14(F-D)$=+0.05.
The large negative polarization in the u-quark sea could account
for the spin deficit.

The relation between the quark flavor and spin can be discussed
in the meson models, for example in the chiral quark model \cite{EHQ,CL}.
As explained in section \ref{CHIRAL}, the model
produces the $\bar d$ excess over $\bar u$. 
In Ref. \cite{CL}, kaons, eta, and eta prime mesons are included
in addition to the pions. Suppression factors are introduced
for heavier meson emissions in comparison with the pion case:
$\epsilon $ for kaons, $\delta $ for eta, and $\zeta $ for eta prime mesons. 
Then, after the emission of a meson from the initial proton ($2u+d$),
the $\bar u-\bar d$ number becomes
\begin{equation}
\bar{u}-\bar{d} = \left[ \frac{2\zeta + \delta }{3} - 1 \right] a
\ .
\label{eqn:club-db}
\end{equation}
The proton spin-up state is 
$\left| p_{+}\right\rangle = (1/\sqrt{6}) \left( 
              2\left| u_{+}u_{+}d_{-}\right\rangle 
              -\left| u_{+}u_{-}d_{+}\right\rangle 
              -\left| u_{-}u_{+}d_{+}\right\rangle \right) $.
This equation suggests the naive quark model prediction 
$\Delta u=u_+ - u_- =4/3$,
$\Delta d=d_+ - d_-=-1/3$, and $\Delta s=0$ with
$u_+=5/3$, $u_-=1/3$, $d_+=1/3$, and $d_-=2/3$.
Next, we discuss corrections to the quark polarization
due to the meson emissions. They are
expressed by the probability $P(q_+ \rightarrow q_-)$
for the splitting process $q_+ \rightarrow q_-$.
The probability $a$ is assigned for the process $u_+\rightarrow \pi^+  d_-$.
$P\left( d_{+}\rightarrow s_{-}\right) =\epsilon ^{2}a$ is the
probability of a spin-up $d$ quark flipping into a spin-down $s$ quark
through the emission of $K^{+}$.
In this way, the probabilities for the $u_+$ splitting processes are
$P(u_{+}\rightarrow (u\overline{d}) _{0}d_{-})=a$,
$P(u_{+}\rightarrow (u\overline{s}) _{0}s_{-})=\epsilon ^{2}a$,
$P(u_{+}\rightarrow (u\overline{u}) _{0}u_{-})=
                [ (\delta +2\zeta +3)/{6} ]^{2} a$,
$P(u_{+}\rightarrow (d\overline{d}) _{0}u_{-})=
                [ (\delta +2\zeta -3)/{6} ]^{2} a$, and
$P(u_{+}\rightarrow (s\overline{s}) _{0}u_{-})=
                [ (\delta -\zeta)/3 ]^{2} a$.
Substituting these equation and similar ones for other states,
we obtain
\begin{align}
\Delta u =& \frac{4}{3}\left[ 1-\Sigma P \right] 
           +\frac{1}{3}P\left(u_{-}\rightarrow u_{+}\right) 
           +\frac{2}{3}P\left( d_{-}\rightarrow u_{+}\right) 
           -\frac{5}{3}P\left( u_{+}\rightarrow u_{-}\right) 
           -\frac{1}{3}P\left( d_{+}\rightarrow u_{-}\right) 
\nonumber \\
         =& \frac{4}{3}-\frac{21+4\delta ^{2}+8\zeta ^{2}
            +12\epsilon ^{2}}{9}a 
\ , \nonumber \\
\Delta d =& -\frac{1}{3}\left[ 1-\Sigma P\right] 
            +\frac{1}{3}P\left(u_{-}\rightarrow d_{+}\right) 
            +\frac{2}{3}P\left( d_{-}\rightarrow d_{+}\right) 
            -\frac{5}{3}P\left( u_{+}\rightarrow d_{-}\right) 
            -\frac{1}{3}P\left( d_{+}\rightarrow d_{-}\right) 
\nonumber \\
         =&-\frac{1}{3}-\frac{6-\delta ^{2}-2\zeta ^{2}
            -3\epsilon ^{2}}{9}a
\ , \nonumber \\
\Delta s =& \frac{1}{3}P\left( u_{-}\rightarrow s_{+}\right) 
           +\frac{2}{3}P\left( d_{-}\rightarrow s_{+}\right) 
           -\frac{5}{3}P\left( u_{+}\rightarrow s_{-}\right) 
           -\frac{1}{3}P\left( d_{+}\rightarrow s_{-}\right)
\nonumber \\
         =&-\epsilon ^{2}a
\ .
\label{eqn:cl-dq}
\end{align}
The parameter values, $\epsilon$, $\delta$, and  $\zeta$, have to
be determined in order to evaluate above quantities numerically.
We may simply take the same suppression factors for the $K$ and $\eta$
production processes. Because $\eta ^{\prime }$ is heavier, a smaller
value may be taken for $\zeta$. For a rough estimate, we assume
$\epsilon =\delta =-2\zeta$.
These parameter values are fixed by Eq. (\ref{eqn:club-db}) so as to
explain the NMC $\bar u-\bar d$, and they are
$\epsilon =\delta =-2\zeta=0.6$ and $a=0.15$.
These are substituted into Eq. (\ref{eqn:cl-dq}) to obtain
$\Delta u$=0.87, $\Delta d=-$0.41, and $\Delta s=-$0.05.
The total quark spin content is $\Delta u+\Delta d+\Delta s$=0.4 which
is close to the recent measurements. 
Comparing these with the naive quark model predictions
$\Delta u=4/3$, $\Delta d=-1/3$, and $\Delta s=0$, we find that
the u quark polarization is significantly reduced due to the meson
emission mechanism.
Although the above calculation
is a very rough one, the results indicate that the chiral quark model
could also explain the proton spin problem. 
In this way, we find that the flavor asymmetry problem is closely
connected with the proton spin issue.

\subsection{Comment on effects of quark mass and transverse motion}
\label{KINE}

Although the parton model is considered to be valid in
the Bjorken scaling limit, there could be some 
corrections from finite quark masses and transverse motion.
Such corrections are estimated in Ref. \cite{SV},
and the obtained result indicates $\delta I_G =-$0.01 to $-$0.02
at $Q^2$=4 GeV$^2$. 
Later, a more careful analysis indicates a slightly larger 
correction $\delta I_G =-$0.029 to $-$0.051
at $Q^2$=3 GeV$^2$ \cite{MS96}.
It is interesting to find that
the discrepancy between the NMC result and the sum
becomes smaller, but it is not large enough to explain
the NMC deficit.
Because the correction becomes smaller:
$\delta I_G =-$0.009 to $-$0.017 even at slightly
larger $Q^2$ (=10 GeV$^2$). This kind of simple
interpretation could be tested by future experiments.
For the details of this topics,
the interested reader may look at the original papers \cite{SV,MS96}.

\vfill\eject
\section{{\bf Finding the flavor asymmetry $\bf\bar u-\bar d$
                              in various processes}}\label{OTHEREXP}
\setcounter{equation}{0}
\setcounter{figure}{0}
\setcounter{table}{0}

Because there could be a significant contribution from the small $x$ region
to the Gottfried sum rule, it is important to test the NMC flavor asymmetry 
by independent experiments. 
We discuss various processes in probing the $\bar u-\bar d$ distribution.
First, the Drell-Yan experiment is explained.
It should be the best candidate, in fact, existing
Drell-Yan data are used for investigating the flavor asymmetry.
We also discuss other processes such as W charge asymmetry,
quarkonium production, charged hadron production, 
and neutrino reaction.

\subsection{Drell-Yan process}\label{DRELLYAN}

\begin{floatingfigure}{7.0cm}
   \begin{center}
      \mbox{\epsfig{file=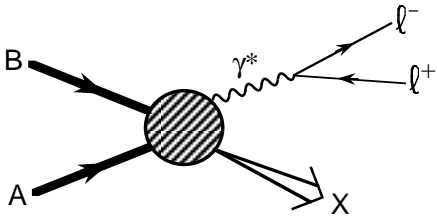,width=6.2cm}}
   \end{center}
 \vspace{-0.5cm}
\caption{\footnotesize Drell-Yan process.} 
\label{fig:dy}
\end{floatingfigure}
\quad
The Drell-Yan is a lepton-pair production process in hadron-hadron
collisions $A+B\rightarrow \ell^+\ell^- X$, where $\ell$ is for example
the muon, as shown in Fig. \ref{fig:dy}.
The Drell-Yan experiments have been used in determining quark
distributions in a hadron, in particular sea-quark distributions.
Therefore, it is ideal for examining the flavor dependence 
in the light antiquark distributions, even though
there is an undetermined $K$-factor in the experimental analysis.
The $K$-factor is the ratio between the measured cross section and
the leading-order prediction.
The theoretical studies of $\alpha_s$ and $\alpha_s^2$ corrections
revealed that the $K$-factor could be rather well explained
by the higher-order QCD corrections \cite{KFACT}.

Its cross section is given by \cite{MUTA}
\begin{multline}
d\sigma = \frac{1}{4\sqrt{(P_A\cdot P_B)^2-M_A^2 M_B^2}} \ 
                 \overline{\sum_{pol}} \sum_X 
                  \, (2\pi)^4 \, \delta(P_A+P_B-k_1-k_2-P_X)
 \\
        \times \    | {\mathcal M} (AB\rightarrow \ell^+ \ell^- X) |^2    \,
               \frac{d^3 k_1}{(2\pi)^3 2 k_{10}}  \,
               \frac{d^3 k_2}{(2\pi)^3 2 k_{20}}
\ \ \ ,
\end{multline}
where $k_1$ and $k_2$ are $\ell^-$ and $\ell^+$ momenta,
spin summation is taken for the final state particles,
and spin average is taken for the initial hadrons.
Because the matrix element is given by
\begin{equation}
{\mathcal M} (AB\rightarrow \ell^+ \ell^- X)
 \, = \, \bar u(k_1,\lambda_1)e\gamma_\mu v(k_2,\lambda_2)   \, 
                               \frac{g^{\mu\nu}}{(k_1+k_2)^2}  \,
                                <X|e J_\nu(0) |AB>
\ \ \ ,
\end{equation}
the cross section is written in terms of lepton and hadron tensors:
\begin{equation}
d\sigma = \frac{4\pi M \, e^4}{\sqrt{[s-(M_A^2+M_B^2)]^2-4M_A^2 M_B^2}}  \,
                \frac{L^{\mu\nu} W_{\mu\nu}}{(k_1+k_2)^4}      \,
               \frac{d^3 k_1}{(2\pi)^3 2 k_{10}}               \,
               \frac{d^3 k_2}{(2\pi)^3 2 k_{20}}
\ \ \ ,
\end{equation}
where $s$ is the center-of-mass energy squared $s=(P_A+P_B)^2$.
The tensors are 
\begin{align}
L^{\mu\nu} &= \frac{1}{2} \sum_{\lambda_1,\lambda_2} 
        \,  \left [ \, \bar u(k_1,\lambda_1) \gamma^\mu 
                            v(k_2,\lambda_2) \, \right ] ^* \,
            \left [ \, \bar u(k_1,\lambda_1) \gamma^\nu 
                            v(k_2,\lambda_2) \, \right ]
      \nonumber \\
           &= 2 \, \left ( \, k_1^\mu k_2^\nu + k_1^\nu k_2^\mu 
                            - k_1 \cdot k_2 g^{\mu\nu} \, \right )
\ \ \ ,
\end{align}
and 
\begin{align}
W_{\mu\nu} &= \frac{1}{4\pi M}
               \sum_X \, (2\pi)^4 \, \delta(P_A+P_B-k_1-k_2-P_X)
                \, \overline{\sum_{pol}} 
                 <AB| J_\mu(0) |X> \, <X| J_\nu(0) |AB> 
      \nonumber \\
           &=  \frac{1}{4\pi M}  \overline{\sum_{pol}}
               \int d^4 \xi \, e^{-i(k_1+k_2)\cdot \xi}
                 <AB \, | \, J_\mu(\xi) J_\nu(0) \, | \, AB>
\ \ \ .
\end{align}
Because the hadron tensor contains the currents with
two-nucleon state, the analysis in the deep inelastic scattering
is not directly applied. 
The papers \cite{MUTA} and \cite{JAFFE96}
discuss a dominant contribution to the cross section and
factorization of the amplitude into short-distance and long-distance physics.
The leading light-cone singularity comes from the process
that a quark radiates a virtual photon which splits
into the $\ell^+\ell^-$ pair. However, the process does
not dominate the cross section because the quark,
which radiates the massive photon, has to be far off-shell.
It has been shown that the following parton-parton fusion processes are the
dominant ones in the Drell-Yan cross section.

\begin{floatingfigure}{7.0cm}
   \begin{center}
      \mbox{\epsfig{file=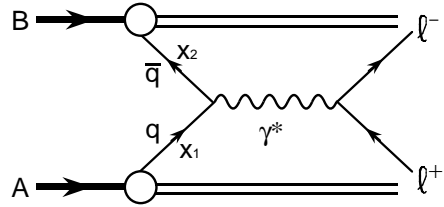,width=6.2cm}}
   \end{center}
 \vspace{-0.5cm}
\caption{\footnotesize
          Dominant contribution to the Drell-Yan cross section.} 
\label{fig:qqbar}
\end{floatingfigure}
\quad
In the leading order, the Drell-Yan is described by the quark-antiquark
annihilation process $q+\bar q\rightarrow \ell^+ +\ell^-$.
For example, Fig. \ref{fig:qqbar} indicates 
that a quark with the momentum fraction $x_1$ in the hadron A 
annihilates with an antiquark with $x_2$ in the hadron B.
Considering the color factor $3 \cdot (1/3)^2=1/3$, 
we obtain the LO Drell-Yan cross section 
\begin{equation}
s \frac{d\sigma}{d\sqrt{\tau}dy} =
\frac{8\pi\alpha^2}{9\sqrt{\tau}} \sum_i \, e_i^2 \,
[ \, q_i^A(x_1,Q^2) \, \bar q_i^B(x_2,Q^2)
  + \bar q_i^A(x_1,Q^2) \, q_i^B(x_2,Q^2) \, ]
\ \ \ ,
\label{eqn:DYCROSS}
\end{equation}
where $Q^2$ is the dimuon mass squared: $Q^2=m_{\mu\mu}^2$, 
and $\tau$ is given by $\tau=m_{\mu\mu}^2/s=x_1 x_2$.
The rapidity $y$ is defined by dimuon longitudinal momentum
$P_L^*$ and dimuon energy $E^*$ in the c.m. system:
$y=(1/2)ln[(E^*+P_L^*)/(E^*-P_L^*)]$.
The momentum fractions $x_1$ and $x_2$ can be written 
by these kinematical variables:
$x_1=\sqrt{\tau} e^y$ and $x_2=\sqrt{\tau} e^{-y}$.

According to Eq. (\ref{eqn:DYCROSS}), the process
can be used for measuring the antiquark distributions if 
the quark distributions in another hadron are known.
For finding the flavor asymmetry $\bar u-\bar d$,
the difference between p-p and p-n (practically p-d)
Drell-Yan cross sections is useful. 
Considering the rapidity point $y$=0
and retaining only the valence-sea annihilation terms, we have \cite{ES}
\begin{align}
\sigma^{pp} &= \frac{8\pi\alpha^2}{9\sqrt\tau} \,
                 \left [ \, \frac{8}{9} u_v (x) \bar u (x)
                        +\frac{2}{9} d_v (x) \bar d (x)
                 \, \right ]
\ \ \ , \nonumber \\
\sigma^{pn} &= \frac{8\pi\alpha^2}{9\sqrt\tau} \,
       \left [ \, \frac{4}{9} \left \{ u_v (x) \bar d (x) 
                                   +d_v (x) \bar u (x) \right \}
             + \frac{1}{9} \left \{ d_v (x) \bar u (x) 
                                   +u_v (x) \bar d (x) \right \} 
        \, \right ]
\ \ \ \text{at $y=0$}
\ \ \ ,
\end{align}
for the proton-proton and proton-neutron cross sections.
All the above distributions are at $x=\sqrt\tau$
because of $y=0$.
From these equations, the p-n asymmetry becomes
\begin{align}
A_{DY} &= \frac{\sigma^{pp}-\sigma^{pn}}{\sigma^{pp}+\sigma^{pn}}
              \nonumber \\
       &= \frac{[4u_v(x)-d_v(x)][\bar u(x)-\bar d(x)]
                +[u_v(x)-d_v(x)][4\bar u(x)-\bar d(x)]}
                {[4u_v(x)+d_v(x)][\bar u(x)+\bar d(x)]
                +[u_v(x)+d_v(x)][4\bar u(x)+\bar d(x)]}
\ \ \ \text{at $y=0$}
\ \ \ .
\label{eqn:ADYY0}
\end{align}
This quantity is very sensitive to the $\bar u-\bar d$ distribution.
However, antiquark-quark annihilation processes also contribute to
the above equation at the rapidity point $y$=0. 
In this sense, it is better to take large $x_F$ ($\equiv x_1-x_2$)
data so that the antiquarks in the projectile do not affect 
the asymmetry \cite{KL}:
\begin{equation}
A_{DY} =  \frac {[4u(x_1)-d(x_1)][\bar u(x_2)-\bar d(x_2)]}
                {[4u(x_1)+d(x_1)][\bar u(x_2)+\bar d(x_2)]}
\ \ \ \text{at large $x_F$}
\ \ \ .
\label{eqn:ADYKL}
\end{equation}
The above discussions are, of course, based on the LO cross section.
On the other hand, the higher-order corrections are rather large
as represented by the $K$-factor. 
Therefore, it is important to investigate whether the $K$-factor
cancels out in the asymmetry $A_{DY}$.

The Drell-Yan process has been already used for studying the flavor asymmetry.
There are existing data by the Fermilab-E288, the Fermilab-E772,
and the CERN-NA51. 
Furthermore, detailed studies of the Drell-Yan asymmetry are
in progress at Fermilab by the E866 collaboration.

\begin{figure}[h]
\parbox[b]{0.46\textwidth}{
   \begin{center}
      \mbox{\epsfig{file=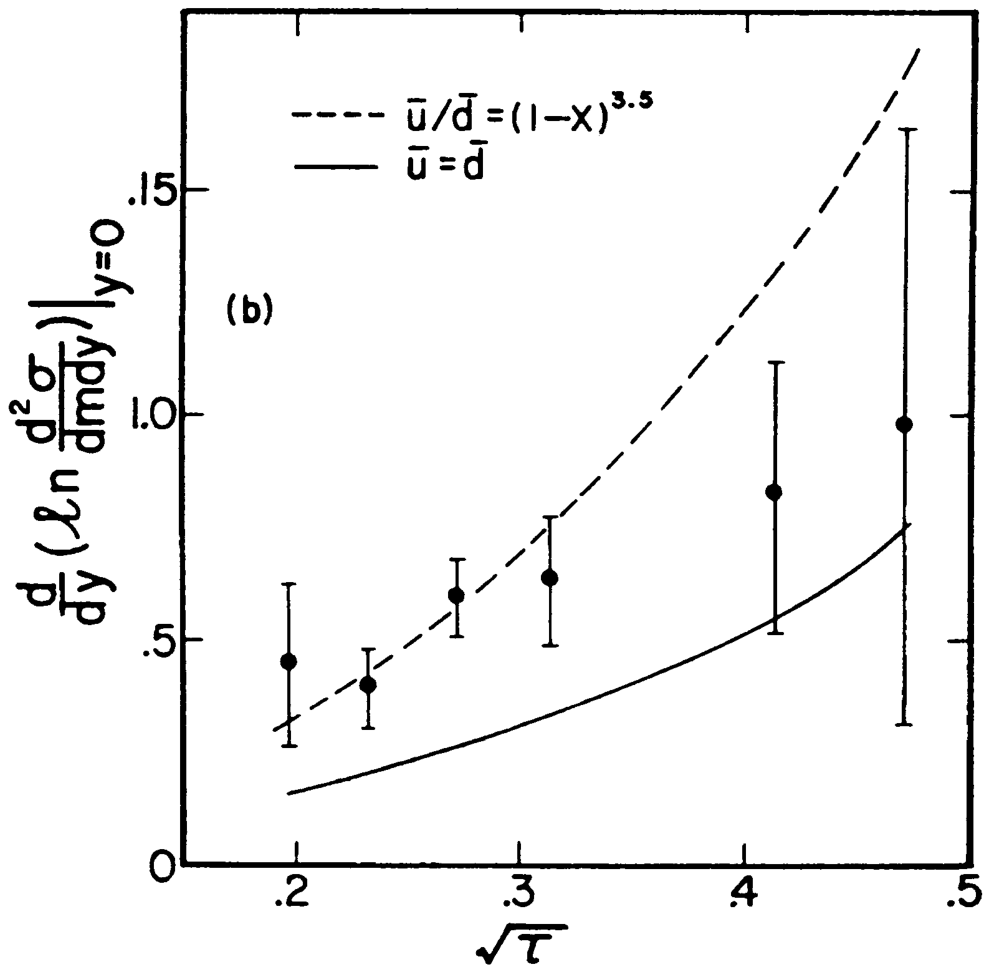,width=6.0cm}}
   \end{center}
 \vspace{-0.8cm}
\caption{\footnotesize Slope of rapidity distribution at $y$=0 
                       in the E288 experiment 
                       (taken from Ref. {\normalsize\cite{E288}}). 
                       Flavor symmetric and asymmetric
                       results are shown by the solid and dashed curves.}
\label{fig:e288}
}\hfill
\parbox[b]{0.46\textwidth}{
   \begin{center}
      \mbox{\epsfig{file=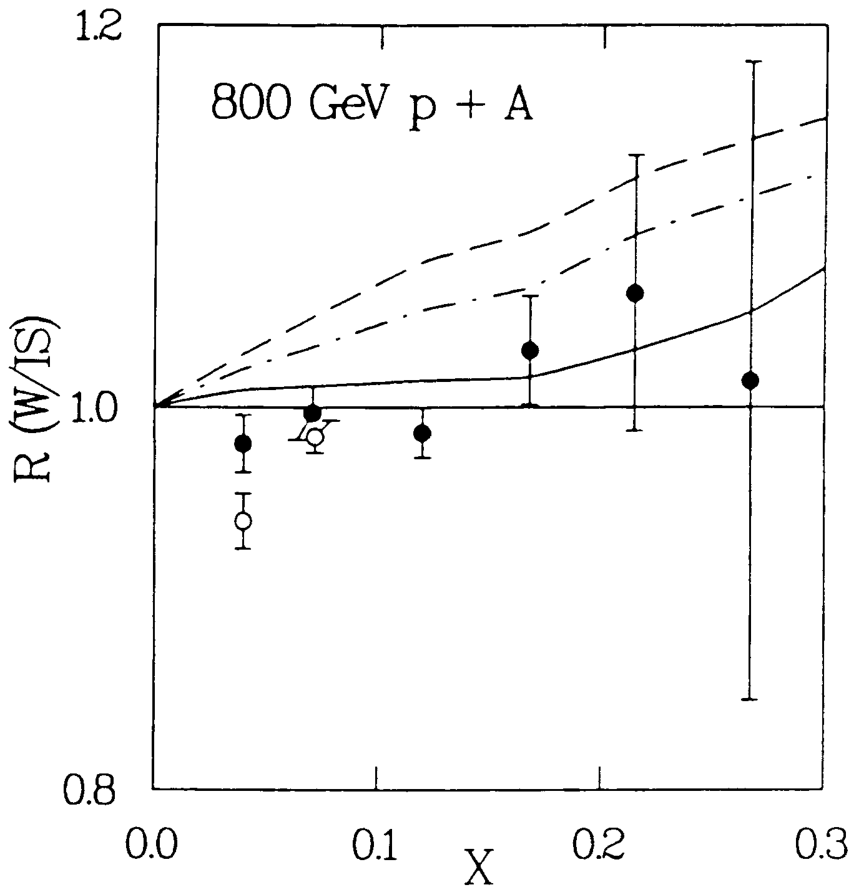,width=6.0cm}}
   \end{center}
 \vspace{-0.8cm}
\caption{\footnotesize Drell-Yan cross section ratio $\sigma_W/\sigma_{IS}$ 
             in the E772 experiment 
             (taken from Ref. {\normalsize\cite{E772}}).
             Open circles are the data without shadowing correction.
             Three theoretical asymmetry results 
             are shown by the solid, dashed, and dot-dashed curves.}
\label{fig:e772}
}
\end{figure}

The Fermilab-E288 collaboration measured dileptons produced
in proton-nucleus collisions. Proton-beam energies are
200, 300 and 400 GeV, and targets are beryllium, copper and platinum.
The dimuon data are taken in the mass region $m_{\mu\mu}$=4$-$17 GeV,
and they are analyzed by using Eq. (\ref{eqn:DYCROSS}).
The isospin symmetry is assumed for parton distributions in the proton and
neutron. No nuclear correction is made except for the
Fermi motion correction. We note that the E288 paper was published
before the finding of the EMC effect \cite{EMCF2}. 
$Q^2$ dependent $F_2^p$ data from electron and muon scattering
are used together with a fit $F_2^n/F_2^p=1.0-0.8x$ and
parametrized antiquark distributions.
The antiquark part is assumed to be $Q^2$ independent,
and they are determined by fitting their data:
$\bar d=0.548 (1-x)^{7.62}$ and $\bar u=0.548 (1-x)^{11.1}$.
Calculated DY cross sections with these $\bar u$ and $\bar d$ distributions
are shown by the dashed curve in Fig. \ref{fig:e288},
and flavor-symmetric ones are shown by the solid curve.
The E288 data favor a $\bar d$ excess over $\bar u$: 
$\bar u=\bar d(1-x)^{3.48}$.

\vspace{-1.0cm}
\noindent

Next, the Fermilab-E772 collaboration showed their Drell-Yan results
on the flavor asymmetry in 1992 \cite{E772}. 
The Drell-Yan experiments are done
for isoscalar targets, deuteron and carbon,
and for tungsten which has a large neutron excess.
The proton beam energy is 800 GeV.
Forward production of the dileptons is dominated by beam-quark
annihilation with a target antiquark in Eq. (\ref{eqn:ADYKL}).
Therefore, the cross sections in the region $x_F\ge 0.1$
can be used for investigating the antiquark distributions in the target.
If sea-quark distributions in a nucleus are just
the summation of proton and neutron contributions without
nuclear modification, the cross-section ratio
for the tungsten (W) and the isoscalar targets (IS)
becomes
\begin{equation}
R_A(x)\equiv \frac{\sigma_A(x)}{\sigma_{IS}(x)}
\approx 1 + \frac{N-Z}{A} 
            \frac{\bar d(x)-\bar u(x)}{\bar d(x)+\bar u(x)}
\ \ \ ,
\end{equation}
where $A$, $Z$, and $N$ are the atomic weight, atomic number,
and number of neutrons in the target nucleus.
The $d\bar d$ annihilation is neglected because
the $u\bar u$ dominates the cross section.
Shadowing correction is applied to the tungsten data at $x<0.1$
in the following way. 
First, A-dependent shadowing factor $\alpha_{sh}$ is determined from
EMC, NMC, and E665 shadowing data for D, C, and Ca.
Then, the tungsten cross sections with $x<0.1$ are corrected by
$\sigma_{_A}=\sigma_{_N} A^{\alpha_{sh}}$.
The obtained ratios are shown in Fig. \ref{fig:e772}
together with three theoretical expectations.
The solid curve is a pion-model prediction \cite{INDIANA,KL} 
in section \ref{MESON-1}. 
The dashed one is a simple parametrization in section \ref{PARAMET} 
for explaining the NMC data: 
$\bar d-\bar u=A(1-x)^b$ with $A=0.15(1+b)$ and $b$=9.6 \cite{ES}.
The dot-dashed one is a chiral model result \cite{EHQ,EHQDY}
in section \ref{CHIRAL}.
As it is obvious from the figure, the data do not reveal
significant flavor asymmetry.
Although the E772 data are consistent with the flavor asymmetric
model predictions, they could be also explained by
the flavor symmetric sea by considering experimental errors.
The tungsten is a heavy nucleus, so that there could be
a significant nuclear effect on the flavor distribution
(see section \ref{NUCLEI}).
The E772 collaboration also showed the $x_F$ distribution
of p+d data in connection with the flavor asymmetry.
However, there is also no evidence for the asymmetry.
These results are somewhat in conflict with the NMC result and other
Drell-Yan data.  

There are Drell-Yan data for various nuclear targets; however,
the NA51 data point at $x$=0.18 \cite{NA51} is the only existing one
which is extracted from the p-p and p-d Drell-Yan experiments.
Considering the $x$ range of the NA51 measurements, we neglect
shadowing correction in the deuteron. 
Then, the asymmetry can be written
as $A_{DY}=2\sigma^{pp}/\sigma^{pd}-1$.
The NA51 collaboration used 450 GeV primary proton beam from 
the CERN-SPS. The targets are liquid hydrogen and deuterium.
The accepted rapidity range is from $-$0.5 to 0.6,
and the muon mass region $M_{\mu\mu}\ge$4.3 GeV is
used for the analysis. 
The obtained asymmetry is
\begin{equation}
A_{DY}= -0.09 \pm 0.02 \, (stat.) \pm 0.025 \, (syst.)
\ \ \ .
\label{eqn:NA51-2}
\end{equation}
As the valence-quark value $\lambda_V=u_v/d_v$ at $x$=0.18, they
take $\lambda_V$=2.2 averaged over the parton
distributions, MRS-S$_0 '$, MRS-D$_0 '$, and MRS-D${\_} '$,
GRV-HO, and CTEQ-2M. 
From Eqs. (\ref{eqn:ADYY0}) and (\ref{eqn:NA51-2}),
the observed asymmetry becomes
\begin{equation}
\frac{\bar u}{\bar d} = 0.51 \pm 0.04 \, (stat.) \pm 0.05 \, (syst.)
\ \ \ \ \ at\ x=0.18
\ \ \ .
\label{eqn:NA51}
\end{equation}
This is a clear indication of the flavor asymmetry 
in the light antiquark distributions.
There is an excess of $\bar d$-quarks over $\bar u$
in the nucleon, and the NA51 result agrees with the 
tendency obtained by the NMC.
Unfortunately, the only one data point at $x=0.18$
is available in the NA51.
More complete Drell-Yan experiments at the Fermilab (E866)
should give a clearer answer 
to the flavor symmetry problem \cite{E866}.
At this stage, the preliminary E866 data
seem to show the NMC type asymmetry in addition. 

The E772 and NA51 data are compared with various model predictions in
Refs. \cite{E772,KRE,ES,KL,EHQDY,HGMP,MRSDY,JULDY,MSGDY}.
The theoretical works are done mainly to compare the mesonic
calculations with the Drell-Yan data. At this stage, 
the mesonic models could be consistent not only with the NMC result
but also with the E772 and NA51 Drell-Yan data.

\vfill\eject
\subsubsection{Fermilab-E866 results}\label{E866}

The Fermilab-E866 collaboration reported high statistical
experimental results for the ratio of the Drell-Yan cross sections
$\sigma_{pd}/2\sigma_{pp}$ \cite{E866}.
The dimuons are measured in 800 GeV proton scattering on
the liquid hydrogen or liquid deuterium target.
Six month data were collected until March of 1997,
and 350,000 Drell-Yan events were obtained.
The goal of the experiment is to measure the cross-section
ratios with 1\% accuracy in the $x$ range $0.03 < x < 0.15$.
The accuracy becomes worse in the larger $x$ region.
At large $x_{_F}$, the ratio is approximated as
$[\sigma_{pd}/2\sigma_{pp}]_{x_{_F}\gg \, 0} \approx 
                [1+\bar d(x)/\bar u(x)]/2$,
so that it is possible to extract the distribution
ratio $\bar d(x)/\bar u(x)$ from the large $x_{_F}$ data
of the cross sections.
They compared the obtained ratios $\sigma_{pd}/2\sigma_{pp}$
with the flavor symmetric distributions [CTEQ4M($\bar u=\bar d$)
and MRS(S0)] and asymmetric ones [CTEQ4M($\bar u \ne \bar d$),
GRV, and MRS(G)].
The symmetric CTEQ4M($\bar u=\bar d$) curve is obtained
by modifying the distributions as $\bar u=\bar d=(\bar u+\bar d)/2$.
The accurate part of the data in the region $0.03<x<0.15$
tends to agree with the asymmetric distributions, in particular
with the CTEQ4M($\bar u \ne \bar d$). However, it is also interesting
to find that the data seem to deviate from the present asymmetric
parametrizations in the larger $x$ region.
Although their results are still preliminary, the accurate
E866 data confirm the flavor asymmetry conclusions of
the NMC and NA51.

\vfill\eject
\subsection{W and Z production}\label{W}

Instead of the virtual photon production in the Drell-Yan case,
weak boson production could also have information on
the antiquark distributions  \cite{BS93,DHKS,PJ}.
There are existing CDF data for the charged lepton asymmetry
(or W charge asymmetry) in the $p+\bar p$ reaction: 
$p\bar p\rightarrow W^\pm X\rightarrow (\ell^\pm \nu_\ell)X$ \cite{CDF95}.
They are first analyzed in Ref. \cite{MRS90} in connection with
the $\bar u-\bar d$ distribution. 
Although the CDF data constrain the $u/d$ ratio in the region
of $x=M_W/\sqrt{s}=0.045$, they are consistent with the symmetric 
sea $\bar u=\bar d$.
However, the $p+\bar p$ reaction is not
very sensitive to the sea-quark distributions as we discuss
in this section. Therefore, future $p+p$ colliders
such as Relativistic Heavy Ion Collider (RHIC), rather than the $p+\bar p$, 
are crucial for investigating the flavor asymmetry
in W and Z production processes.
We discuss the sensitivity of $W^\pm$ and $Z^0$ production
cross sections on the $\bar u/\bar d$ asymmetry 
based on Ref. \cite{PJ} in the following.

We show how the W production processes in the $p+p$ collider
could be used for probing the flavor asymmetry.
The $W^+$ production cross section is given by parton-subprocess ones
together with parton distributions in the colliding hadrons \cite{BPBOOK}:
\begin{equation}
\sigma(p+p\rightarrow W^+ X) = 
\frac{1}{3} \, \int_0^1 dx_1 \int_0^1 dx_2
\sum_{q, \bar q'} q(x_1,M_W^2) \, \bar q'(x_2,M_W^2) \,  
\hat \sigma(q\bar q'\rightarrow W^+)
\ \ \ ,
\end{equation} 
where 1/3 is the color factor $3 \cdot (1/3)^2=1/3$.
The subprocess cross section is given by
\begin{equation}
d\hat\sigma(q\, \bar q'\rightarrow W^+) = \left ( \frac{1}{2} \right )^2
          \frac{1}{2\hat s} \, \sum_{pol}
 \left | {\mathcal M} ( q \, \bar q' \rightarrow W^+ ) \right |^2
 \, (2\pi)^4 \, \delta^{4}(p_1+p_2-p) \, \frac{d^3 p}{2E_p (2\pi)^3}
\ ,
\end{equation}
where $p_1$, $p_2$, and $p$ are $\bar q'$, $q$, and $W^+$
momenta respectively, and $\hat s$ is given by $\hat s=(p_1+p_2)^2$. 
The matrix element is
\begin{equation}
{\mathcal M}(q \, \bar q'\rightarrow W^+) = 
   -i \, V_{qq'} \, \frac{g}{\sqrt{2}} \, {\varepsilon_\alpha^\lambda}^* (p)
     \, \bar v(p_1) \, \frac{1}{2} \, \gamma^\alpha (1-\gamma_5) \, u(p_2)
\ \ \ ,
\end{equation}
and the Cabibbo mixing is used in our calculation:
$V_{ud}=\cos\theta_c$, $V_{us}=\sin\theta_c$,
$V_{cd}=-\sin\theta_c$, and $V_{cs}=\cos\theta_c$.
Taking the spin summation, we obtain
\begin{equation}
\sum_{pol}  \left | {\mathcal M} 
         \left ( q \, \bar q' \rightarrow W^+ \right ) \right |^2
= \frac{8}{\sqrt{2}} \, G_F \, M_W^4 \, |V_{qq'}|^2 
\ \ \ ,
\end{equation}
with the Fermi coupling constant $G_F/\sqrt{2}=g^2/(8M_W^2)$.
Noting $\delta^{4}(p_1+p_2-p)d^3 p/(2E_p)=\delta(\hat s-M_W^2)$
and $dx_1 dx_2 = d\hat s \, dx_F/[(x_1+x_2)s]$, we have the
$W^+$ production cross section in the $p+p$ reaction in terms of
the parton distributions:

\begin{align}
\frac{d \sigma_{p+p\rightarrow W^+}}{dx_F}  
                 = \frac{\sqrt 2 \pi}{3} \, G_F
\left(    \frac{x_1 x_2}{x_1 + x_2}   \right)
& \left\{  \, \cos^2 \theta_c  \,
[u(x_1) \bar d(x_2) + \bar d(x_1) u(x_2)]  \right. \nonumber \\
& \left. + \sin^2 \theta_c  \,
[u(x_1) \bar s(x_2) + \bar s(x_1) u(x_2)] \, \right\}
\ \ \ .
\end{align}
The dominant processes of producing $W^+$ are
$u(x_1)+\bar d(x_2)\rightarrow W^+$ and 
$u(x_2)+\bar d(x_1)\rightarrow W^+$; however,
the first one becomes much larger than the second at large $x_F$.
Therefore, the cross section is
sensitive to the $\bar d$ distribution at large $x_F$. 
On the other hand,
the cross section for the $W^-$ production is given 
in the same way:
\begin{align}
\frac{d \sigma_{p+p\rightarrow W^-}}{dx_F} 
                 = \frac{\sqrt 2 \pi}{3} \, G_F
\left( \frac{x_1 x_2}{x_1 + x_2}\right)
& \left\{ \, \cos^2 \theta_c \,
[\bar u(x_1) d(x_2) +  d(x_1) \bar u(x_2)] \right. \nonumber \\
& \left. + \sin^2 \theta_c \,
[\bar u(x_1) s(x_2) + s(x_1) \bar u(x_2)] \, \right\}
\ \ \ .
\end{align}
At large $x_F$, it is sensitive to the $\bar u$ distribution
instead of the $\bar d$ in the $W^+$ case.
This difference makes it possible to find the difference
$\bar u-\bar d$.
Because the Cabbibo angle is small, the $\sin^2 \theta_c$ terms
are neglected for simplicity in the following discussions.
The $W^\pm$ production ratio is then given by
\begin{equation}
R_{p+p}(x_F) \equiv 
\frac{d \sigma_{p+p\rightarrow W^+}/ dx_F}
     {d \sigma_{p+p\rightarrow W^-}/ dx_F } =
\frac {u(x_1) \bar d(x_2) +  \bar d(x_1) u(x_2)}
      {\bar u(x_1) d(x_2) +  d(x_1) \bar u(x_2)}
\ \ \ .
\label{eqn:PPW}
\end{equation}
At large $x_F$ (large $x_1$), the antiquark distribution
$\bar q(x_1)$ is very small, so that the above equation
becomes
\begin{equation}
R_{p+p}(x_F \gg 0) \approx
\frac{u(x_1)}{d(x_1)} \, \frac{\bar d(x_2)}{\bar u(x_2)}
\ \ \ ,
\end{equation}
which is directly proportional to the ratio $\bar d/\bar u$.

\begin{wrapfigure}{r}{7.0cm}
   \begin{center}
      \mbox{\epsfig{file=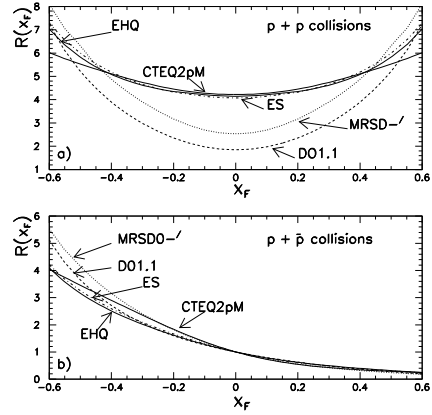,width=6.2cm}}
   \end{center}
 \vspace{-0.8cm}
\caption{\footnotesize
          $W^\pm$ production ratios a) in $p+p$
                        and b) in $p+\bar p$ 
                (taken from Ref. {\normalsize\cite{PJ}}).} 
\label{fig:pj}
\end{wrapfigure}

The situation is very different in the $p+\bar p$ reaction case.
Replacing the parton distributions in Eq. (\ref{eqn:PPW})
by $q(x_2)\rightarrow \bar q(x_2)$ and $\bar q(x_2)\rightarrow q(x_2)$,
we obtain the ratio
\begin{equation}
R_{p+\bar p}(x_F) = 
\frac{u(x_1) d(x_2) +  \bar d(x_1) \bar u(x_2)}
                      {\bar u(x_1) \bar d(x_2) +  d(x_1) u(x_2)}
\begin{CD}
@>>{x_F \gg 0}>
\end{CD}  
 \frac{u(x_1)}{d(x_1)} \, \frac{d(x_2)}{u(x_2)}
\ \ \ .
\label{eqn:PPBARW}
\end{equation}
In the $p+\bar p$ reaction,
the ratio is no more sensitive to the $\bar u/\bar d$ asymmetry.
How about the $x_F\approx 0$ region?
We find from Eq. (\ref{eqn:PPBARW})
that the $p+\bar p$ ratio is independent: $R_{p+\bar p}(x_F=0)=1$,
even though the $p+p$ ratio is still sensitive to the
flavor asymmetry at $x_F=0$: 
$R_{p+p}(x_F=0)=[u(x)/d(x)][\bar d(x)/\bar u(x)]$.
From these discussions, it is more appropriate to
use a $p+p$ collider in finding the $\bar u/\bar d$ asymmetry
from W production data. This fact is numerically shown in
Fig. \ref{fig:pj}.

\vfill\eject
\quad
The ratios in the $p+p$ and $p+\bar p$ reactions
are evaluated at $\sqrt s$=500 GeV
in Fig. \ref{fig:pj} by using various 
parametrizations for the parton distributions \cite{PJ}.
The distributions are evolved to the scale $Q^2=M_W^2$.
The figures a) and b) show the $p+p$ and $p+\bar p$ results
respectively. 
The dashed curve indicates the results of using the flavor symmetric 
($\bar u=\bar d$) DO1.1 distributions. 
The others are the results for flavor asymmetric distributions 
(MRSD$_-$$'$, CTEQ2pM, ES, EHQ).
Because the NA51 result ruled out the MRSD$_-$$'$ distribution,
the small difference between the flavor asymmetric MRSD$_-$$'$
and the symmetric DO1.1 should not be taken seriously.
As we expected, the $p+p$ reaction is sensitive to 
the parton-distribution models, in particular
the light antiquark flavor asymmetry,
not only in the large $|x_F|$ region but also
in the $x_F\approx 0$ region.
On the other hand, the $p+\bar p$ reaction is almost insensitive 
to the asymmetry. The model dependence appears only in the very small
$x_F$.

The W production processes in the $p+p$ and $p+d$ reactions could
also be used for studying the flavor asymmetry.
The cross-section ratio is 
\begin{equation}
R^\prime (x_F) \equiv 2 \,
\frac{d \sigma_{p + p \rightarrow W^+} / dx_F } 
     {d \sigma_{p + d \rightarrow W^+} / dx_F} \approx
\frac{u(x_1) \bar d(x_2) +  \bar d(x_1) u(x_2)} 
{u(x_1) \, [\bar u(x_1) + \bar d(x_2)] 
 + \bar d(x_1) \, [u(x_2) + d(x_2)]}
\ \ \ ,
\end{equation}
by neglecting nuclear corrections in the deuteron.
Although it is independent of the sea distributions at small $x_F$:
$R^\prime (x_F \ll 0)=1+[u(x_2)-d(x_2)]/[u(x_2)+d(x_2)]$,
large $x_F$ data are useful:
\begin{equation}
R^\prime (x_F) \approx 1 - \frac{\bar u(x_2) - \bar d(x_2)}
{\bar u(x_2) + \bar d(x_2)} 
\ \ \ \text{at $x_F \gg 0$}
\ \ \ .
\end{equation}

In the similar way, $Z^0$ production data in the $p+p$ and $p+d$ reactions
are valuable. The $Z^0$ production cross section in the $p+p$
is 
\begin{multline}
\frac{d \sigma_{p+p\rightarrow Z^\circ} }{dx_F} 
             = \frac{\pi}{3 \sqrt{2}} \, G_F
\left( \frac{x_1 x_2}{x_1 + x_2} \right) \left\{  \,
(1- \frac{8}{3}\chi_w + \frac{32}{9} \chi_w^2) \,
[u(x_1) \bar u(x_2) + \bar u(x_1) u(x_2)]  \right. 
\\
 \left. + (1-\frac{4}{3}\chi_w + \frac{8}{9} \chi_w^2) \,
[d(x_1) \bar d(x_2) + \bar d(x_1) d(x_2) +
s(x_1) \bar s(x_2) + \bar s(x_1) s(x_2)] \, \right\}
\ ,
\end{multline}
where $\chi_w$ is given by the Weinberg angle 
as $\chi_w=\sin^2 \theta_W$. Taking into account the
dominant $u\bar u$ contribution at large $x_F$, we obtain
the ratio
\begin{equation}
R''(x_F) \equiv 2 \ \frac {d \sigma_{p + p \rightarrow Z^\circ} / dx_F}
                        {d \sigma_{p + d \rightarrow Z^\circ} / dx_F }
   \approx 1 +  \frac{\bar u(x_2) - \bar d(x_2)}{\bar u(x_2) + \bar d(x_2)} 
\ \ \ \text{at $x_F \gg 0$}
\ \ \ .
\end{equation}

We find that not only the $W^\pm$ production but also the $Z^0$ 
production could be used in determining the $\bar u-\bar d$ distribution.
Because $p+p$ and $p+d$ reactions are very sensitive to the asymmetry,
future colliders such as RHIC should
be able to find the antiquark asymmetry by the $W^\pm$ and $Z^0$ production
measurements.

\subsection{Quarkonium production at large $x_{_F}$}\label{QQBAR}

We discuss the possibility of finding the $\bar u-\bar d$ distribution
in quarkonium production processes.
J/$\psi$ production data have been used in extracting gluon 
distributions in the nucleon and in nuclei.
Because the dominant process is the gluon fusion 
$gg\rightarrow c\bar c\rightarrow J/\psi$, $q\bar q$ annihilation
is in general a small effect.
However, the $q\bar q$ process could become important at large $|x_F|$.

The mechanism of producing the quarkonium is a strong interaction,
which makes the description more model-dependent than the electromagnetic
Drell-Yan case. A popular description is a color-singlet 
(and recent color-octet) model,
which includes $gg$, $gq$, $g\bar q$, and $q\bar q$ fusion 
up to $\alpha_s^3$. 
Instead of stepping into the detailed production mechanism,
we discuss general features by selecting a simpler one,
the semi-local duality model.
The quarkonium production processes are analyzed in 
this model, and the results are related to 
the flavor asymmetry in Ref. \cite{PJC}.

The cross section for a $Q\overline Q$ pair production is given
by parton subprocess cross sections multiplied by
the corresponding parton distributions
\begin{multline}
\frac{d \sigma_{Q\overline Q}}{dx_F d\tau} = \frac{2\tau}
{\sqrt{x_{F}{^2} + 4\tau^2}} \,
\biggl [ \, G(x_1) G(x_2) \, \sigma(gg \rightarrow
Q \overline{Q}; m^2) \, 
\\
  + \sum_{\rm i=u,d,s}  \bigl\{ \, q^{\,i}(x_1)
\overline{q}^{\,i}(x_2) + \overline{q}^{\,i}(x_1)
q^{\,i}(x_2) \, \bigr\} \,
\sigma(q \overline{q} \rightarrow Q \overline{Q}; m^2) \, \biggr ]
\ \ .
\end{multline}
The only $gg$ and $q\bar q$ type subprocesses are taken into 
account in the above expression, and 
$\sigma(gg \rightarrow Q \overline{Q}; m^2)$ and
$\sigma(q \overline{q} \rightarrow Q \overline{Q}; m^2)$
are the corresponding cross sections.
The variables $x_1$ and $x_2$ are fractional momenta carried
by the projectile parton and by the target one, and
$x_F$ and $\tau$ are given by $x_F=x_1-x_2$ and $\tau=m/\sqrt{s}$
with the invariant mass of the $Q\bar Q$ pair $m$.
The subprocess cross sections are 
\begin{align}
\sigma (q \bar q \to Q \bar Q; m^2) &= 
\frac{8 \pi \alpha_s^2}{27m^6} \, (m^2 +2m_Q^2) \, \lambda
\ \ \ ,    \nonumber \\
\sigma (g g \to Q \bar Q; m^2) &= 
\frac{\pi \alpha_s^2}{3m^6} \,
 \biggl [ \, (m^4 + 4m^2 m_Q^2 +m_Q^4) \, 
             \ln(\frac{m^2+\lambda}{m^2-\lambda})
\nonumber \\
& \ \ \ \ \ \ \ \ \ \ \ \ 
- \, \frac{1}{4} \, (7m^2 +31 m_Q^2) \, \lambda \, \biggr ]
\ \ \ ,
\end{align}
where $m_Q$ is a quark mass and $\lambda$ is given by
$\lambda = \sqrt{m^4 - 4m^2 m_{Q}{^2}}$.
According to the semi-local duality model, the quarkonium
production cross section is obtained by integrating 
the subprocesses cross section from the $Q\bar Q$ threshold
to the open charm (beauty) threshold:
\begin{equation}
\frac{d\sigma_{p+p\rightarrow J/\psi \, (\Upsilon)}}{dx_F}
= F \int_{2m_Q/\sqrt\tau}^{2m_{D(B)}/\sqrt\tau} d\tau \,
\frac{d\sigma_{Q\overline Q}}{dx_F d\tau}
\ \ \ ,
\end{equation}
where $F$ is the probability of a J/$\psi$ ($\Upsilon$)
creation from the $Q\overline Q$ state.

We hope to find the antiquark flavor asymmetry from
these quarkonium production processes.
The gluon-gluon fusion process dominates the cross section
in general. However, the $q\bar q$ processes could become
more important in certain kinematical regions.
The $q\bar q$ fusion contributions in the $p+p$ collision are
\begin{equation}
d\sigma_{p+p} \propto u(x_1) \bar u(x_2) +\bar u(x_1) u(x_2)
                    + d(x_1) \bar d(x_2) +\bar d(x_1) d(x_2)
\ \ \ .
\label{eqn:ppcc}
\end{equation}
In order to find the $\bar u-\bar d$ distribution, the $p+d$ reaction
has to be studied in addition:
\begin{equation}
d\sigma_{p+d} \propto [u(x_1) +d(x_1)] [\bar u(x_2) +\bar d(x_2)]
                    + [\bar u(x_1) +\bar d(x_1)] [u(x_2) +d(x_2)]
\ \ \ ,
\label{eqn:pdcc}
\end{equation}
where the isospin symmetry is assumed. The cross-section ratio
\begin{equation}
R(x_F) = 2 \,
\frac{d\sigma (p + p \to J/\psi (\Upsilon)) /dx_F} 
{d\sigma (p + d \to J/\psi (\Upsilon)) /dx_F}
\ \ \ ,
\end{equation}
should be sensitive to $\bar u-\bar d$ particularly at large $x_F$.
From Eqs. (\ref{eqn:ppcc}) and (\ref{eqn:pdcc}), it is obvious that
the ratio is $R(x_F)=1$ in the flavor symmetric case $\bar u=\bar d$.
Therefore, the deviation from unity is a signature of a finite
$\bar u-\bar d$ distribution.

\begin{wrapfigure}{r}{7.0cm}
   \begin{center}
      \mbox{\epsfig{file=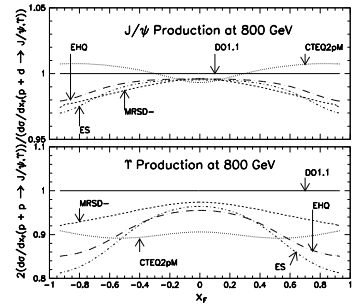,width=5.0cm}}
   \end{center}
 \vspace{-0.5cm}
\caption{\footnotesize J/$\psi$ and $\Upsilon$ production ratios    
                      (taken from Ref. {\normalsize\cite{PJC}})} 
\label{fig:pjc}
\end{wrapfigure}
\quad
The ratio $R(x_F)$ is evaluated for the 800 GeV proton beam
in Fig. \ref{fig:pjc}, where the upper (lower) figure shows
the $J/\psi$ ($\Upsilon$) production results.
Input parton distributions are the same in Fig. \ref{fig:pj}.
The ratio is unity if the sea is flavor symmetric, and it is shown
by the solid lines DO1.1 in Fig. \ref{fig:pjc}.
As we expected, effects of the flavor asymmetry become conspicuous 
at large $|x_F|$. This is because the gluon distribution $G(x)$ is 
much smaller than the quark one $q(x)$ at large $x$, and the cross 
section is dominated by the $q\bar q$ fusion processes.
The parton-distribution dependence is more evident in the $\Upsilon$ 
production. Because $\Upsilon$ is more massive than J/$\psi$, 
we have $(x_1 x_2)_{\Upsilon} \sim 2m_B/\sqrt{s} > 2m_C/\sqrt{s}$. 
The $\Upsilon$ process is sensitive to the larger $x$ region, so that
the flavor-asymmetry effects become more conspicuous. 

The above results show interesting contributions from the $\bar u/\bar d$
asymmetry. 
Therefore, measurements of the $J/\psi$ and $\Upsilon$ production
cross sections for the proton and deuteron targets at large $x_F$ 
should also be able to clarify the $\bar u-\bar d$
distribution problem.
However, the semi-local duality model is probably too simple to explain the
quarkonium-production processes. 
It is now revealed that the prediction of a more sophisticated color-singlet
model is inconsistent with the $J/\psi$ and $\psi '$ production data by
the CDF. The discrepancy could be understood by the color-octet mechanism.
Therefore, a better model analysis is necessary in comparing
the theoretical results with future experimental data.

\subsection{Charged hadron production}\label{CHARGED}

Semi-inclusive reactions in the electron or muon scattering could 
be used for finding the antiquark distributions.
In particular, charged-hadron production could have 
information on the flavor asymmetry.
The hadron-production cross section is written by 
the lepton and hadron tensors in the same way with
the inclusive one in section \ref{GOTT}.
However, in spite of the fact that the semi-inclusive process
is dominated by the light-cone region,
the operator product expansion cannot be applied \cite{JAFFE96}.
It is because the summation on $X$ in the final state $| \, X;p_h,s_h>$
cannot be taken independently from the hadron state $| \, p_h,s_h>$. 
Therefore, we discuss the theoretical analysis 
on a relation between the charged-hadron-production cross section and
the distribution $\bar u-\bar d$ by using a quark-parton model \cite{LMS}.

In the parton picture, the semi-inclusive cross section is given by
\cite{FEC}
\begin{equation}
\frac{1}{\sigma_N(x)} \, \frac{\partial \sigma_N^h(x,z)}{\partial z} =
\frac{\sum_i e_i^2 \, f_i(x) \, D_i^h(z)}{\sum_i e_i^2 \, f_i(x)}
\ \ \ ,
\end{equation}
where $f_i(x)$ is the quark distribution with flavor $i$
and momentum fraction $x$,
and $D_i^h(z)$ is the $i$-quark to $h$-hadron fragmentation
function with $z=E_h/\nu$.
The numerator for charged hadron production is
\begin{align}
N^{Nh^\pm} &\equiv \sum_i e_i^2 \, f_i(x) \, D_i^{h^\pm}(z)    
   \nonumber \\
       &= \frac{4}{9} \, u \, D_u^\pm 
        + \frac{4}{9} \, \bar u \, D_{\bar u}^\pm
        + \frac{1}{9} \, d \, D_d^\pm 
        + \frac{1}{9} \, \bar d \, D_{\bar d}^\pm 
        + \frac{1}{9} \, s \, D_s^\pm 
        + \frac{1}{9} \, \bar s \, D_{\bar s}^\pm 
\ \ \ .
\end{align}
Assuming the isospin symmetry in the parton distributions, 
we consider a combination of proton and neutron cross sections:
\begin{align}
R(x,z) &= \frac{(N^{p+}-N^{n+})+(N^{p-}-N^{n-})}
                {(N^{p+}-N^{n+})-(N^{p-}-N^{n-})} \nonumber \\
       &= \frac{u(x)-d(x)+\bar u(x)-\bar d(x)}
                {u(x)-d(x)-\bar u(x)+\bar d(x)} \cdot
           \frac{4 \, D_u^+(z)+4 \, D_{\bar u}^+(z)
                 -D_d^+(z)- D_{\bar d}^+(z)}
                {4 \, D_u^+(z)-4 \, D_{\bar u}^+(z)
                 -D_d^+(z)+ D_{\bar d}^+(z)}
\ .
\label{eqn:RXZ}
\end{align}
Here, $N^{p+}$ ($N^{p-}$) and $N^{n+}$ ( $N^{n-}$)
correspond to the production processes of
positively (negatively) charged hadrons
from the proton and the neutron respectively.
If the denominator and numerator are integrated over $x$
individually, the Gottfried sum is obtained from the numerator integral,
and the denominator becomes a sum for the valence quarks:
\begin{align}
Q(z) &= \frac{\int dx \, \{ \, (N^{p+}-N^{n+})+(N^{p-}-N^{n-}) \, \} }
             {\int dx \, \{ \, (N^{p+}-N^{n+})-(N^{p-}-N^{n-}) \, \} }
\nonumber \\
       &= 3 \, I_G \, 
           \frac{4 \, D_u^+(z)+4 \, D_{\bar u}^+(z)
                 -D_d^+(z)- D_{\bar d}^+(z)}
                {4 \, D_u^+(z)-4 \, D_{\bar u}^+(z)
                 -D_d^+(z)+ D_{\bar d}^+(z)}
\ \ \ .
\label{eqn:hpm}
\end{align}

According to this equation, if the Gottfried sum rule is violated,
it should appear in the charged-hadron-production asymmetry.
Available EMC data \cite{EMCHPM} are analyzed by using Eq. (\ref{eqn:hpm})
\cite{LMS}. Contributions from pion, kaon, and (anti)proton production 
processes are taken into account in evaluating the fragmentation function,
for example
\begin{equation}
D_u^+ = D_u^{\pi^+} + D_u^{K^+} + D_u^{p} 
\ \ \ .
\end{equation}
Isospin and charge conjugation invariance reduces the number
of fragmentation functions for the pion:
\begin{align}
& D\equiv D_{u}^{\pi^+} = D_{\bar d}^{\pi^+} 
          = D_{d}^{\pi^-} = D_{\bar u}^{\pi^-}  
\ \ \ , \nonumber \\
& \widetilde D \equiv D_{d}^{\pi^+} = D_{\bar u}^{\pi^+} 
                      = D_{u}^{\pi^-} = D_{\bar d}^{\pi^-}  
\ \ \ .
\end{align}
In the kaon case, the reflection symmetry along the V-spin axis
($D_d^{K^+}=D_d^{K^-}$)
is used in addition to the isospin and charge conjugation invariance:
$D^K \equiv D_{\bar u}^{K^-} = D_{u}^{K^+}$,
$\widetilde D^K \equiv D_{u}^{K^-} = D_{\bar u}^{K^+}$,
${\widetilde D}^{' K} \equiv D_{d}^{K^+} = D_{\bar d}^{K^-} 
                             =D_{\bar d}^{K^+} = D_{d}^{K^-}$.
Furthermore, $\widetilde D^K$ and ${\widetilde D}^{' K}$ are assumed
equal. Similar equations are taken for proton and antiproton
production:
$D^p \equiv D_{u}^{p} = D_{d}^{p} 
             = D_{\bar u}^{\bar p} = D_{\bar d}^{\bar p}$,
$\widetilde D^p \equiv D_{\bar u}^{p} = D_{\bar d}^{p} 
             = D_{u}^{\bar p} = D_{d}^{\bar p}$.
Experimental information is provided for these fragmentation
functions. In particular, we use parametrizations fitted to the EMC data
\cite{EMCPARA}:
\begin{alignat}{3}
\frac{\widetilde D(z)}{D(z)} &= \frac{1-z}{1+z} \ \ \ , & \ & 
\nonumber \\
\frac{D^K(z)}{D(z)} &= 0.35\, z+0.15 \ \ , \ \ \ \ \
    &\frac{\widetilde D^K(z)}{D(z)} &= 0.45 \, z \, \frac{1-z}{1+z}
\ \ \ , \nonumber \\
\frac{D^p(z)}{D(z)} &= 0.20 \ \ , \ \ \ \ \
    &\frac{\widetilde D^p(z)}{D(z)} &= 0.12 \, \frac{1-z}{1+z}
\ \ \ .
\end{alignat}
With these experimental parametrizations for the fragmentation functions,
the ratio of the cross sections $Q(z)$ becomes
\begin{equation}
z \, Q^{ch}(z)= 3 \, I_G \, z \, 
            \frac{0.50 \, z^2+3.1 \, z+7.6}{3.2 \, z^2+11 \, z+0.84}
\ \ \ .
\label{eqn:zQ}
\end{equation}

\begin{wrapfigure}{l}{7.2cm}
   \begin{center}
      \mbox{\epsfig{file=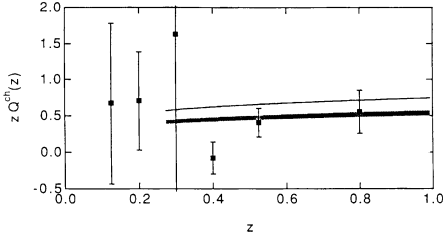,width=6.0cm}}
   \end{center}
 \vspace{-0.6cm}
\caption{\footnotesize Charged-hadron-production ratio $zQ^{ch}(z)$
                       (taken from Ref. {\normalsize\cite{LMS}}).} 
\label{fig:lms}
\end{wrapfigure}
\quad
Experimental data are given for $(d\sigma_N^h/dz)/\sigma_N$,
so that $F_1(x)$ is multiplied in getting $N^{Nh}$.
The experimental data $zQ^{ch}$ obtained in this way 
are compared with Eq. (\ref{eqn:zQ}) in Fig. \ref{fig:lms}.
The upper curve is obtained by assuming the Gottfried sum $I_G=1/3$
in Eq. (\ref{eqn:zQ}). On the other hand, the hatched area is based
on the 1991 NMC result. It is interesting to find the difference
between the two results in the semi-inclusive processes.
However, as it is obvious from the figure, we cannot judge
whether or not the sea is $\bar u/\bar d$ symmetric 
from the data. 

The recent HERMES preliminary data seem to be accurate enough
to find the $\bar u/\bar d$ asymmetry \cite{HERMES}.
The following $\pi^+$ and $\pi^-$ production ratio is related to the 
function $R(x,z)$ for the pion by
\begin{equation}
r(x,z) = \frac{N^{p\pi^-} - N^{n\pi^-}}{N^{p\pi^+} - N^{n\pi^+}}
       = \frac{R_\pi (x,z) -1}{R_\pi (x,z)+1}
\ \ \ .
\end{equation}
The obtained data of $r(x,z)$ in the range $0.1 < x < 0.3$
agree well with the NMC flavor asymmetry, and they are 
significantly different from the symmetric expectation.
The HERMES results will be submitted for publication
in the near future.

\subsection{Neutrino scattering}\label{NEUTRINO}

Neutrino interactions are useful for determining
the valence-quark distributions by taking advantage of parity-violation terms. 
On the other hand, neutrino-induced dimuon data
are used for determining the $s$ and $\bar s$ distributions, so that
the neutrino interactions could be valuable also for determining
the light antiquark distributions $\bar u$ and $\bar d$.
We discuss what kind of cross-section combination is appropriate
for finding the $\bar u-\bar d$ distribution.
The neutrino reaction via the charged current is given by 
the amplitude \cite{RGRBOOK,BPBOOK}  
\begin{equation}
{\mathcal M}(\nu_\ell p\rightarrow \ell X) \, = \,
 \frac{G_F/\sqrt{2}}{1+Q^2/M_W^2}
      \, \bar u(k') \gamma^\mu (1-\gamma_5) u(k)
       < X \, | \, J_\mu^{weak}(0) \, | \, p,\sigma >
\ \ ,
\end{equation}
so that the differential cross section becomes
\begin{equation}
d\sigma = \frac{M}{s-M^2} \, 
            \frac{G_F^2}{(2\pi)^2 \, (1+Q^2/M_W^2)^2} \, 
            \ell^{\mu\nu} \, W_{\mu\nu} \,     
            \frac{d^3 k'}{E'} 
\ \ \ .
\end{equation}
The leptonic tensor is given by
\begin{align}
\ell^{\mu\nu} &=  \overline\sum_{\lambda, \lambda '}
                  \ [ \, \bar u(k',\lambda') \gamma^\mu (1-\gamma_5)
                          u(k,\lambda) \, ]^*
                  \ [ \, \bar u(k',\lambda') \gamma^\nu (1-\gamma_5)
                          u(k,\lambda) \, ]    
\nonumber \\
              &= 2\, ( \, k^\mu {k'} ^\nu + {k'} ^\mu k^\nu 
                      - k\cdot k' g^{\mu\nu}
             + i \varepsilon^{\mu\nu\rho\sigma} k_\rho k'_\sigma \, )
\ \ \ ,
\end{align}
where $\varepsilon^{\mu\nu\rho\sigma}$ is an antisymmetric tensor
with $\varepsilon^{0123}=+1$.
The last term does not appear in the electron or muon scattering
because it is associated with the parity violation in weak interactions.
This term makes it possible to probe new structure in the target
hadron. There exists an antisymmetric term under
the $\mu\leftrightarrow \nu$ exchange in addition to the hadron
tensor in Eq. (\ref{eqn:HADRON}):
\begin{align}
W_{\mu\nu} = - W_1 \left ( g_{\mu\nu} - \frac{q_\mu q_\nu}{q^2} \right )
             &+ \frac{1}{M^2} \, W_2 
                      \left ( p_\mu - \frac{p\cdot q}{q^2} q_\mu \right )
                      \left ( p_\nu - \frac{p\cdot q}{q^2} q_\nu \right )
\nonumber \\
             &- \frac{i}{M} \, W_3 \, \varepsilon_{\mu\nu\rho\sigma} \, 
                        p^\rho q^\sigma
\ \ \ .
\end{align}
The $W_3$ structure function is proportional to the difference
between left- and right-transverse cross sections for the $W$ boson.
With these structure functions, the cross section becomes
\begin{equation}
\frac{d\sigma^\pm}{d\Omega \, dE'} =
             \frac{G_F^2 \, {E'}^2}{2\pi^2 \, (1+Q^2/M_W^2)^2} \,
 \left [ \, 2 \, W_1 \, \sin^2 \frac{\theta}{2}
           + W_2 \, \cos^2 \frac{\theta}{2}
        \mp \frac{E+E'}{M} \, W_3 \, \sin^2 \frac{\theta}{2} \, 
     \right ]
\ ,
\label{eqn:NU123}
\end{equation}
where $\pm$ indicates $W^\pm$ in the reaction.
Structure functions $F_1$, $F_2$, and $F_3$ are defined by
$F_1=MW_1$, $F_2=\nu W_2$, and $F_3=\nu W_3$.

On the other hand, the charged-current process is described 
by neutrino-quark interactions with the current
\begin{align}
J_\mu &= \bar u(x) \, \gamma_\mu \, (1-\gamma_5) \, 
                 [ \, d(x) \, \cos\theta_c +s(x) \, \sin\theta_c \, ]
\nonumber \\
      &+ \bar c(x) \, \gamma_\mu \, (1-\gamma_5) \, 
                 [ \, s(x) \, \cos\theta_c -d(x) \, \sin\theta_c \, ]
\ \ \ .
\end{align}
Comparing a calculated cross section in the parton model
with Eq. (\ref{eqn:NU123}), we express 
the structure functions in terms of quark distributions
\begin{align}
F_1 &= F_2/2x                                     \ \ \ , \nonumber \\
F_2^{\nu p}      &= 2 \, x \, (d+s+\bar u+\bar c) \ \ \ , \nonumber \\
F_2^{\bar\nu p}  &= 2 \, x \, (u+c+\bar d+\bar s) \ \ \ , \nonumber \\
xF_3^{\nu p}     &= 2 \, x \, (d+s-\bar u-\bar c) \ \ \ , \nonumber \\ 
xF_3^{\bar\nu p} &= 2 \, x \, (u+c-\bar d-\bar s)  
\ \ \ .
\label{eqn:NUF123}
\end{align}
For the time being, we discuss only LO contributions without
NLO corrections from the coefficient functions.
Because the antiquarks have negative parity, there are negative signs
in the $F_3$ structure functions.
Combining the $\nu$ and $\bar\nu$ $F_3$ structure functions,
we obtain the valence quark distribution
$(F_3^{\nu p}+F_3^{\bar\nu p})/2=u_v+d_v$ with the assumptions
$s=\bar s$ and $c=\bar c$.
Therefore, it is the advantage of neutrino reactions that
the valence distribution can be determined. 
However, they are also used for studying antiquark distributions.
In fact, neutrino-induced dimuon data enable us to determine 
the $\bar s$ distribution difference from $(\bar u+\bar d)/2$
by assuming the charm-quark production scenario:
$\nu_\mu+s\rightarrow \mu^-+c$, $c\rightarrow s+\mu^+ +\nu_\mu$.
We discuss the possibility of extracting the $\bar u-\bar d$ distribution
from neutrino data.

\begin{wrapfigure}{l}{6.5cm}
   \begin{center}
      \mbox{\epsfig{file=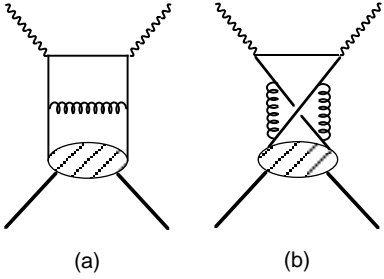,width=5.0cm}}
   \end{center}
 \vspace{-0.4cm}
\caption{\footnotesize Contributions to the $\bar u -\bar d$
                   distribution in Eq. (\ref{eqn:udnu1}) or (\ref{eqn:udnu2})
         (a) from quarks and (b) from antiquarks.}
\label{fig:f13}
\end{wrapfigure}
\quad
From Eq. (\ref{eqn:NUF123}), it is possible to combine
the $F_2$ and $F_3$ structure functions for the proton and
the deuteron in order to get the flavor asymmetry \cite{INDIANA}
\begin{equation}
\bar u-\bar d = 
           \frac{1}{2} \, (F_2^{\nu p}/x - F_3^{\nu p} )
          -\frac{1}{4} \, (F_2^{\nu d}/x - F_3^{\nu d} ) \  ,
\label{eqn:udnu1}
\end{equation}
by neglecting nuclear effects in the deuteron.
It may also be defined in terms of $F_1$ and $F_3$ structure functions
\cite{RS}
\begin{equation}
\bar u-\bar d = \frac{1}{2} \, (F_1^{\nu p} -F_1^{\bar\nu p})
                   -\frac{1}{4} \, (F_3^{\nu p} -F_3^{\bar\nu p})
\ \ \ ,
\label{eqn:udnu2}
\end{equation}
if the $\bar s$ and $\bar c$ distributions can be neglected!
These expressions are correct only in the LO.
However, they are practically convenient because
the charged reactions are easier to be measured experimentally. 
Unfortunately, present data are not accurate enough to be used for finding
the $\bar u-\bar d$ distribution.
To be precise, Eqs. (\ref{eqn:udnu1}) and (\ref{eqn:udnu2}) are
not appropriate in defining the $\bar u-\bar d$ distribution
if NLO effects are taken into account.
For example, the coefficient functions are different in $F_1$ and $F_3$
structure functions, so that 
$ (F_1^{\nu p}-F_1^{\bar\nu p})/2-(F_3^{\nu p}-F_3^{\bar\nu p})/4$
has a contribution, already of the order of $\alpha_s$,
from the quark in Fig. \ref{fig:f13}(a)
in addition to the antiquark in Fig. \ref{fig:f13}(b).
Although it is useful in getting experimental information
on $\bar u-\bar d$ from $F_1$ and $F_3$ in the charged reactions,
it is not a precise definition if higher-order corrections
are taken into account.

A consistent way is to use 
both the charged and neutral current reactions \cite{RS}.
The neutrino-quark interaction via the neutral current is described by 
\cite{RGRBOOK}
\begin{equation}
J_\mu =\sum_i \frac{1}{2} \, \bar q_i(x) \, \gamma_\mu \, 
         \left [ \, g_{Li} (1-\gamma_5) +g_{Ri} (1+\gamma_5) 
                 \, \right ] \, q_i(x)
\ \ \ ,
\end{equation}
where $g_{Li}$ and $g_{Ri}$ are defined by the Weinberg angle $\theta_W$
\begin{alignat}{3}
g_{Li} &= 1-\frac{4}{3} \, \sin^2\theta_W \ \ \ , \ \ \ 
g_{Ri} &=  -\frac{4}{3} \, \sin^2\theta_W \ \ \ \ \ \  
       &\text{for i=u, c} \ \ \ , \nonumber \\
       &= -1+\frac{2}{3} \, \sin^2\theta_W \ \ \ , \ \ \ 
       &=  +\frac{2}{3}  \, \sin^2\theta_W \ \ \ \ \ \ 
       &\text{for i=d, s} \ \ \ .
\end{alignat}
With the neutral-current observables, the distribution $\bar u-\bar d$ can be 
defined by only one type of the structure functions, for example $F_1$.
The neutral-current structure functions $F_1$ for neutrino and
electron reactions become
\begin{multline}
F_1^{\nu p\rightarrow \nu X}  =  g^2 \, \sec^2\theta_W \, 
      \left [ \, \left ( \frac{1}{4} - \frac{2}{3} \sin^2\theta_W
                                     + \frac{8}{9} \sin^4\theta_W \right ) \, 
                \left \{ u(x)+\bar u(x) \right \}
\ \ \ 
\right.  \\
    \left.  + \left ( \frac{1}{4} - \frac{1}{3} \sin^2\theta_W
                                     + \frac{2}{9} \sin^4\theta_W \right ) \, 
                \left \{ d(x)+\bar d(x) \right \} \, \right ]
\ \ \ ,
\end{multline}
\begin{equation}
F_1^{e p\rightarrow e X}  = g^2 \, \sin^2\theta_W \, 
           \left [ \, \frac{4}{9} \, \left \{ u(x)+\bar u(x) \right \} 
                     +\frac{1}{9} \, \left \{ d(x)+\bar d(x) \right \} \, 
           \right ]
\ \ \ .
\end{equation}
According to the definition of Ref. \cite{RS}, it is given by
combining the charged and neutral current $F_1$ structure functions as
\begin{equation}
\widehat F(x) = \frac{1}{2} \, [ \, \widetilde F_1(x)-
                 \{  F_1(x)_{\bar\nu p\rightarrow e^+ X}
                         -F_1(x)_{    \nu p\rightarrow e^- X} \} \, ]
          =\bar u(x)-\bar d(x) 
\label{eqn:DEFUD}
\ \ \ ,
\end{equation}
where $\widetilde F_1(x)$ is defined by the structure functions
in the neutral current reactions:
\begin{equation}
\widetilde F_1(x) = 
    \frac{5}{ \left ( \frac{2}{3} \, sin^2 \theta_{_W} - 
                      \frac{3}{4} \right )
                          sec^2 \theta_{_W} }                 
         F_1(x)_{\nu p \rightarrow \nu X}                   
  - \frac{ 9 \left ( \frac{1}{2} -sin^2 \theta_{_W} 
                   + \frac{11}{9} \, sin^4 \theta_{_W} \right ) }
        { \left ( \frac{2}{3} \, sin^2 \theta_{_W} - \frac{3}{4} \right )
                      sin^2 \theta_{_W} }
         F_1(x)_{ep\rightarrow eX}
\ .
\end{equation}
Because the $\bar s$ and $\bar c$ distributions are neglected
in the above discussion \cite{RS} and in Eq. (\ref{eqn:udnu2}),
it is necessary to subtract out these contributions
by combining Eqs. (\ref{eqn:DEFUD}) and (\ref{eqn:udnu2})
with the deuteron $F_1$ structure functions.

It is impossible to obtain the $\bar u/\bar d$ asymmetry from
present neutrino data. However, we hope that much better data
will enable us to extract the flavor asymmetry distribution.

\subsection{Experiments to find 
            isospin symmetry violation}\label{ISOSPINEX}

We discussed in section \ref{ISOSPIN} that the violation
of the Gottfried sum rule could be due to 
the isospin-symmetry violation instead of the flavor asymmetry.
These two mechanisms cannot be distinguished at this stage. 
The isospin-violation effects are believed to be very small in the
structure functions. However, it is important to confirm this 
common sense experimentally.
Various processes are discussed in the following
for finding the effects of the isospin-symmetry violation in the
antiquark distributions \cite{MSG}.

The $F_2$ structure functions in neutrino interactions are
useful in distinguishing between the two mechanisms \cite{MSG}. 
The difference between proton and neutron structure functions is
\begin{align}
I_{ISV} &= \int \frac{dx}{x} \, \frac{1}{2} \, 
           [ \, F_2^{\nu p}(x) +  F_2^{\bar \nu p}(x)
            -F_2^{\nu n}(x) -  F_2^{\bar \nu n}(x) \, ]
               \nonumber \\
        &= 2 \int dx \, 
           \left [ \, \{ \, \bar u(x)+\bar d(x)+\bar s(x)+\bar c(x) \, \}_p
                     -\{ \, \bar u(x)+\bar d(x)+\bar s(x)+\bar c(x) \, \}_n
                     \, \right ]
\ \ \ .
\end{align}
If the failure of the Gottfried sum is entirely due to the flavor asymmetry,
the integral vanishes $I_{ISV}=0$. On the other hand.
if it is entirely due to the isospin violation
and if the $\bar s$ and $\bar c$ terms can be neglected, 
the integral is $I_{ISV}=-0.336\pm 0.058$.
Because the flavor asymmetry does not contribute, the sum $I_{ISV}$
should give a clue in finding an isospin-violation sign.

Isospin-violation effects on the Drell-Yan processes
are also discussed in Ref. \cite{MSG}.
In the pion scattering case $\pi^\pm A \rightarrow \ell^+ \ell^- X$,
we consider the difference of nuclear cross sections at large $x_\pi$:
\begin{equation}
R_{sea} = \frac{ 4 \, [ \, \sigma(\pi^+ A_1)-\sigma(\pi^+ A_0)\, ]
            + [ \, \sigma(\pi^- A_1)-\sigma(\pi^- A_0) \, ]} 
            {\sigma(\pi^+ A_0)-\sigma(\pi^- A_0)}
\ \ \ .
\end{equation}
The $A_0$ and $A_1$ denote different nuclear species, but
we may choose $A_0$ as an isoscalar nucleus and
$A_1$ as a neutron-excess nucleus.
With the isospin symmetry assumption, it becomes
\begin{equation}
R_{sea} = \frac{10 \, (\epsilon_1 -\epsilon_0) \, (\bar u-\bar d)}
           { u_V + d_V}
\ \ \ ,
\end{equation}
where $\epsilon$ is a neutron excess parameter $\epsilon=N/A-1/2$.
On the other hand, if the sea is flavor symmetric with
the isospin violation, the Drell-Yan ratio becomes
\begin{equation}
R_{sea} = \frac{50 \, (\epsilon_1 -\epsilon_0) \, (\bar q^{\ p}-\bar q^{\ n})}
           { 3 \, (u_V + d_V)}
\ \ \ ,
\end{equation}
where the isospin symmetry is assumed for the valence-quark distributions.
Similar equations are obtained for proton Drell-Yan cross sections.
The p-n cross section asymmetry is given in the isospin symmetry case as
\begin{align}
A_{DY} &= \frac{\sigma^{pp}-\sigma^{pn}}{\sigma^{pp}+\sigma^{pn}}
            \nonumber \\
       &= \frac{(4u_V-d_V)(\bar u-\bar d)+(u_V-d_V)(4\bar u-\bar d)}
            {(4u_V+d_V)(\bar u+\bar d)+(u_V+d_V)(4\bar u+\bar d)}
\ \ \ .
\end{align}
On the other hand, it is given in the isospin-violation case as
\begin{equation}
A_{DY} = \frac {(4u_V-d_V) \, 5 \, (\bar q^{\, p}-\bar q^{\, n})/3
           +(u_V-d_V)(\bar q^{\, p}+8\bar q^{\, n})/3} 
            {9 \, (\sigma^{pp}+\sigma^{pn})}
\ \ \ .
\end{equation}
The details of the Drell-Yan cross sections and the asymmetry
are discussed in section \ref{DRELLYAN}.
From these equations, we find that the Drell-Yan cross sections
could be interpreted in principle either by the flavor asymmetry or
by the isospin violation.
Both effects are taken into account to explain the NA51 result 
in Ref. \cite{STNA51}.
The obtained result indicates that the ratio $\bar u/\bar d$ could be larger
than the NA51 value at the cost of isospin symmetry violation.
However, it is not possible to separate these two contributions clearly. 

It is shown in Ref. \cite{LMS} that the flavor asymmetry 
could be found in semi-inclusive leptoproduction of charged hadrons. 
The number of produced $h$ hadrons in the lepton-nucleon scattering
at Bjorken $x$ and $z=E_h/\nu$ is given by
$N^{Nh}(x,z)=\sum_i e_i^2 \, q_i^N(x) \, D_i^h(z)$,
where $D_i^h$ is the fragmentation function.
The details of the charged-hadron production are discussed
in section \ref{CHARGED}.
The following equation is obtained for finding the flavor asymmetry:
\begin{align}
Q(z) &= \frac{N^{p+} - N^{n+} + N^{p-} - N^{n-}}
         {N^{p+} - N^{n+} - N^{p-} + N^{n-}} 
          \nonumber \\
  &= 3 \, I_G \, \frac{0.50z^2+3.1z+7.6}{3.2z^2+11z+0.84}
\ \ \ ,
\end{align}
where the notations $N^{p+}$, $N^{n+}$, $N^{p-}$, and $N^{n-}$
are the same as those in Eq. (\ref{eqn:RXZ}).
The isospin symmetry is assumed in the above equation. 
If the sea is flavor symmetric and if the isospin symmetry is violated,
the above quantity becomes \cite{MSG}
\begin{align}
Q(z) &=  \frac{ 4 \, [ \, D_u^+(z)+D_{\bar u}^+(z) \, ] \, 
                                 (1-2 \delta\bar q)
                    - [ \, D_d^+(z)+D_{\bar d}^+(z) \, ] \, (1+2\delta\bar q)}
                { 4 \, [ \, D_u^+(z)-D_{\bar u}^+(z) \, ] 
                    - [ \, D_d^+(z)-D_{\bar d}^+(z) \, ]}
          \nonumber \\
     &= 3 \, I_G  \, \frac{0.80z^2+3.37z+7.63}{3.2z^2+11z+0.84}
\ \ \ ,
\end{align}
where $\delta \bar q=\bar q^{\, p}-\bar q^{\, n}$. 
Both expressions have different $z$ dependence, so that we should be
able to distinguish the mechanisms if experimental data are accurate.
At the present stage, charged-hadron production data are not
accurate enough for finding the discrepancy.

The experimental studies would be difficult because
the isospin effects are considered to be very small theoretically.
However, accurate experimental data are desperately
needed in order to shed light on the 
isospin-symmetry violation in the antiquark distributions.

\vfill\eject
\section{Related topics on antiquark distributions}\label{COMMENT}
\setcounter{equation}{0}
\setcounter{figure}{0}
\setcounter{table}{0}

As a topic of flavor asymmetry, we have discussed the 
light-antiquark-distribution difference in the nucleon.
There are other important issues 
on the antiquark distributions.
We briefly comment on related topics.

First, the $\bar u/\bar d$ asymmetry in hyperons could be studied
if charged hyperon beam becomes available in future.
For example, a possibility to find the asymmetry in $\Sigma^\pm$
is investigated in Ref. \cite{AH}.
In a naive quark model, they consist of $\Sigma^+(uus)$
and $\Sigma^-(dds)$.
The Pauli-blocking and meson-cloud models predict 
$\bar d$ excess over $\bar u$ in $\Sigma^+$ and 
$\bar u$ excess over $\bar d$ in $\Sigma^-$.
It is an interesting test of the theoretical models in section 
\ref{THEORY}. 
The Drell-Yan cross section for the $\Sigma^+ p$ reaction
at $y=0$ is given by
\begin{equation}
\sigma^{\Sigma^+ p} \approx \frac{8\pi\alpha^2}{9\sqrt\tau} \, 
          \left [ \, \frac{4}{9} \, \{ \, u_p (x) \bar u_\Sigma (x)
                                       +u_\Sigma (x) \bar u_p (x) \, \}
                    +\frac{1}{9} \, \{ \, u_p (x) \bar d_\Sigma (x)
                                       +s_\Sigma (x) \bar s_p (x) \, \}
                 \right ]
\ ,
\end{equation}
where only valence-sea annihilation terms are retained, and
$q_\Sigma$ denotes the distribution in $\Sigma^+$ 
($q_\Sigma \equiv q_{\Sigma^+}$).
In calculating cross sections for other reactions $\Sigma^- n$,
$\Sigma^+ n$, and $\Sigma^- p$,
we assume isospin symmetry, $u_p=d_n$, $\bar u_p=\bar d_n$,
$u_{\Sigma^+} =d_{\Sigma^-}$, 
$\bar u_{\Sigma^+} =\bar d_{\Sigma^-}$, together with
the assumption $s_{\Sigma^+} =s_{\Sigma^-}$.
From the Drell-Yan cross sections with $\Sigma^\pm$ beams 
on the proton and deuteron targets, we take the ratio
\begin{align}
R(x) &\equiv \frac{(\sigma^{\Sigma^+ p}-\sigma^{\Sigma^- n})
               +\bar r_p(x) \, (\sigma^{\Sigma^- p}-\sigma^{\Sigma^+ n})}
              {(\sigma^{\Sigma^+ p}-\sigma^{\Sigma^+ n})
               +4 \, (\sigma^{\Sigma^- p}-\sigma^{\Sigma^- n})}
\nonumber \\
     &= \frac{\bar r_\Sigma (x) \, [r_p(x)-\bar r_p(x)] 
                                -[1-\bar r_p(x) r_p(x)]}
              {5 \, [r_p(x)-1]}
\ \ \ \text{at $y=0$}
\ \ \ ,
\end{align}
where $r_p\equiv u_p/d_p$, $\bar r_p\equiv \bar u_p/\bar d_p$,
and $\bar r_\Sigma \equiv \bar u_\Sigma /\bar d_\Sigma$.
In this way, if $r_p$ and $\bar r_p$ are known from other experiments,
$\bar r_\Sigma$ could be measured by the hyperon Drell-Yan experiments.

Second, we mentioned the $\bar s$-quark distribution difference
from the $(\bar u+\bar d)/2$. It has been measured
experimentally by the neutrino induced opposite-sign dimuon events.
This topic is studied within the meson-cloud models.
For example, because the pions do not contain the valence $\bar s$ quark,
their contributions to $\bar s$ and $(\bar u+\bar d)/2$ in the proton
are different.
It is particularly important in discussing the size 
of the $\pi NN$ form factor and its relation to nuclear potentials.
This topic is also discussed in the chiral field theory, so that
the interested reader may look at the meson-model papers
in the reference section. 

Third, the difference between $s$ and $\bar s$ is also
important \cite{DIFFS}. Because there is no net strangeness
in the proton, the integral of the difference has to vanish:
$\int dx (s-\bar s)=0$. However, $x$ dependence of
both distributions could be different.
In fact, the proton virtually decays into for example
$K^+(u\bar s) \Lambda(uds)$, $K^+(u\bar s) \Sigma^0 (uds)$, 
and $K^0(d\bar s) \Sigma^+(uus)$. Within the three decay modes,
the valence $\bar s$ is contained in the kaons and
$s$ is in the hyperons.
Because the hyperon masses are larger than those of the kaons,
the $\bar s$ distribution is distributed in the outer side.
It means that the $\bar s$ distribution is softer than
that of the $s$-quark one.
Of course, the $s$ and $\bar s$ distributions should
be dominated by the perturbative contributions.
However, these could be canceled out by taking the
difference $s-\bar s$.
It is impossible to find this kind of small
effect at this stage \cite{CCFR}.  We hope to have much
accurate data in future.

Fourth, flavor asymmetry in polarized antiquark distributions
should become an exciting topic in the near future.
As far as the model is concerned, we have explained
the flavor dependence in section \ref{SPIN}.
However, because we do not have a variety of polarized data
at this stage, it is very difficult to find the difference between
$\bar u$, $\bar d$, and $\bar s$ distributions
from experimental data.
In any case, there is an attempt to study the flavor decomposition
by including semi-inclusive data in Ref. \cite{BT}.
Future experimental programs for the polarized flavor asymmetry
are for example the RHIC-SPIN \cite{RHICSPIN}
and the Common Muon and Proton Apparatus
for Structure and Spectroscopy (COMPASS) \cite{COMPASS}.
In the similar way with the unpolarized case in
section \ref{W}, the $W^\pm$ production measurements
by the RHIC-SPIN collaboration should enable us to find 
$\Delta \bar u$ and $\Delta \bar d$ distributions.
The strange polarization and other polarized valence and sea 
distributions will be measured in semi-inclusive reactions
by the COMPASS collaboration.
Much progress is expected on the flavor dependence
of the polarized antiquark distributions in the next several years.

Fifth, there is a similar sum rule to the Gottfried 
in the spin-dependent structure function $b_1$ for spin-one hadrons.
This new structure function is related to quadrupole structure
of the spin-one hadrons. 
Its sum rule was proposed in Ref. \cite{CKSUM} as
\begin{equation}
\int dx b_1 (x) = \lim_{t\rightarrow 0}
              - \frac{5}{3} \frac {t}{4M^2} F_Q (t)
           + \delta Q_{sea} 
\ \ \ ,
\label{eqn:b1}
\end{equation}
where $F_Q (t=0)$ is the quadrupole moment in the unit
of $e/M^2$ for a spin-one hadron with the mass $M$.
The second term $ \delta Q_{sea}$ is the sea-quark tensor
polarization defined, for example, 
$\delta Q_{sea}^D= \int dx  
          [8 \delta \bar u (x) +2 \delta \bar d (x)
           +\delta s (x) +\delta \bar s (x)]^D /9$
for the deuteron.
The distribution $\delta q$ is given by
$\delta q=[q^0-(q^{+1}+q^{-1})/2]/2$, 
where the superscript indicates the hadron helicity
in an infinite momentum frame.
The Gottfried sum 1/3 corresponds to the first term
$\lim_{t\rightarrow 0}
              - \frac{5}{3} \frac {t}{4M^2} F_Q (t) =0$.
Because the valence-quark number depends on flavor,
the finite sum 1/3 is obtained in the Gottfried.
However, it does not depend on spin, so that  
the first term vanishes in the $b_1$ case.
The second term in Eq. (\ref{eqn:b1}) corresponds 
to $\int dx (\bar u-\bar d)$ in Eq. (\ref{eqn:GINT}).
Therefore, a deviation from the sum $\int dx b_1(x)=0$
should suggest the sea-quark tensor polarization
as the Gottfried sum rule violation suggested
the finite $\bar u-\bar d$ distribution.
Recent studies indicate that the diffractive-nuclear-shadowing
and pion-excess mechanisms produce a tensor polarization, which
leads to violation of the $b_1$ sum rule \cite{NSEPW}.

\vfill\eject
\section{{\bf Summary and outlook}}\label{SUMMARY}
\setcounter{equation}{0}
\setcounter{figure}{0}
\setcounter{table}{0}

The light antiquark distributions $\bar u$ and $\bar d$
had been assumed equal for a long time.
The Gottfried sum rule can be derived with this assumption.
Even though there were some experimental efforts to test the sum
rule and the flavor asymmetry $\bar u-\bar d$, it was not possible
to draw a reliable conclusion.
However, recent accurate experimental measurements made it possible to
find the difference between the $\bar u$ and $\bar d$ distributions.
The NMC finding of the Gottfried-sum-rule violation and the $\bar u/\bar d$
asymmetry motivated us to study theoretical mechanisms and 
different experimental possibilities. The flavor asymmetry is now
confirmed by the NA51 Drell-Yan experiment, and it is also suggested
by the preliminary HERMES and E866 data.
On the other hand, future experimental facilities should be able
to pin down the $\bar u$ and $\bar d$ distributions.
For example, Drell-Yan and $W$-production measurements 
at RHIC should be very useful.
In testing the Gottfried sum itself, we need to accelerate
the deuteron at HERA. 

On the theoretical side, the perturbative corrections to the sum
are very small. Therefore, the violation should be explained
by a nonperturbative mechanism.
Within the proposed models, the mesonic model is a strong candidate
in the sense that it can explain the major part of the violation.
The Pauli blocking effect is smaller than that of the mesonic model
according to the naive counting estimate. 
Furthermore, if the antisymmetrization
is considered in addition to the Pauli principle, 
both mechanisms could produce a $\bar u$ excess over $\bar d$.
Because there are other theoretical candidates as explained in this paper,
we should investigate more details of these models in order to 
find a correct explanation.
The flavor asymmetry studies provide us an important clue
to understand nonperturbative aspects of
nucleon substructure. Future experimental and theoretical
efforts on this topic are important for understanding
internal structure of hadrons.

\vspace{0.7cm}
\section*{{\bf Acknowledgments}}
\addcontentsline{toc}{section}{\protect\numberline{\S}{Acknowledgments}}

This research was partly supported by the Grant-in-Aid for
Scientific Research from the Japanese Ministry of Education,
Science, and Culture under the contract number 06640406.
S. K. thanks the Institute for Nuclear Theory at the University of
Washington for its hospitality and the US Department of Energy
for partial support.
He thanks the Elsevier Science, A. S. Ito, K. F. Liu,
W. Melnitchouk, J. C. Peng, and W. J. Stirling
for permitting him to quote some figures directly from their
publications.
He thanks H.-L. Yu for his hospitality in staying in the
Academia Sinica of Taiwan, where this manuscript is partially
written.

\vfill\eject
\addcontentsline{toc}{section}{\protect\numberline{\S}{References}}


\end{document}